\documentclass[11pt,3p,review,authoryear]{elsarticle}

\journal{International Journal of Forecasting}
\biboptions{longnamesfirst}

\usepackage{amssymb}
\usepackage{amsfonts}
\usepackage{afterpage}
\usepackage{dsfont}
\usepackage{bbm}
\usepackage{amsmath}
\usepackage{graphicx}
\usepackage{amsthm}
\usepackage{array}
\usepackage{arydshln}
\usepackage{verbatim}
\usepackage{pifont}
\usepackage{eurosym}
\usepackage{multirow}							
\usepackage{arydshln}							
\usepackage{adjustbox}
\usepackage{dsfont}
\usepackage{subcaption}
\usepackage{float}
\usepackage{algorithm,algcompatible}
\usepackage[section]{placeins} 
\usepackage{longtable} 
\usepackage{booktabs} 
\usepackage{rotating, hyperref} 
\makeatletter
\AtBeginDocument{%
 \expandafter\renewcommand\expandafter\subsection\expandafter{%
   \expandafter\@fb@secFB\subsection
 }%
 }
\makeatother

\usepackage[font=footnotesize]{caption}

\algnewcommand\INPUT{\item[\textbf{Input:}]}%
\algnewcommand\OUTPUT{\item[\textbf{Output:}]}%

\makeatletter
\def\@biblabel#1{}
\makeatother

\makeatletter
\let\@fnsymbol\@arabic
\makeatother

\usepackage{color}
\definecolor{kit-blue}{RGB}{70,100,170}

\begin{document}

\begin{frontmatter}

\title{
Predicting Value at Risk for Cryptocurrencies With Generalized Random Forests  
}

\author[aa]{Rebekka Buse}
\address[aa]{Karlsruhe Institute of Technology}
\author[bb]{Konstantin G\"{o}rgen }
\address[bb]{Allianz Suisse, Switzerland}
\author[aa]{Melanie Schienle\corref{cor}}

\cortext[cor]{Corresponding author}
 \ead{schienle@kit.edu}

\begin{abstract}
We study the prediction of Value at Risk (VaR) for cryptocurrencies. In contrast to classic assets, returns of cryptocurrencies are often highly volatile and characterized by large fluctuations around single events. Analyzing a comprehensive set of 105 major cryptocurrencies, we show that Generalized Random Forests (GRF) \citep{Athey2019} adapted to quantile prediction have superior performance over other established methods such as quantile regression, GARCH-type and CAViaR models. This advantage is especially pronounced in unstable times and for classes of highly-volatile cryptocurrencies. Furthermore, we identify important predictors during such times and show their influence on forecasting over time. Moreover, a comprehensive simulation study also indicates that the GRF methodology is at least on par with existing methods in VaR predictions for standard types of financial returns and clearly superior in the cryptocurrency setup.
\end{abstract}

\begin{keyword}
Generalized Random Forests \sep Value at Risk \sep Quantile Prediction \sep Backtesting \sep Cryptocurrencies 
\end{keyword}

\footnotetext[2]{We thank the audiences of the HKMetrics workshops and the COMPSTAT Conference for valuable comments and acknowledge excellent research assistance by Niklas Korn and Jonas Meirer.}
\footnotetext[3]{The numerical results presented in this manuscript were reproduced on the 13th of December 2024.}

\end{frontmatter}


\section{Introduction}\label{Sec:Introduction}
Cryptocurrencies are an important and rising part of today’s digital economy. Currently, the market capitalization of the top 10 cryptocurrencies in the world is close to \$2 trillion and growing\footnote{See e.g. \url{https://coinmarketcap.com/charts/}; accessed at 22nd March 2022.}.
The use of cryptocurrencies in terms of daily volume exploded from 2016 to 2018, which not only attracts individuals but also business users such as hedge funds or merchants as well as long-term investors such as crypto-focused and traditional investment funds \citep{vigliotti2020rise}. The crypto asset market, however, remains highly volatile where e.g. an investment in Bitcoin in 2013 would have seen a return of roughly $20,000\%$ in 2017, but an investment in 2017 would have led to a performance of -75\% in 2019\footnotemark[1]. Consequently, there is a need to predict and monitor the risks associated with cryptocurrencies.
To address this, we find that classic approaches such as historical simulation, GARCH-type, or CaViaR methods are too restrictive. More general non-linear methods provide more flexibility to account for such non-standard time series behaviour that might be attributed to a large extend to speculators.
 
In this paper, we propose a novel flexible way for out-of-sample prediction of the Value at Risk for cryptocurrencies. We use a quantile version of Generalized Random Forests \citep[GRF,~see][]{Athey2019}, which builds upon mean random forests \citep{Breiman2001} now tailored to quantiles. This framework shows to be especially promising when dealing with more volatile classes of cryptocurrencies due to the non-linear structure of their returns. In a comprehensive out-of-sample scenario using more than 100 of the largest cryptocurrencies, GRF outperforms other established methods such as CAViaR \citep{Engle2004}, quantile regression \citep{koenker2001quantile} or GARCH-models \citep{Bollerslev1986,GLOSTEN_1993} over a rolling window, particularly in unstable times. This can be attributed to the non-parametric approach of random forests that is flexible and adaptable considering important factors and non-linearity. We further analyze performance in different important subperiods, consider different classes of cryptocurrencies, and employ different sets of covariates with the forest-based methods and the benchmark procedures.

Previous studies have confirmed that there exist speculative bubbles \citep{CHEAH2015,Hafner2020}, and we find that our approach assesses risks especially well during such times. Moreover, we account for a large number of covariates that describe volatility, liquidity, and supply \citep{Liu2020}. It can be seen that variable importance differs substantially depending on time, where long-term measures of standard deviation, that are an important predictor in stable times, are not relevant predictors for VaR in unstable, volatile times. Furthermore, only few of the additional covariates beside lagged standard deviations and lagged returns are relevant. We find that for other, less volatile classes of cryptocurrencies such as stablecoins, especially GJR-GARCH models and quantile regression can compete with GRF.
A simulation study also highlights that the proposed GRF methodology is at least on par in prediction of VaR also for standard-type financial returns with clear advantages in the crypto-currency type case.

Our paper contributes to the growing literature on cryptocurrencies. Analyses performed in the past include GARCH models \citep{chu2017garch} as well as ARMA-GARCH models \citep{platanakis2019portfolio}, approaches using RiskMetrics \citep{Pafka2001} and GAS-models \citep{liu2020forecasting}, application of extreme value theory \citep{gkillas2018application}, vine copula-based approaches \citep{trucios2020value}, Markov-Switching GARCH models \citep{maciel2020cryptocurrencies}, non-causal autoregressive models \citep{hencic2015noncausal} and also some machine learning-based approaches \citep[see~e.g.][]{Takeda2008}. Additionally, cryptocurrencies can be used for diversification in investment strategies with other, traditional assets \citep[see~e.g.][]{Trimborn2020,Petukhina2021}, as the correlation between them and more established assets tends to be low \citep{elendner2018cross,platanakis2019portfolio}. This again poses the question of assessing the risks of cryptocurrencies, where new methods of addressing the above mentioned challenges need to be explored.

The paper is structured as follows. Section \ref{Sec:Data} presents the underlying data and cryptocurrencies with descriptive statistics and standard Box-Jenkins time-series checks. In Section \ref{Sec:grf_method}, we introduce the main random forest-type techniques for conditional quantiles and present the employed evaluation tests and framework. The comprehensive simulation study in Section \ref{Sec:Simulation} demonstrates the performance of the different methods under various data generating processes. The empirical results are contained in Section \ref{Sec:Results}, where we present aggregate prediction results for all currencies in coverage performance (Section \ref{sec:crypto_results_backetesting}) as well as in pairwise comparison tests (Section \ref{sec:crypto_results_cpa}). In Section \ref{sec:crypto_results_case_study} we provide extensions to the main results. The section comprises a detailed analysis of representative important currencies, a study of the importance of specific factors and results for the most recent but shorter post-pandemic period. Finally, we conclude in Section \ref{Sec:Conclusion}. All data and replication materials can be found in the Github repository \url{https://github.com/KITmetricslab/crypto-VaR-predictions}.

\section{Data}\label{Sec:Data}
We use daily log-returns of 105 of the largest cryptocurrencies\footnote{The data was obtained on 8th April 2024 from \url{https://docs.coinmetrics.io/} using the community data set, which can be downloaded from a public Github repository at \url{https://github.com/coinmetrics/data/}.} from coinmetrics by market capitalization\footnote{All currencies have a maximum market capitalization of more than 15 million USD each.} in US-Dollar (USD) at time of retrieval, in the period from 07/2010 to 04/2024. The time to define a day is based on Coordinated Universal Time (UTC). The coinmetrics data includes spot-market information from 30 different exchanges, such as Binance, ZB.COM, FTX, OKX, Coinbase, KuCoin, or Kraken\footnote{See \url{https://docs.coinmetrics.io/exchanges/all-exchanges} for an overview of all exchanges included.}. Depending on the currency, the number of available observations varies between 261 and 5012. \footnote{For the start of the availability period, we refer to the first recorded trade date on the exchanges.} 

The distribution of the data over time and assets is summarized in Table \ref{Tab:Desc}. In the cross-section, cryptocurrency returns are characterized by large positive and negative spikes and substantial excess-kurtosis in over 75\% of all assets. Furthermore, they also show substantial skewness, indicating asymmetry in the distribution of log-returns and heterogeneity of in the cross-section of considered assets. Remarkably, both positive and negative skewness occurs in the cross-section, and skewness can be classified in more than 10\% of the considered assets as high.

We do not detect any stochastic non-stationarities in the data, which is supported by Augmented Dickey-Fuller (ADF) tests and Kwiatkowski-Phillips-Schmidt-Shin (KPSS) tests \citep{Kwiatkowski1992}. With Alpha Finance Lab (alpha), Polymath (poly), and Synthetix (snx), KPSS tests against level stationarity seem slightly significant, while trend KPSS tests and ADF tests suggest stationarity. With Algorand (algo), Binance Coin (bnb), Curve DAO Token (crv), FTX Token (ftt), Internet Computer (icp), Aave (lend), OMG Network (omg), SushiSwap (sushi), and Monero (xmr), KPSS tests against trend stationarity are slightly significant, while ADF tests and level KPSS tests again suggest stationarity. All results of the stationarity tests can be found in Table \ref{tab:crypto_tests} in the appendix.

\begin{table}[ht]
\centering
\caption{Descriptive Statistics of Log-Returns of Cryptocurrencies}
\label{Tab:Desc}
\resizebox{\textwidth}{!}{%
\begin{tabular}{rrrrrrrrrrrr}
  \hline
 {} &    Min &     1\% &     5\% &  Median &    95\% &    99\% &    Max &  Skewness &  Excess-Kurtosis &  Std. Dev. &  Observations \\
        \hline
Min\% & -1.402 & -0.437 & -0.169 & -0.006 & 0.001 & 0.002 & 0.008 & -4.030 & 3.214 & 0.001 & 260.00 \\ 
  1\% & -1.378 & -0.306 & -0.158 & -0.004 & 0.001 & 0.002 & 0.015 & -3.852 & 3.412 & 0.001 & 352.40 \\ 
  5\% & -1.104 & -0.246 & -0.128 & -0.003 & 0.002 & 0.006 & 0.044 & -1.214 & 4.950 & 0.003 & 1044.00 \\ 
  25\% & -0.635 & -0.189 & -0.105 & -0.001 & 0.078 & 0.160 & 0.398 & -0.207 & 7.968 & 0.056 & 1397.75 \\ 
  50\% & -0.506 & -0.168 & -0.094 & -0.000 & 0.094 & 0.196 & 0.505 & 0.364 & 14.195 & 0.066 & 2024.00 \\ 
  75\% & -0.378 & -0.146 & -0.081 & 0.001 & 0.106 & 0.215 & 0.738 & 1.094 & 26.293 & 0.074 & 2416.25 \\ 
  95\% & -0.041 & -0.007 & -0.002 & 0.001 & 0.137 & 0.316 & 1.374 & 2.349 & 120.236 & 0.094 & 3591.25 \\ 
  99\% & -0.019 & -0.002 & -0.001 & 0.002 & 0.196 & 0.466 & 1.431 & 3.345 & 320.802 & 0.135 & 4009.15 \\ 
  Max\% & -0.006 & -0.002 & -0.001 & 0.002 & 0.202 & 0.541 & 1.462 & 3.619 & 326.429 & 0.141 & 5012.00 \\ 
   \hline
\end{tabular}
}
\caption*{\scriptsize For each cryptocurrency in the sample, we calculate the sample statistics indicated by the column headers. The rows show the quantiles of each of these statistics in the cross-section thus marking key points of the cross-sectional distribution of the respective statistics in each column. For example, over the cross-section of cryptocurrencies, the median number of available time series observations is 2024.}
\end{table}

\begin{figure}[htb]
    \centering
    \includegraphics[width=\textwidth]{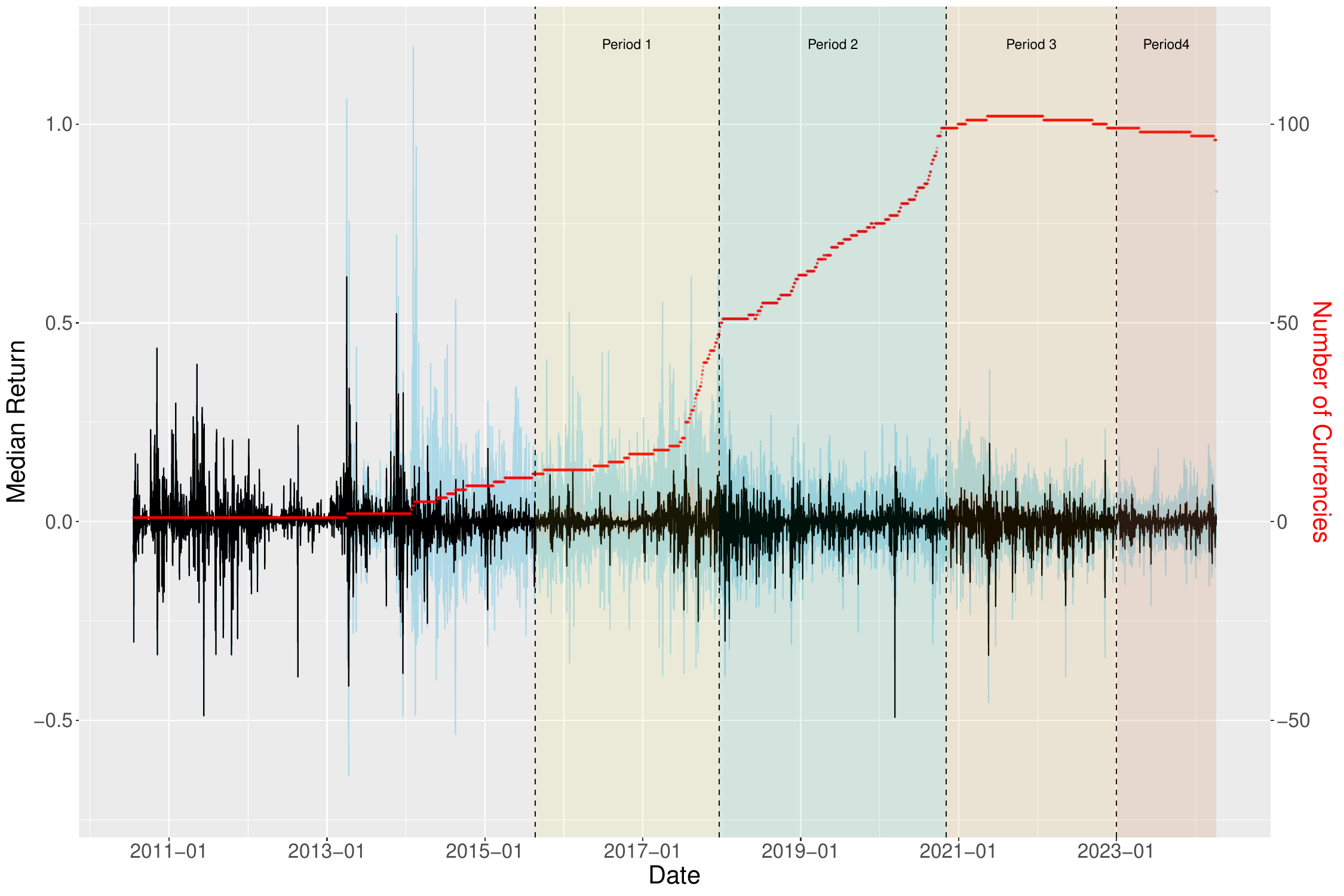}
    \caption{For each point in time, the figure displays pointwise median returns over all currencies in black with $5\%$ and $95\%$ sample quantiles in blue. The number of available currencies with non-zero returns is indicated in red. The shaded areas in yellow, green, and orange correspond to the three periods of the main study in Section \ref{Sec:Results}. The substantially shorter red shaded time period is analyzed in the Section Extensions.}
    \label{fig:median_rets}
\end{figure}

Figure \ref{fig:median_rets} illustrates the median returns (black) over all cryptocurrencies by date. We can see that in the beginning of the time period, only one currency, namely Bitcoin, was present in the data set. From 2014, we see an incremental increase (red line), while there is a jump up in 2017 and an consecutively faster increase in available cryptocurrencies. 
 
Given the hugely dynamic evolution of the cryptocurrency market, we distinguish between  three main systematically different periods in terms of actively traded currencies and market capitalization and treat them separately. The first period ranges from August 2015 (2015-08-22) where the crypto market is still rising with a relatively small number of available currencies to the end of 2017 (2017-12-21). It is characterized by a few, rapidly changing currencies and extreme returns and volatilities. The subsequent second period is less extreme in returns and shows a strong increase in median market caps with a few dozens of market components. It lasts until November 2020 (2020-11-05) with an increasing number of currencies capturing the beginning of the COVID-19 pandemic. The third period ranges from November 2020 until December 2022 (2022-12-31) with a large number of cryptocurrencies containing more than four times the number of currencies in comparison to the second period. The third period is also marked by the endemic stage of the COVID-19 pandemic. In an extension in Subsection \ref{subsec:period4} we also study the substantially shorter Period 4 (2023-01-01 until 2024-04-06) that is characterized by lower cryptocurrency volatility in a post pandemic setting influenced by the political and economic consequences of the war in Ukraine.\footnote{The specific cutoff-dates for the periods account for the required training time in model-fitting for each currency and the emergence of new assets making sure that we capture a maximum number of cryptocurrencies in each time period.}

To capture the autoregressive and time series structure of the asset processes, we include the classic one period lagged return in our analysis and the lagged 3, 7, 30, and 60 day return standard deviation. Additionally, we employ information specific to each cryptocurrency in 7 external covariates. The external covariates are the number of unique active daily addresses (Active\_Users), the number of unique addresses that hold any amount of native units of that currency or at least 10 or 100 USD equivalent (Total\_Users, Total\_Users\_USD10,Total\_Users\_USD100), the supply equality ratio (SER), i.e. the ratio of supply held by addresses with less than $1/10^7$ of the current supply to the top 1\% of addresses with the highest current supply, the number of initiated transactions (Transactions), and the velocity of supply in the current year (Velocity), which describes the the ratio of current supply to the sum of the value transferred in the last year. See also Table \ref{tab:crypto_cov_descr} in the appendix for details on the covariates.

\begin{table}[ht]
\centering
\caption{Summary of Covariates for Different Time Periods} 
\label{tab:sum_covariates}
\resizebox{\textwidth}{!}{
\begin{tabular}{lrrrrrrrrrrrr|r}
  \hline
Quantile & Ret & Active\_Users & User & User$>100\$$ & User$>10\$$ & SER & Transactions & Velocity & sd\_3 & sd\_7 & sd\_30 & sd\_60 & Volume \\ 
  \hline

    \multicolumn{14}{l}{\large\textit{Period 1: 5 Currencies}}\\
  5\% & -0.090 & 9059 & 266499 & 18640 & 40353 & 0.002 & 3781 & 6.898 & 0.007 & 0.014 & 0.022 & 0.025 & 21 \\  
  Median & -0.000 & 156216 & 4934148 & 670847 & 1482211 & 0.019 & 56929 & 32.161 & 0.034 & 0.041 & 0.049 & 0.051 & 9155 \\
  95\% & 0.097 & 333644 & 12730286 & 3867047 & 6670343 & 0.036 & 179828 & 110.209 & 0.130 & 0.125 & 0.118 & 0.117 & 188918 \\
  \multicolumn{14}{l}{\large\textit{Period 2: 14 Currencies}}\\
  5\% & -0.085 & 5147 & 137498 & 11914 & 25839 & 0.001 & 19971 & 3.823 & 0.007 & 0.014 & 0.022 & 0.025 & 109 \\ 
  Median & -0.000 & 96620 & 5585463 & 483593 & 1113547 & 0.013 & 200836 & 17.823 & 0.034 & 0.040 & 0.047 & 0.049 & 9458 \\
  95\% & 0.092 & 209115 & 15533203 & 2831599 & 6082908 & 0.055 & 745090 & 83.331 & 0.123 & 0.118 & 0.112 & 0.111 & 107684 \\ 
     \multicolumn{14}{l}{\large\textit{Period 3: 64 Currencies}}\\
  5\% & -0.084 & 3103 & 731930 & 20990 & 57769 & 0.003 & 31838 & 4.014 & 0.008 & 0.015 & 0.023 & 0.026 & 255 \\ 
  Median & -0.000 & 48931 & 2302405 & 201705 & 481099 & 0.008 & 158681 & 16.833 & 0.034 & 0.040 & 0.046 & 0.049 & 4368 \\ 
  95\% & 0.085 & 118964 & 5377667 & 870485 & 1911853 & 0.020 & 461557 & 68.267 & 0.115 & 0.110 & 0.105 & 0.101 & 30618 \\
     \multicolumn{14}{l}{\large\textit{Period 4: 83 Currencies}}\\
  5\% &  -0.085 & 2497 & 559538 & 18741 & 48212 & 0.002 & 24963 & 5.183 & 0.008 & 0.016 & 0.023 & 0.027 & 395 \\
  Median & -0.000 & 31607 & 1795653 & 132859 & 293588 & 0.006 & 120609 & 15.492 & 0.035 & 0.041 & 0.047 & 0.050 & 3408 \\
  95\% & 0.086 & 79546 & 3793313 & 588142 & 1245011 & 0.014 & 355595 & 61.036 & 0.114 & 0.110 & 0.105 & 0.100 & 25221 \\ 
     \multicolumn{14}{l}{\large\textit{Full Data: 106 Currencies}}\\
  5\% & -0.083 & 1980 & 434899 & 15289 & 38671 & 0.002 & 19505 & 4.926 & 0.008 & 0.015 & 0.022 & 0.026 & 406 \\
  Median & -0.000 & 30350 & 1525012 & 129358 & 304196 & 0.005 & 105990 & 15.432 & 0.034 & 0.040 & 0.046 & 0.049 & 3538 \\
  95\% & 0.085 & 75129 & 3452320 & 522625 & 1145034 & 0.013 & 311205 & 79.758 & 0.114 & 0.111 & 0.105 & 0.100 & 21640 \\
   \hline
\end{tabular}
}
\caption*{\scriptsize The table displays summary statistics for all included covariates in the different time periods. Please see Table \ref{tab:crypto_cov_descr} for the definitions. Displayed values are means of quantiles over all assets contained in the specific time period. User corresponds to the variable Total\_Users, User$>100\$$ marks Total\_Users\_USD100, and User$>10\$$ displays the variable Total\_Users\_USD10. Additionally in the last column volume, we provide information on the  market cap in Mio. USD.}
\end{table}

Table \ref{tab:sum_covariates} gives an overview of the employed covariates and their values in the different time periods. We can see that in period 1, we have the most extreme returns on average as well as the most extreme lagged standard deviations. This is not surprising looking at the first period (853 days), which arguably marks the most volatile period, with many new currencies being created, as well as the second longest period. In the following period (1050 days), the average median market cap reaches a high as well as the number of users invested in the currencies, indicating that the market is growing while stabilizing more. This is followed by a sharp drop in the market cap for the last, shortest period (500 days), which starts at the beginning of the COVID-19 pandemic. There, the number of active users as well as the SER decreases, indicating that more smaller addresses are pushed off the market, while it is the period with the most currencies.


\section{Methodology}\label{Sec:grf_method}
For the VaR-prediction of cryptocurrencies, we advocate the use of tailored non-linear machine learning based techniques. In this way, we intend to accommodate the documented large share of speculation \citep{Ghysels2019,Baur2018,Selmi2018,Glaser2014} and resulting frequent changes in unconditional volatility which make predictions in this market peculiar. In particular, we focus on generalized random forest methods that are tailored for conditional quantiles of returns and thus allow to forecast the VaR. The flexible but interpretable non-linearity of the approach allows for a direct comparison to standard linear and (G)ARCH-type models. We also argue that the difference in forecasting performance can be employed to detect periods of bubbles and extensive speculation. 

Recall that for daily log returns $r_t$ the $VaR_t$ at level $\alpha\in (0,1)$ conditional on some covariates $x_{t-1}$ is defined as  
\begin{equation}\label{var}
    VaR^\alpha_t(x_{t-1}) = \sup_{r_t}\left(F(r_t|x_{t-1})< \alpha \right) \ ,
\end{equation}
where $F$ marks the distribution of $r_t$ conditional on $x_{t-1}$. Generally, the conditioning variables could consist of past lagged returns, standard deviations, but also of external (market) information or other assets. We employ these as covariates that are explained in Section \ref{Sec:Data}. 

We propose a specific random forest-based technique which directly models the conditional VaR in~\eqref{var} and evaluate its prediction performance relative to a different random forest-type benchmark and standard parametric time series models. We closely build on existing techniques by  \citet{Athey2019} and \cite{Meinshausen2006} that were suggested and have so far only been used in a cross-sectional setting, where performance of both technologies has been similar. For forecasting financial time series and in particular time series of cryptocurrencies, however, we show by comprehensive simulations and empirically that a tailored specification of one approach can yield advantages over existing models while the other procedure yields only less favorable results. Both considered random forest methodologies rely on the classic random forest idea \citep{Breiman2001} that produces an ensemble of (decorrelated) decision trees \citep[see~e.g.][]{Hastie2009} for the mean of $r_t$. In a decision tree, each outcome $r_t $ is sorted into leafs of the tree by binary splits. These splits are performed based on different $x_{t-1}$ components falling above or below specific adaptive threshold values that need to be calculated, for example by the Gini Impurity or MSE-splitting in Classification and Regression Trees (CART) \citep{breiman1984classification}, or using other criteria. Finally, the prediction for a new $r_{t}$ is a weighted version of each tree prediction.

In our suggested methodology we rely on the generalized random forest (GRF) from \citet{Athey2019}. In this technique, the random forest split criterion is adapted to mimic the task of quantile regression rather than minimizing a standard mean squared loss criterion for mean regression tasks. Intuitively, the splits in each leaf are conducted by minimizing the Gini-loss, which separates the returns $r_t$ as best as possible at different quantiles. To transform the minimizing problem in the splits into a classification task, the response variable $r_t$ is transformed in each split to obtain pseudo-outcomes $\rho_t=\sum_{m=1}^M 1\{r_t>\theta_{qm} \}$, where $1\{r_t>\theta_{qm} \}$ is 1 for all $t$ with $r_t>\theta_{qm}$ and zero otherwise and $\Theta=(\theta_{q1},\dots,\theta_{qM})$ describe a set of $M$ pilot-quantiles of $r_t$ in the parent node. These quantiles with levels $\tau=q_1,\dots,q_M$ are then used to calibrate the split. In \citet{Athey2019}, this is formally motivated by moment conditions and gradient approximations, but practically, $r_t$ is relabeled to a nominal scale depending on the largest quantile it does not exceed. In a final step, the optimal split on a variable component $p$ of $x_{t-1}$ and $j=1,\dots,J$ observations in the parent node is then based on minimizing the above-mentioned Gini impurity criterion for classification. For a separation into two possible leaf sets $v=l,r$, the Gini impurity for one leaf $v$ is $G_p^{v}=1-\sum_{k=1}^K p_{k,v}^2$, where $p_{k,v}=\sum_{j=1}^J 1\{\rho_j=k \textit{ and } \rho_j \in v\} / \vert v\vert$ is the proportion of $\rho_j$ in group $v$ with value $k=1,\dots,K$. The full loss is then an average weighted by leaf size, yielding 
\begin{align} \label{eq:gini}
    G_p=(\vert l \vert G_p^{l} + \vert r \vert G_p^{r})/(\vert l \vert +\vert r \vert) \ .
\end{align}
In our tailoring of the GRF for cryptocurrencies we feature the Gini-loss since it is fast and in general produces purer nodes than for example using entropy as a splitting criterion \citep[see~e.g][]{breiman1996}. This can be particularly helpful when dealing with changing variance (and thus time-varying quantiles) of returns, where we would like to detect single extreme events. For our specific case of $\alpha=0.05$, this implies that values larger than $\theta_{0.05}$ in the parent node are given the value $1$, while others are $0$. Algorithm \ref{Algorithm_GRF} briefly summarizes the tree building algorithm from \citet{Athey2019} for the quantile version of GRF, where the main differences with regard to the splitting regime in comparison to a classic CART occur in every step the tree is grown.

As a benchmark, we employ the quantile regression forest (QRF) based on \cite{Meinshausen2006}. This random forest, however, obtains the target $\alpha$-quantile by a  weighted average of the empirical CDF $\hat{F}(r_{t}\rvert x_{t-1})=E[1_{\{r_{t}\}|x_{t-1}}]$ rather than from averaging minimiziers of the direct moment condition which can suffer from robustness problems. Intuitively, log returns $r_t$ that have similar $x_{t-1}$ in comparison to a new observation $x_{v}$ receive higher weight in the empirical CDF. Similarity weights $w_t(x_{v})$ are measured as the relative frequency on how often $x_{v}$ falls in the same terminal leaf as $x_{t-1}$, for $t=1,\dots,T$, and averaged over all trees for each $x_{t-1}$. This last step was originally introduced by \citet{Meinshausen2006} for random forests.  QRF employs the same splitting regime as the original CART random forest and therefore does not account explicitly for situations where the variance and therefore the quantile changes, as splits are conducted based on a mean-squared error criterion. Since such volatility changes are a feature of our cryptocurrency data, we expect GRF to perform better than QRF, but still include both in the analysis to see potential differences in predictions. Furthermore, GRF uses so-called ``honest'' trees, meaning that different data (usually the subsampled data for each tree is split again in half) is used for building and ``filling'' each of the trees with values.

For both forest type techniques we use the automatic simple cross-validation choice for the hyperparameters based on a pre-training set with $K=1$ and $\tau=\alpha$ in the GRF case of Algorithm \ref{Algorithm_GRF} and the automatic build-in procedure for QRF in the respective R package QRF. In both cases, the minimum node size is set to 20.

As benchmarks we further include two types of standard time series methods. We use the CAViaR (CAV) methodology by \citet{Engle2004} and standard quantile regression \citep{koenker2001quantile}. Both make use of quantile regression (QR) techniques \citep{Koenker1978} that do not minimize the squared error as in ordinary regression, but use the check function $\rho_{\alpha}(u)=u\left(\alpha-1\{u\leq 0\}\right)$ to minimize $L_{\alpha}(f_{\alpha}(\cdot),x_t)=\sum_{t=1}^T \rho_{\alpha}\left(r_t-f_{\alpha}(r_t,x_{t})\right)$. For CAV, we use a symmetric absolute value (SAV) component for $f_{\alpha}(\cdot)$, i.e $f_{\alpha}(r_t, x_t)=\beta_1+\beta_2 f_{\alpha}(x_{t-1},r_{t-1}) + \beta_3 \vert r_{t-1}\vert$, and $f_{\alpha}(r_{t},x_t)=\beta_4 x_{t-1}+\beta_5 r_{t-1}$ for the quantile regression. We furthermore use an asymmetric CAViaR specification with $f_{\alpha}(r_t, x_t)=\gamma_1+\gamma_2 f_{\alpha}(x_{t-1},r_{t-1}) + \gamma_3 \max(r_{t-1},0) - \gamma_4 \min(r_{t-1},0)$ (CAV\_ASY). In contrast to the former methods, they can only capture parametric (non)-linear effects which limits their flexibility. For comparison, we use the asymmetric GJR-GARCH(1,1) model \citep{GLOSTEN_1993}, a GARCH(1,1) \citep{Bollerslev1986} model, a simple historical simulation (Hist), meaning that we predict $VaR^\alpha_{t+1}$ at level $\alpha$ as the sample $\alpha$-quantile of the preceding returns in a window of length $K$, i.e. $(r_{t-K+1},\dots,r_t)$, and one that fits a normal distribution to the sample data and uses the theoretical fitted $\alpha$-quantile as the prediction for $VaR^\alpha_{t+1}$ (NormFit). We do not expect the latter to perform well as we have high skewness and excess-kurtosis in the data (see Table \ref{Tab:Desc} in Section \ref{Sec:Data}).

For the proposed random forest-type and all benchmark procedures that allow for additional covariates (GRF, QRF, QR) we include lagged standard deviations (SD) in addition to the lagged level $r_{t-1}$ in the model. We expect these to capture the strongly varying levels of unconditional volatility in particular for the cryptocurrencies in the non-linear structure. Additionally, we also employ the above methods and the GARCH(1,1) model using both the latter covariates as well as the 7 external covariates described in Section \ref{Sec:Data}. These methods are GRF-X, QRF-X, QR-X, and GARCH-X. 
To establish a fair common ground in used model complexity for the QR, QRF, GARCH-X, and GRF, we select a common set of different SD lags as covariates from an additional Monte-Carlo study. For this, we use the simple SAV-model from Section \ref{Sec:Simulation} as a baseline where unconditional volatilities vary over time. The model is essentially linear autoregressive of order 1 in the VaR and thus directly yields the VaR as outputs. With this, it is possible to select the variables which maximize the p-value of the DQ-test for the subsequent simulations and data analysis. For a realistic specification of the SAV-model, we allow for regime changes. These regime changes are obtained by testing for regime shifts in the realized variances of all three considered currencies in Subsection \ref{sec:crypto_results_case_study} with a Chow-test and then using the empirical distribution of all detected change points as points for regime shifts in the model initializing a new draw of $\sigma_t$.  
Table \ref{tab:MSE_sim_sd} summarizes the results of this short simulation study and motivates the use of 3, 7, 30, and 60 day lagged SD as covariates. 

\begin{table}[htp]
    \centering
    \caption{DQ-Test p-values for Different Covariate Combinations}
    \resizebox{\textwidth}{!}{%
    \begin{tabular}{r|ccccccccc}
    \hline
         Lagged SD (in days) & 3 & 7 & 30 & 3 and 7 & 3 and 30 & 7 and 30 & 3, 7 and 30 & 3, 7, 30, and 60 \\ \hline
         DQ-test p-value & 0.205 & 0.160 & 0.174 & 0.249 & 0.254 & 0.244 & 0.298 & \textbf{0.306} \\
         \hline
    \end{tabular}
    }
    \caption*{\scriptsize{DQ-Test p-values for a simulated SAV-model as in Section \ref{Sec:Simulation} for GRF. The depicted values are averages over 100 iterations. The maximum p-value is marked in bold.}}
    \label{tab:MSE_sim_sd}
\end{table}

For the highly speculative cryptocurrency market, we focus on one-step ahead predictions that are practically the most meaningful. Note that an extension to multistep-ahead predictions with the considered random forest type techniques would not be straightforward. Since the nonlinear GRF and QRF are trained for and therefore tailored to one-step ahead predictions, further forecast horizons would require changing the fitted models for each forecast horizon and retraining them.  Otherwise the required volatility and other inputs would not be available for producing forecasts from the estimated model. Plugging in the last available observation for these quantities might not be valid in the highly nonlinear RF-structure. For a fair comparison, retraining RFs would also require to adapt the parametric benchmark models accordingly.  

To compare the performance of the above methods, we use two types of evaluation approaches. First, we test how well each model predicts the conditional $\alpha$-VaR over the entire out-of-sample horizon using three different sets of evaluation techniques. The simplest way of checking whether a model predicts $VaR^{\alpha}$ correctly over a time horizon is to look at its coverage meaning the number of times $r_t$ is smaller than the predicted $VaR^{\alpha}_t$. Ideally, this should be exactly $\alpha T$ times. This measure is called Actual over Expected Exceedances (AoE) and is computed as $AoE_{\alpha}=(\alpha T)^{-1}\sum_{t=1}^T 1\{r_t<VaR^{\alpha}_t\}$. To test this intuition formally, we employ three tests, the DQ-test\footnote{We use the implementation from the \texttt{GAS}-package in \texttt{R} \citep{Ardia2019} including 4 lagged Hit-values, a constant, the VaR-forecast, and the squared lagged (log-)return in the regression model.} from \citet{Engle2004}, the Christoffersen-test \citep{Christoffersen1998} and the Kupiec-test \citep{Kupiec1995}. All three tests assume that the forecasts have correct coverage under the null hypothesis. The Kupiec-test is simply the formalization of the above intuition, the Christoffersen-test is robust against serial correlation by assuming that $g_t=1\{r_t<VaR^{\alpha}_t\}\sim Bern(\alpha)$, and the DQ-test additionally accounts for problems with conditional coverage due to clustering of the hits exceedance sequences $g_t$ with a regression-based approach. In the empirical analysis, we only report the values of the DQ-test, which is the strictest of the tests, for reasons of clarity. Results for the other tests do not differ substantially and are available upon request from the authors.

Secondly, for comparing the forecast performance of two models 1 and 2 directly, we implement the one-step ahead test for conditional predictive ability (CPA) from \citet{Giacomini2006}, Theorem 1, that assumes under the null hypothesis that forecasts of model 1 and model 2 have on average equal predictive ability conditional on previous information. As suggested by \citet{Giacomini2005}, we use the quantile loss function $L_{\alpha}$ for the test. This tests assumes under the null hypothesis that $H_0: \ E\left[\Delta L_{t}\vert \mathcal{F}_{t-1}\right] = E\left[L_{\alpha}(f_{\alpha}^{(1)},r_{t}) - L_{\alpha}(f_{\alpha}^{(2)},r_{t})\vert \mathcal{F}_{t-1}\right]\equiv E\left[h_{t-1}\Delta L_{t}\right] =0 $ and that this loss difference is a martingale difference sequence, where $\mathcal{F}_{t-1}$ contains all information up to time $t-1$ and $f_{\alpha}^{(1)}$ and $f_{\alpha}^{(2)}$ are two competing forecasts. The test statistic is computed using a Wald-type test with a set of factors $h_{t-1}$ that can possibly predict the loss difference $\Delta L_{t}$ and is $\chi^2_{q}$-distributed under $H_0$. More specifically, we choose $h_{t-1}=(1,\Delta L_{t-1})$ (i.e. $q=2$), i.e. using the lagged loss difference and an intercept as predictors in a linear regression with parameter $\beta_0$ for the simulation and application.

\section{Simulation}\label{Sec:Simulation}
In this section, we study the finite sample forecast performance in quantiles for true DGPs of standard stock-type dynamics as well as for set-ups of cryptocurrency type. Overall, we find that, as expected, GRF performs well even in simple settings where the oracle parametric models should actually have an advantage.  For cases which are designed to mimic the cryptocurrency behaviour over time with changing volatilities, GRF outperforms the competitor models. For QRF, the performance is entirely different as it is outperformed by most other models. These results are also strongly confirmed by pairwise CPA-tests. 

We study three different types of DGPs. The first one is a standard GARCH(1,1) process, i.e
\begin{align}
    r_t&= z_t \sigma_t\\
    \sigma^2_t &= \omega + \beta_0\varepsilon^2_{t-1} + \beta_1\sigma^2_{t-1} \ ,
\end{align}
where the parameters are estimated on the full Bitcoin data to mimic the behavior of established cryptocurrencies (with $z_t$ either follows a normal distribution or an asymmetric, skewed-$t$ distribution). We denote these settings as \textit{GARCH Bitcoin Norm} and \textit{GARCH Asym-$t$}. Moreover, we also consider the specification with $\beta_0=0.1$, $\beta_1=0.8$, $\omega=10^{-4}$ and $z_t \sim N(0,1)$ (\textit{ GARCH}) and with $z_t$ as $t_5$-distributed (\textit{GARCH t}), which corresponds to standard stock index data. Secondly, for the \textit{SAV-Model} setting, we fit a symmetric absolute value (SAV) model to normal returns, i.e. 
\begin{align}
    \textit{VaR}_{t+1}=\gamma_0+\gamma_1 \textit{VaR}_t+ \gamma_2 \vert r_t^{(\textit{init})} - \gamma_3 \vert \ ,
\end{align}
with $r_t^{(\textit{init})}\sim N(0,\sigma_t^2)$ and $\dfrac{\sigma_t}{65} \sim \chi^2_2$, where new draws of $\sigma_t$ are only taken every 100 observations, keeping $\sigma_t$ constant meanwhile. We then generate the final return as $r_t \sim N\left(0,\dfrac{\widehat{VaR}_t}{\Phi(\alpha)^{-1}}\right)$ from the fitted SAV-model, where $\Phi(\alpha)^{-1}$ is the quantile function of a standard normal variable. We do this to obtain returns that have exactly the \textit{VaR} that we obtained from the SAV model before.
Thirdly, we use a simple stochastic volatility model that draws the parameters in the conditional volatility specification from a normal innovation and uses pre-estimates of the two lag one parameters obtained from an average of the mostly emerging crypto-currencies contained in the Group\_high of Table. \ref{tab:crypto_dif_rel}. We denote this setting as \textit{Varying Vola GARCH}.

\begin{table}[htbp]
\centering
\caption{Simulation: $5\%$ VaR }
\resizebox{\textwidth}{!}{ \def\arraystretch{0.7}
\begin{tabular}{lcccc|lcccc}
        \hline   \textit{Rolling Window}       & \multicolumn{4}{c}{$l=500$}   &  \multicolumn{4}{c}{$l=1000$}     \\ \hline 
 & DQ & Kupiec & Christoffersen & AoE & DQ & Kupiec & Christoffersen & AoE \\ \hline \hline
 \multicolumn{4}{l}{\textit{GARCH Normal}}                    &     &  &      &                &     \\
 QRF & 0.960 $( 0.006 )$ & 0.330 $( 0.196 )$ & 0.210 $( 0.270 )$ & 1.184 & 0.710 $( 0.068 )$ & 0.190 $( 0.336 )$ & 0.110 $( 0.349 )$ & 1.159 \\ 
  GRF & 0.470 $( 0.184 )$ & 0.000 $( 0.568 )$ & 0.000 $( 0.544 )$ & 1.043 & 0.270 $( 0.316 )$ & 0.030 $( 0.537 )$ & 0.050 $( 0.491 )$ & 1.031 \\ 
  QR & 0.730 $( 0.047 )$ & 0.080 $( 0.451 )$ & 0.100 $( 0.463 )$ & 1.094 & 0.310 $( 0.261 )$ & 0.060 $( 0.555 )$ & 0.020 $( 0.518 )$ & 1.038 \\ 
  Hist & 0.860 $( 0.048 )$ & 0.020 $( 0.539 )$ & 0.190 $( 0.361 )$ & 1.045 & 0.630 $( 0.114 )$ & 0.080 $( 0.478 )$ & 0.220 $( 0.358 )$ & 1.031 \\ 
  NormFit & 0.760 $( 0.069 )$ & 0.080 $( 0.566 )$ & 0.190 $( 0.322 )$ & 1.012 & 0.590 $( 0.124 )$ & 0.100 $( 0.449 )$ & 0.210 $( 0.334 )$ & 1.009 \\ 
  CAV & 0.610 $( 0.134 )$ & 0.010 $( 0.557 )$ & 0.050 $( 0.541 )$ & 1.044 & 0.200 $( 0.317 )$ & 0.050 $( 0.492 )$ & 0.040 $( 0.457 )$ & 1.034 \\ 
  CAV\_ASY & 0.760 $( 0.049 )$ & 0.010 $( 0.484 )$ & 0.030 $( 0.496 )$ & 1.076 & 0.270 $( 0.249 )$ & 0.090 $( 0.475 )$ & 0.070 $( 0.464 )$ & 1.047 \\ 
  GARCH(1,1) & 0.470 $( 0.182 )$ & 0.040 $( 0.521 )$ & 0.190 $( 0.385 )$ & 1.057 & 0.250 $( 0.320 )$ & 0.080 $( 0.517 )$ & 0.060 $( 0.433 )$ & 1.029 \\ 
  GJR-GARCH & 0.500 $( 0.151 )$ & 0.040 $( 0.497 )$ & 0.140 $( 0.383 )$ & 1.070 & 0.280 $( 0.288 )$ & 0.070 $( 0.506 )$ & 0.080 $( 0.417 )$ & 1.040 \\

  &&&&&&&\\
  \multicolumn{4}{l}{\textit{GARCH Bitcoin Asym-t }}                  &     &  &      &                &     \\
  
    QRF & 0.640 $( 0.086 )$ & 0.300 $( 0.236 )$ & 0.170 $( 0.317 )$ & 1.174 & 0.350 $( 0.243 )$ & 0.140 $( 0.386 )$ & 0.110 $( 0.390 )$ & 1.150 \\ 
  GRF & 0.350 $( 0.303 )$ & 0.030 $( 0.609 )$ & 0.060 $( 0.559 )$ & 1.005 & 0.170 $( 0.443 )$ & 0.050 $( 0.518 )$ & 0.030 $( 0.490 )$ & 1.004 \\ 
  QR & 0.470 $( 0.209 )$ & 0.080 $( 0.427 )$ & 0.080 $( 0.397 )$ & 1.102 & 0.200 $( 0.419 )$ & 0.060 $( 0.489 )$ & 0.090 $( 0.475 )$ & 1.064 \\ 
  Hist & 0.990 $( 0.003 )$ & 0.160 $( 0.425 )$ & 0.690 $( 0.111 )$ & 1.081 & 0.920 $( 0.021 )$ & 0.290 $( 0.302 )$ & 0.590 $( 0.134 )$ & 1.074 \\ 
  NormFit & 0.960 $( 0.008 )$ & 0.500 $( 0.166 )$ & 0.780 $( 0.054 )$ & 0.788 & 0.900 $( 0.024 )$ & 0.550 $( 0.168 )$ & 0.760 $( 0.056 )$ & 0.731 \\ 
  CAV & 0.270 $( 0.305 )$ & 0.040 $( 0.517 )$ & 0.030 $( 0.508 )$ & 1.043 & 0.110 $( 0.507 )$ & 0.070 $( 0.498 )$ & 0.080 $( 0.454 )$ & 1.028 \\ 
  CAV\_ASY & 0.350 $( 0.263 )$ & 0.080 $( 0.500 )$ & 0.050 $( 0.495 )$ & 1.080 & 0.160 $( 0.459 )$ & 0.110 $( 0.500 )$ & 0.060 $( 0.453 )$ & 1.047 \\ 
  GARCH(1,1) & 0.420 $( 0.256 )$ & 0.220 $( 0.314 )$ & 0.350 $( 0.219 )$ & 0.859 & 0.270 $( 0.313 )$ & 0.240 $( 0.315 )$ & 0.250 $( 0.246 )$ & 0.826 \\ 
  GJR-GARCH & 0.430 $( 0.247 )$ & 0.160 $( 0.352 )$ & 0.310 $( 0.253 )$ & 0.890 & 0.250 $( 0.292 )$ & 0.220 $( 0.324 )$ & 0.250 $( 0.252 )$ & 0.836 \\ 
  
  &&&&&&&\\
\multicolumn{4}{l}{\textit{SAV-Model}}                  &     &  &      &                &     \\

  QRF & 0.970 $( 0.005 )$ & 0.300 $( 0.192 )$ & 0.160 $( 0.270 )$ & 1.186 & 0.710 $( 0.074 )$ & 0.170 $( 0.328 )$ & 0.130 $( 0.378 )$ & 1.175 \\ 
  GRF & 0.470 $( 0.157 )$ & 0.010 $( 0.568 )$ & 0.020 $( 0.589 )$ & 1.047 & 0.230 $( 0.321 )$ & 0.040 $( 0.511 )$ & 0.050 $( 0.533 )$ & 1.061 \\ 
  QR & 0.930 $( 0.014 )$ & 0.050 $( 0.431 )$ & 0.050 $( 0.455 )$ & 1.102 & 0.360 $( 0.228 )$ & 0.050 $( 0.485 )$ & 0.060 $( 0.516 )$ & 1.076 \\ 
  Hist & 0.320 $( 0.227 )$ & 0.030 $( 0.564 )$ & 0.070 $( 0.513 )$ & 1.052 & 0.190 $( 0.368 )$ & 0.060 $( 0.515 )$ & 0.070 $( 0.480 )$ & 1.070 \\ 
  NormFit & 0.280 $( 0.344 )$ & 0.030 $( 0.546 )$ & 0.110 $( 0.501 )$ & 1.004 & 0.150 $( 0.435 )$ & 0.060 $( 0.484 )$ & 0.070 $( 0.451 )$ & 1.022 \\ 
  CAV & 0.540 $( 0.086 )$ & 0.010 $( 0.539 )$ & 0.020 $( 0.534 )$ & 1.059 & 0.160 $( 0.295 )$ & 0.030 $( 0.547 )$ & 0.030 $( 0.503 )$ & 1.053 \\ 
  CAV\_ASY & 0.750 $( 0.046 )$ & 0.020 $( 0.481 )$ & 0.020 $( 0.535 )$ & 1.081 & 0.250 $( 0.275 )$ & 0.050 $( 0.494 )$ & 0.070 $( 0.501 )$ & 1.064 \\ 
  GARCH(1,1) & 0.210 $( 0.313 )$ & 0.050 $( 0.582 )$ & 0.050 $( 0.523 )$ & 1.035 & 0.080 $( 0.448 )$ & 0.030 $( 0.513 )$ & 0.030 $( 0.480 )$ & 1.043 \\ 
  GJR-GARCH & 0.350 $( 0.214 )$ & 0.060 $( 0.540 )$ & 0.060 $( 0.479 )$ & 1.048 & 0.060 $( 0.442 )$ & 0.020 $( 0.547 )$ & 0.030 $( 0.484 )$ & 1.050 \\ 
   
 &&&&&&&\\
\multicolumn{4}{l}{\textit{Varying Vola GARCH}}                  &     &  &      &                &     \\
  
  QRF & 0.524 $( 0.124 )$ & 0.270 $( 0.215 )$ & 0.111 $( 0.297 )$ & 1.175 & 0.410 $( 0.218 )$ & 0.190 $( 0.335 )$ & 0.190 $( 0.370 )$ & 1.166 \\ 
  GRF & 0.381 $( 0.269 )$ & 0.000 $( 0.509 )$ & 0.079 $( 0.393 )$ & 1.016 & 0.240 $( 0.323 )$ & 0.110 $( 0.436 )$ & 0.160 $( 0.373 )$ & 0.959 \\ 
  QR & 0.381 $( 0.195 )$ & 0.000 $( 0.538 )$ & 0.143 $( 0.474 )$ & 1.013 & 0.290 $( 0.296 )$ & 0.060 $( 0.440 )$ & 0.110 $( 0.381 )$ & 1.065 \\ 
  Hist & 0.984 $( 0.004 )$ & 0.476 $( 0.218 )$ & 0.810 $( 0.069 )$ & 1.281 & 0.980 $( 0.005 )$ & 0.620 $( 0.163 )$ & 0.910 $( 0.040 )$ & 1.140 \\ 
  NormFit & 0.984 $( 0.003 )$ & 0.476 $( 0.212 )$ & 0.810 $( 0.053 )$ & 1.212 & 0.980 $( 0.003 )$ & 0.580 $( 0.162 )$ & 0.850 $( 0.041 )$ & 0.991 \\ 
  CAV & 0.270 $( 0.315 )$ & 0.079 $( 0.467 )$ & 0.048 $( 0.472 )$ & 1.051 & 0.110 $( 0.427 )$ & 0.120 $( 0.406 )$ & 0.140 $( 0.407 )$ & 1.012 \\ 
  CAV\_ASY & 0.397 $( 0.212 )$ & 0.127 $( 0.436 )$ & 0.016 $( 0.449 )$ & 1.084 & 0.110 $( 0.420 )$ & 0.140 $( 0.449 )$ & 0.150 $( 0.427 )$ & 1.026 \\ 
  GARCH(1,1) & 0.302 $( 0.279 )$ & 0.048 $( 0.458 )$ & 0.254 $( 0.310 )$ & 1.060 & 0.480 $( 0.205 )$ & 0.140 $( 0.397 )$ & 0.340 $( 0.237 )$ & 1.084 \\ 
  GJR-GARCH & 0.317 $( 0.309 )$ & 0.095 $( 0.459 )$ & 0.254 $( 0.348 )$ & 1.074 & 0.500 $( 0.192 )$ & 0.170 $( 0.419 )$ & 0.380 $( 0.235 )$ & 1.093 \\    \hline
 \hline
\end{tabular}
}
\caption*{\scriptsize{ The table displays rejection rates of t-tests of empirical quantile levels against the nominal level of $5\%$ for DQ-, Kupiec- and Christoffersen-tests and mean p-values in parentheses.  Thus lower rejection rates and higher p-values indicate better model performance. The $AoE_{\alpha}=(\alpha T)^{-1}\sum_{t=1}^T 1\{r_t<VaR^{\alpha}_t\}$ should be compared to its optimal value 1. The GARCH case uses predictions from the QMLE-fit of a GARCH(1,1) specification with normally distributed errors and can therefore be seen as an oracle for the GARCH-simulated specifications.}}
\label{tab:add_sim_results}
\end{table}

For all settings, we generate 2000 return observations and forecast the one-step ahead VaR over the different rolling window lengths $l=500,1000$. We repeat this generation process 200 times $\alpha=0.05$. For comparison of the different methods described in Section \ref{Sec:grf_method}, we use the DQ-test, the Kupiec test, the Christoffersen-test, and the AoE. Note that for all tests, we present aggregate results from two-sided t-tests of the empirical versus the nominal coverage. The results are therefore rejection rates of t-tests against the nominal level of 5\%. Therefore, a lower rejection rate and higher mean p-values (in parentheses) indicate better performance. For GRF, QRF, and QR, we use a common set of lagged covariates as described at the end of Section \ref{Sec:grf_method}.

Table \ref{tab:add_sim_results} summarizes the results of the simulation for the 5\% $VaR$. According to the more advanced $DQ$- and Christoffersen-tests for evaluation, GRF is close to on par with models that directly use the simulated model specification and outperforms all other methods in cases where the underlying model deviates from standard GARCH type DGPs as is the case for cryptocurrencies (\textit{Varying Vola GARCH}). This is in sharp contrast to the QRF-model that cannot compete with the other models in all cases. We only display results for the  \textit{GARCH Normal} and \textit{GARCH Bitcoin Asym-t} cases, as \textit{GARCH-t} and \textit{GARCH Normal Bitcoin} yield similar results. A bit of a challenge for GRF is the very simplistic \textit{SAV Model} where the simple models dominate but the GRF performance still improves on advanced parametric models such as CAV, CAV\_asy and GJR-GARCH. QR is dominated completely both, in SAV and standard GARCH-type settings. In the GARCH-bitcoin models, the CAV procedures, perform particularly well, both in the small and large sample case, while the asymmetric specification cannot outperform its symmetric counterpart in most of the cases, specifically for the small window length. This is not surprising, as the asymmetric specification is more complex and therefore needs more data to perform well. In the third time-varying setting \textit{Varying Vola GARCH}, GRF substantially outperforms all other methods when calibrated in the short time interval - but loses this advantage for the larger time period in particular compared to CAV and CAV\_asy. These results suggest that depending on the underlying data-generating process, either GRF or CAV-procedures are best-performing. 

Additionally, we conduct direct pairwise comparison tests between the superior random forest type method GRF against the best performing non-oracle other parametric methods (and a simple baseline) via CPA-tests for each scenario. The respective results are reported in Table \ref{tab:sim_results_CPA}. 
For the standard GARCH-processes, GRF outperforms its competitors on average, however, mean p-values are mostly not significant. As before, we only display results for the \textit{GARCH Normal} case since \textit{GARCH-t} and 
\textit{GARCH Normal Bitcoin} yield similar results. In the other three scenarios, GRF is substantially superior with more significant p-values. Though in the \textit{GARCH Bitcoin Asym-t} and \textit{SAV Model} case, the variance in the p-values over different simulation runs is quite high, marked by many significant test results, but a performance that indicates that both procedures are equally good. Though for \textit{Varying Vola GARCH}, the performance gains of GRF are substantial reaching highly significant test results.  Overall, the results indicate that even for complex data structures, the power of CPA-tests in small samples appears sufficient if compared to the larger sample results. 

\begin{table}[htbp]
\caption{CPA-tests on Predictions of $5\%$ VaR for Different Window Lengths and Different Models}
\centering
\resizebox{0.6\textwidth}{!}{\def\arraystretch{0.7}
\begin{tabular}{lccc|ccc}
        \hline   \textit{Rolling Window}       & \multicolumn{3}{c}{$l=500$}  &  \multicolumn{3}{c}{$l=1000$}     \\ \hline 
                 GRF vs: & QR & Hist & CAV   & QR & Hist & CAV \\ \hline \hline
 \multicolumn{3}{l}{\textit{GARCH Normal}}                     &      &                     \\
Mean P-Value & 0.235 & 0.347 & 0.383 & 0.468 & 0.39 & 0.504 \\ 
  No. P-values < 0.1 & 52 & 23 & 14 & 14 & 23 & 11 \\ 
  GRF-Performance & 0.877 & 0.794 & 0.613 & 0.539 & 0.799 & 0.482 \\ 
 &&&&&&\\
\multicolumn{3}{l}{\textit{GARCH Bitcoin Asym-t}}                   &                      &     \\
Mean P-Value & 0.163 & 0.06 & 0.412 & 0.155 & 0.098 & 0.387 \\ 
  No. P-values < 0.1 & 71 & 85 & 12 & 70 & 76 & 10 \\ 
  GRF-Performance & 0.799 & 0.971 & 0.598 & 0.748 & 0.962 & 0.356 \\
&&&&&&\\
\multicolumn{3}{l}{\textit{SAV-Model}}                   &                      &     \\
Mean P-Value & 0.149 & 0.294 & 0.494 & 0.123 & 0.337 & 0.445 \\ 
  No. P-values < 0.1 & 66 & 31 & 12 & 77 & 29 & 13 \\ 
  GRF-Performance & 0.964 & 0.296 & 0.567 & 0.914 & 0.335 & 0.495 \\ 
&&&&&&\\
\multicolumn{3}{l}{\textit{Varying Vola GARCH}}                   &                      &     \\
Mean P-Value & 0.001 & 0.005 & 0.464 & 0.4 & 0.016 & 0.363 \\ 
  No. P-values < 0.1 & 63 & 62 & 8 & 19 & 97 & 23 \\ 
  GRF-Performance & 0.987 & 0.989 & 0.569 & 0.363 & 0.985 & 0.329 \\ 
&&&&&&\\
\hline
\end{tabular}%
}
\caption*{\scriptsize{The table displays CPA-tests for $5\%$ one-day ahead VaR-forecasts of the best performing random forest type techniques GRF in Table \ref{tab:add_sim_results} versus the best parametric time series models. We report mean p-values, the number of significant p-values over 200 iterations and the rate at which GRF outperforms the competing method (i.e. a value of 0.8 means that GRF has a smaller error loss than the competing method in 80\% of the rolling window forecasts over all runs). Low p-values paired with performance rates larger than 0.5 indicate that GRF outperforms the competing methods.}}
\label{tab:sim_results_CPA}
\end{table}

\section{Results}\label{Sec:Results}
In this section, we highlight the advantages from using specific non-linear machine learning-based methods for forecasting the VaR of cryptocurrencies. In particular, we show for a large cross-section of more than 100 cryptocurrencies that the proposed random forest method GRF yields superior performance across a wide range of different types of cryptocurrencies and different time periods. Investigating the underlying drivers, we illustrate that the non-linear model predictions excel especially for assets that are frequently traded by a large amount of different users, and for more volatile assets and times.

We predict the $5\% \textit{VaR}$ as a key quantity in risk management for our comprehensive set of cryptocurrencies. In an extensive out-of-sample forecasting study, we compare the random forest-based machine learning methods to standard linear time series and GARCH-type models including approaches with exogenous asset information in covariates. The prediction performance is assessed with the DQ-test to obtain an overall aggregate picture on the realized coverages as well as pairwise CPA-tests across different time periods and types of cryptocurrencies.

Based on these findings, we extend our analysis focusing on three important selected currencies detail that comprise Bitcoin (btc) as the largest currency by far regarding market cap, Tether (usdt\_omni) as a stablecoin with lower volatility, and Cardano (ada) as a currency specifically allowing for smart contracts. For these examples, we also consider predicted loss series by CPA-tests and variable importance measures to uncover important drivers. Furthermore we extend the aggregate forecast performance study of the main periods 1-3 to the shorter post-pandemic period 4.

\subsection{Aggregated Forecasting Performance}
In this section, we provide results on aggregate forecast performance of the different modeling approaches over all cryptocurrencies.

\subsubsection{Backtesting} \label{sec:crypto_results_backetesting}
We analyze the three fundamentally distinct periods 2015/08/22 - 2017/12/21, 2017/12/22 - 2020/11/05, and 2020/11/06 - 2022/12/31 of the cryptocurrency market which differ according to the number of actively traded currencies, the market capitalization and the overall market situation (see Section \ref{Sec:Data}). We thus study each period separately, with a focus on period 3 in the aggregate. Given the largely different market situations, we refrain from reporting results for the full time period. For the fundamentally different Post-COVID-19/Ukrainian war situation we provide results up until today in Subsection \ref{subsec:period4}.

We assess the performance of VaR forecasts by their conditional coverage and the dependence structure of exceedances, which we test with the DQ-test (see Section \ref{Sec:grf_method} for details). This test is the strictest of all backtests, conditioning on past information, and therefore most accurately captures the coverage abilities of the procedures\footnote{Detailed results for the other tests are omitted here for reasons of clarity. As in the simulation results they do not differ qualitatively, and are available upon request from the authors.}. 
The results over the different time periods for the 5\% VaR-predictions are shown in Table \ref{tab:median_pvals}. We can see that depending on the time period, the median p-values vary strongly, which is not surprising giving the different characteristics of each period and the increasing number of cryptocurrencies in the later periods. In general, GRF is the only method with median values consistently over the 8\% level, indicating that it is the most consistently calibrated forecasting method. The QRF performs extremely well in the first time period, but rejects the test often in the third period. Adding external covariates as listed in Table~\ref{tab:sum_covariates} does not improve performance and in general leads to substantially lower p-values especially for the QR. Only in the final third period, adding external covariates increases the p-values slightly for the GRF and substantially for the QRF (GRF-X, QRF-X). This indicates that those extra covariates are not necessarily predictive for extreme returns, or rather that the existing measures such as lagged returns and SD comprise the information already quite well.
\begin{table}[!htbp]
\centering
\caption{Medians of P-Values for DQ-Tests in Different Time Periods} 
\label{tab:median_pvals}
\resizebox{\textwidth}{!}{
\begin{tabular}{rrrrrrrrrrrr}
  \toprule
Period & GRF & QRF & QR & CAV & CAV\_ASY & GJR-GARCH & Hist & GRF-X & QRF-X & QR-X & GARCH-X \\ 
  \midrule
   1 & 0.14 & 0.24 & 0.04 & 0.05 & 0.06 & 0.03 & 0.00 & 0.01 & 0.00 & 0.00 & 0.00 \\ 
   2 & 0.29 & 0.04 & 0.14 & 0.13 & 0.31 & 0.19 & 0.11 & 0.16 & 0.07 & 0.00 & 0.00 \\ 
   3 & 0.08 & 0.02 & 0.05 & 0.01 & 0.01 & 0.16 & 0.00 & 0.12 & 0.05 & 0.00 & 0.00 \\ 
   \bottomrule
\end{tabular}
}
\caption*{\scriptsize The table displays medians of p-values for DQ-Tests in different time periods. Higher median p-values indicate correct conditional coverage according to the DQ-test. Periods correspond to the times indicated in Figure \ref{fig:median_rets}.}
\end{table}
Compared to its non-forest counterparts, only GJR-GARCH can partly keep up with GRF. In particular, the GJR-GARCH has higher median p-values than GRF in the third period, which is the endemic COVID period and contains the most currencies. It is also marked by less extreme returns and a large reduction in active users (see Table~\ref{tab:sum_covariates}), which could indicate that the forest methods excel in particular in highly volatile periods, when large shifts in the market are present.  In all periods, CAV and the asymmetric CAV are outperformed by GRF in DQ-test results. This highlights the inability of the parametric methods to adapt to rapidly changing situations such as in Period 1 and 2. QR, Hist, and GARCH-X are all fully dominated by GRF throughout the three time periods as expected, as QR can only incorporate changes linearly, and GARCH-X and Hist serve as simple baselines. Additionally, when checking the unconditional (median) coverage represented as the ratio of exceedances relative to the expected ones, random forest-type methods also show superior performance over standard GARCH-type methods, which have quite low coverage in all (GARCH-X) or the first period (GJR-GARCH)\footnote{Detailed results are available on request.}.
\begin{figure}[!htbp]
    \centering
        \includegraphics[width=0.7\textwidth]{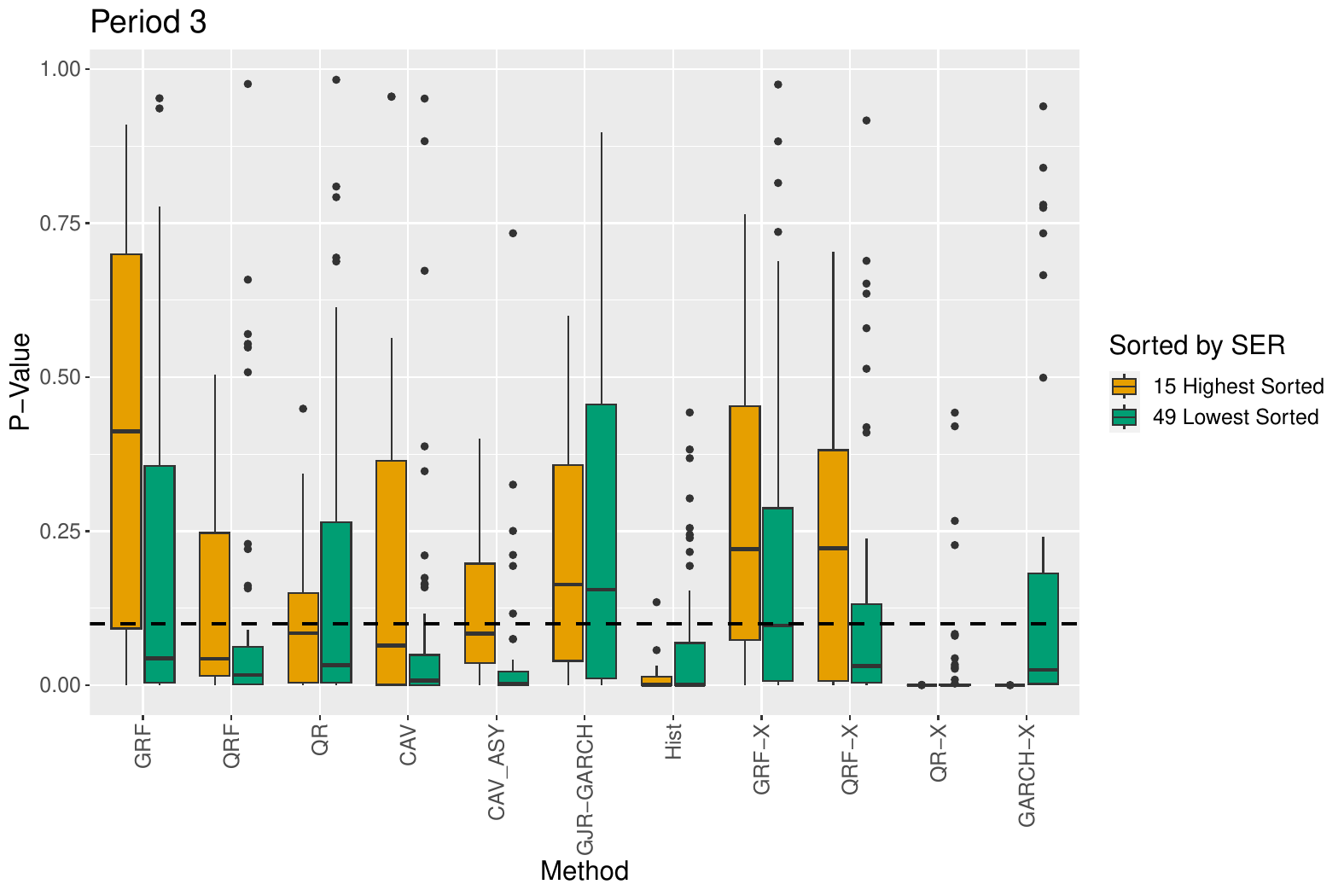}
    \caption{Boxplots of p-values of DQ-tests over all cryptocurrencies contained in Period 3 separated into groups with high vs low SER for every method. For details on the groups, see Table \ref{tab:median_pvals_SER}. The dashed horizontal line depicts a level of 0.1.}
    \label{fig:DQ_boxplot_SER}
\end{figure}
With an LM-type test (see \cite{silvennoinen_testing_2016}) we formally confirm time-variation of the unconditional volatility that is not captured by standard static GARCH-type specifications but leads to advantages of the forest type methods. Across most of the cryptocurrencies, we reject the null hypothesis of constant unconditional variance at a significance level of 5\%  and at even lower levels for the robust version of the test in particular for the first two periods. 

We investigate Period 3, which contains 64 cryptocurrencies, more in detail, especially with regard to how conditional coverage is characterized by external information. For this, we split the currencies into two groups\footnote{Split points correspond to obtaining two groups with mean SER values as homogeneous as possible. They are therefore separated around their steepest decay. The high group always comprises the largest SER sorted quintile and not more than the largest sixtile in each period.} depending on the value of the SER, which characterizes the concentration of supply to users. A low SER consequently implies that most supply is concentrated on a few, large users, often indicating a more stable currency. Figure \ref{fig:DQ_boxplot_SER} depicts boxplots over p-values of DQ-tests for each group and procedure in Period 3. In Period 3, GRF(-X) performs particularly well for the groups with a high SER, which arguably are more prone to speculations during hype/bubble periods due to the large amount of supply held by many small addresses. For the (arguably) more stable addresses with a smaller SER, GJR-GARCH has higher p-values, followed by GRF-X during this period 3. This is in contrast to CAV(\_ASY), QR, or QRF, which are always outperformed by GRF(-X), as well as the simple baselines Hist and GARCH-X. In Period 2, the general market situation is so volatile, that the GRF is superior in DQ performance for the low SER-group where in particular the additional regressors in the GRF-X make a difference.
For details on all time periods and groups according to the different observable factors active users and lagged standard deviation, see Table \ref{tab:median_pvals_sorted} in the appendix.
\begin{table}[!htbp]
\centering
\caption{Medians of P-Values for DQ-Tests Sorted by SER in Different Time Periods} 
\label{tab:median_pvals_SER}
\resizebox{\textwidth}{!}{
\begin{tabular}{lrrrrrrrrrrr}
  \toprule
Group & GRF & QRF & QR & CAV & CAV\_ASY & GJR-GARCH & Hist & GRF-X & QRF-X & QR-X & GARCH-X \\ 
  \midrule
  \multicolumn{12}{l}{\textit{Period 1}} \\
High & 0.17 & 0.81 & 0.00 & 0.00 & 0.00 & 0.12 & 0.00 & 0.01 & 0.00 & 0.00 & 0.00 \\ 
Low & 0.11 & 0.20 & 0.08 & 0.09 & 0.10 & 0.02 & 0.00 & 0.02 & 0.01 & 0.00 & 0.00 \\  
   \multicolumn{12}{l}{\textit{Period 2}} \\
  High& 0.15 & 0.03 & 0.20 & 0.05 & 0.41 & 0.10 & 0.09 & 0.13 & 0.08 & 0.00 & 0.00 \\ 
  Low & 0.31 & 0.04 & 0.08 & 0.17 & 0.29 & 0.27 & 0.17 & 0.45 & 0.07 & 0.00 & 0.11 \\ 
  \multicolumn{12}{l}{\textit{Period 3}} \\
  High & 0.41 & 0.04 & 0.08 & 0.06 & 0.08 & 0.16 & 0.00 & 0.22 & 0.22 & 0.00 & 0.00 \\ 
  Low & 0.04 & 0.02 & 0.03 & 0.01 & 0.00 & 0.15 & 0.00 & 0.10 & 0.03 & 0.00 & 0.02 \\ 
   \bottomrule
\end{tabular}
}
\caption*{\scriptsize The currencies are divided into two groups based on specific covariate values for each time period. The groups are constructed to best separate the sorted covariate values of the currencies. They are therefore separated at their steepest decay to obtain two groups with homogeneous covariate values where the top group in each period contains at least the top quintile but not more than the top sixtile of the data.The remaining details correspond to Table \ref{tab:median_pvals}.}
\end{table}

\subsubsection{Direct Pairwise Forecast Comparisons} \label{sec:crypto_results_cpa}
In addition to assessing the coverage performance, we conduct pairwise CPA-tests for all considered methods in the three different specified time periods over all cryptocurrencies. The results of the tests are contained in Table \ref{tab:CPA_p1_2_grf_base}. Note that the CPA-tests require the rolling window length to be smaller than the out-of-sample forecast window to produce valid results. Therefore by construction, for the large share of newly introduced cryptocurrencies in period 3, the out-of-sample size is too low for the CPA-tests to have high power. Therefore, we additionally look at the direct comparisons of predicted losses (as suggested by \citet{Giacomini2006}, Section 4), where we compare in the loss series how often GRF is better, i.e. has a smaller loss, than its competitors.\footnote{We use the lagged loss difference and an intercept for loss prediction in an autoregressive setup since these are the main drivers of the test statistic in the CPA test.} Note that a value of one thus indicates that GRF has a smaller predicted loss over the full loss series.

In general, GRF performs better than its competitors for a majority of crypotocurrencies over all time periods. Table \ref{tab:CPA_p1_2_grf_base} summarizes the results. We can see that QRF is almost always outperformed, and for around 50\% of cryptocurrencies, losses are even significantly smaller. This is not surprising, as QRF has a similar structure to GRF while not being tuned to predict the quantiles directly. Thus, we expect it to be less sensitive to changes that only affect the quantile of the return distribution, for example large shock events. The same holds for the GARCH-X and Hist, which are clearly outperformed by the GRF, as well as the QR-X and QRF-X. 
Adding exogenous information in covariates as part of the non-linear GRF (i.e. GRF-X) is better especially in later periods (see Figure \ref{fig:CPA_p3_grf-x} in the appendix). This is interesting, since the other methods cannot benefit as much as GRF from additional covariates. Generally, for cryptocurrencies, the non-parametric form of the GRF helps to extract information from exogenous covariates in contrast to standard parametric methods such as QR and GARCH. As GRF accounts specifically for the quantiles in the random forest splitting function, this helps to also favorably integrate additional covariates $X$ in contrast to QRF.
Overall, however, both GRF-procedures seem to perform very similarly, especially in pairwise comparisons. For full results of the GRF-X, see Table \ref{tab:CPA_p1_2_grf-x} and Figure \ref{fig:CPA_p3_grf-x} in the appendix. While CAV(\_ASY), QR, and GJR-GARCH are outperformed over the majority of cryptocurrencies, only 14\% to 24\% of these out-performances reach significance. In Subsection \ref{sec:crypto_results_case_study}, we will focus on specific cryptocurrencies for a more in-depth understanding.

\begin{table}[!htb]
\begin{minipage}{\linewidth}
\centering
\caption{Performance and Significance of CPA-tests Over Different Time Periods for GRF } 
\label{tab:CPA_p1_2_grf_base}
\resizebox{0.83\textwidth}{!}{\def\arraystretch{0.6}
\begin{tabular}{rrrrrrrrrrr}
  \toprule
GRF vs.: & QRF & QR & CAV & CAV\_ASY & GJR-GARCH & Hist & GRF-X & QRF-X & QR-X & GARCH-X \\ 
  \midrule
  \multicolumn{10}{l}{\textit{Share of GRF With Better Performance}}  \\
  Period 1 & 1.00 & 0.80 & 0.60 & 0.80 & 1.00 & 1.00 & 0.60 & 1.00 & 1.00 & 1.00 \\ 
  Period 2 & 1.00 & 0.71 & 0.79 & 0.93 & 0.86 & 1.00 & 0.36 & 1.00 & 1.00 & 1.00 \\ 
  Period 3 & 0.94 & 0.75 & 0.75 & 0.80 & 0.70 & 0.97 & 0.39 & 0.97 & 0.98 & 0.80 \\ 
  Full.Data & 0.91 & 0.70 & 0.73 & 0.75 & 0.58 & 0.94 & 0.33 & 0.92 & 0.98 & 0.75 \\
  \multicolumn{10}{l}{\textit{Share of GRF With Significantly Better Performance}}  \\
  Period 1 & 0.60 & 0.40 & 0.00 & 0.20 & 0.00 & 1.00 & 0.00 & 1.00 & 1.00 & 0.60 \\ 
  Period 2 & 0.79 & 0.29 & 0.21 & 0.50 & 0.29 & 0.79 & 0.00 & 0.93 & 0.93 & 0.71 \\ 
  Period 3 & 0.61 & 0.20 & 0.19 & 0.28 & 0.14 & 0.78 & 0.06 & 0.77 & 0.84 & 0.55 \\ 
  Full Data & 0.50 & 0.14 & 0.19 & 0.24 & 0.12 & 0.71 & 0.05 & 0.66 & 0.74 & 0.53 \\
   \bottomrule
\end{tabular}

}
\vspace{1cm}
\end{minipage}
\begin{subfigure}{0.52\textwidth}
    \centering
    \includegraphics[width=\textwidth]{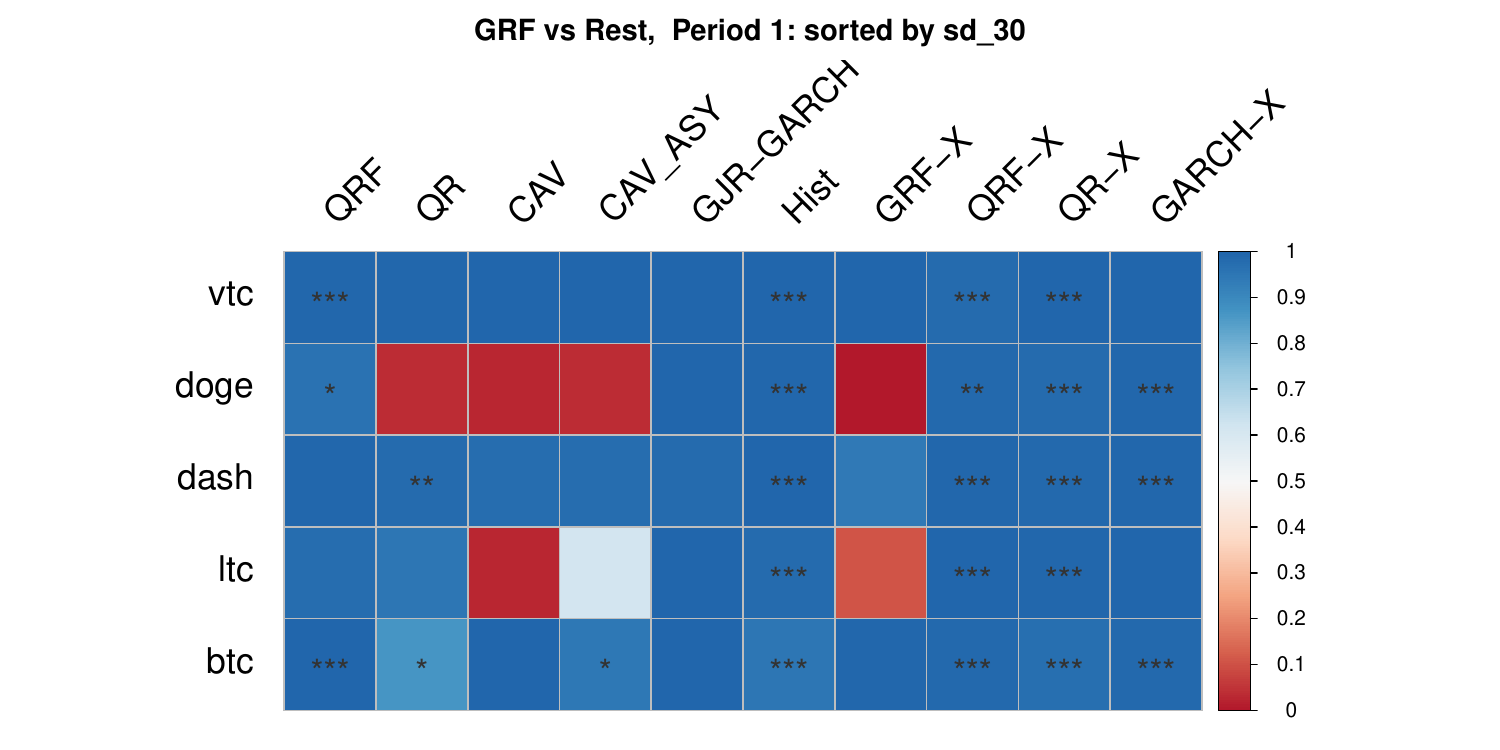}
\end{subfigure}
\begin{subfigure}{0.52\textwidth}
    \centering
    \includegraphics[width=\textwidth]{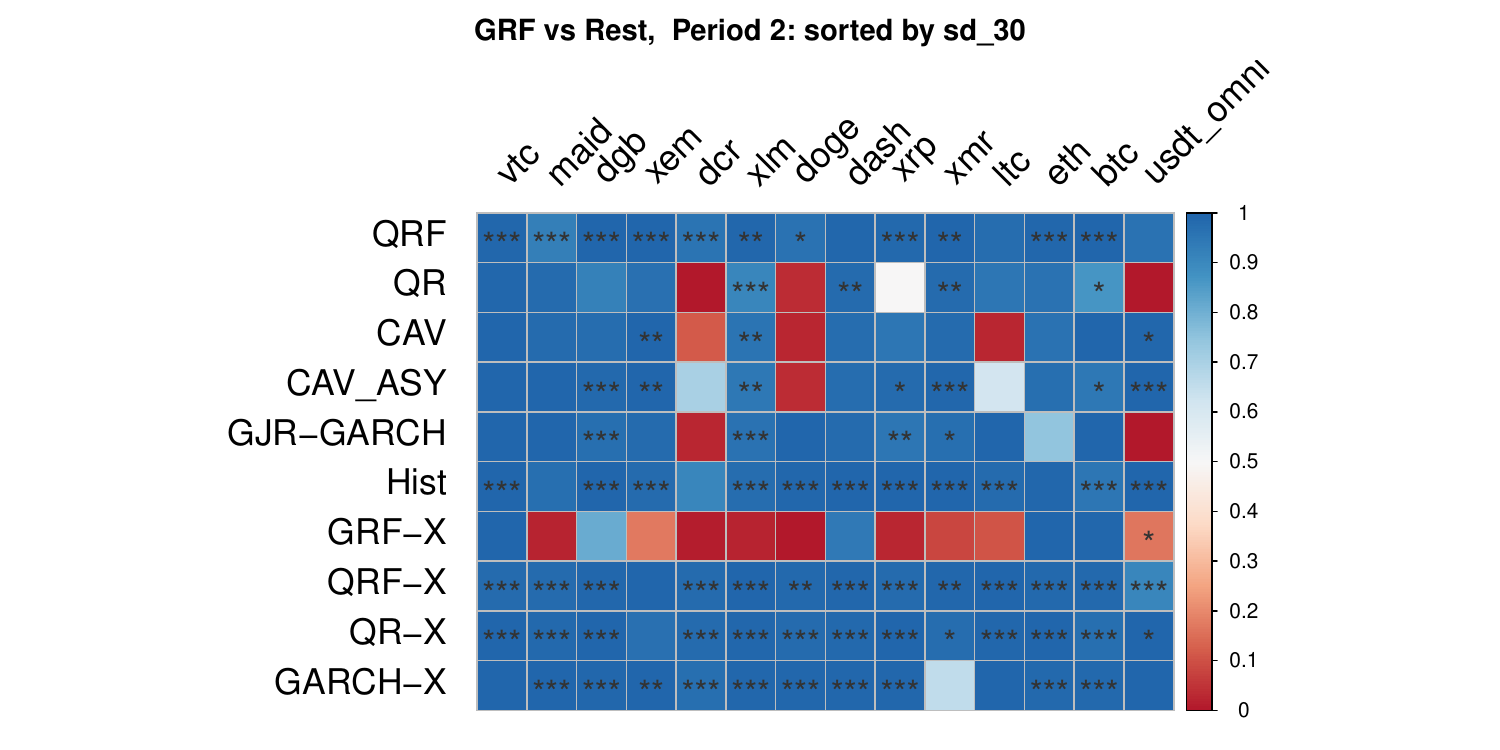}
\end{subfigure}
    \begin{subfigure}[b]{0.49\textwidth}
    \includegraphics[width=\textwidth]{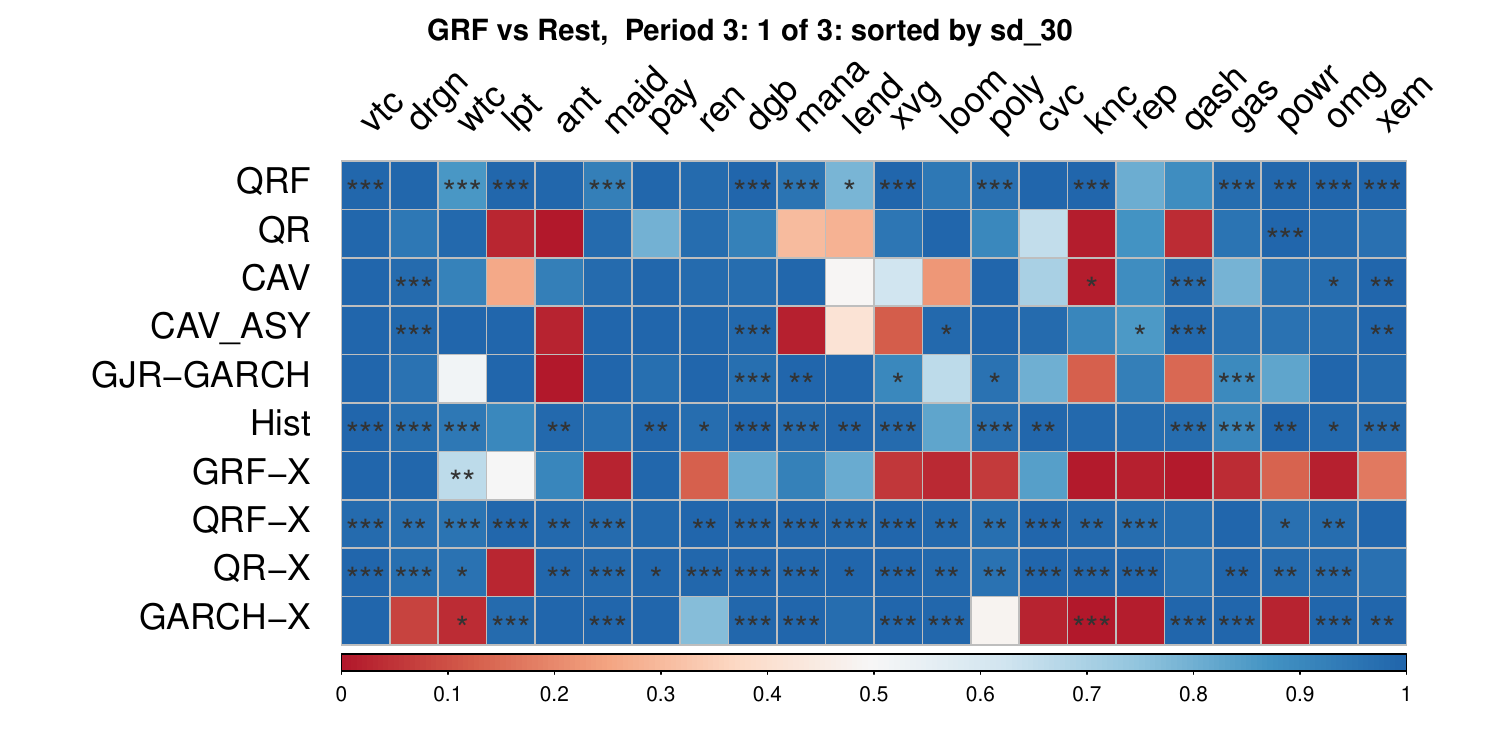}
    \end{subfigure}
    \hfill
    \begin{subfigure}[b]{0.49\textwidth}
    \includegraphics[width=\textwidth]{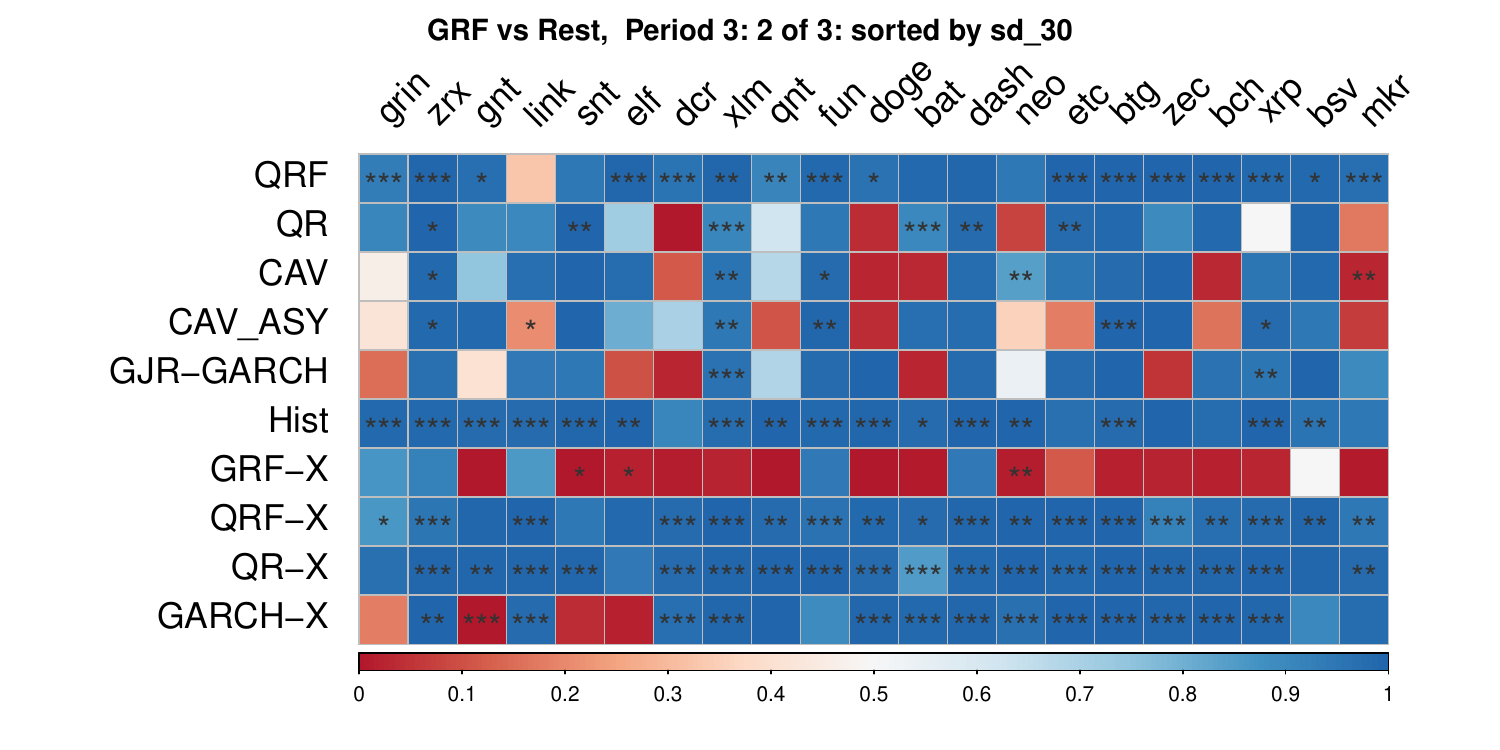}
    \end{subfigure}
        \begin{subfigure}[b]{0.49\textwidth}
    \includegraphics[width=\textwidth]{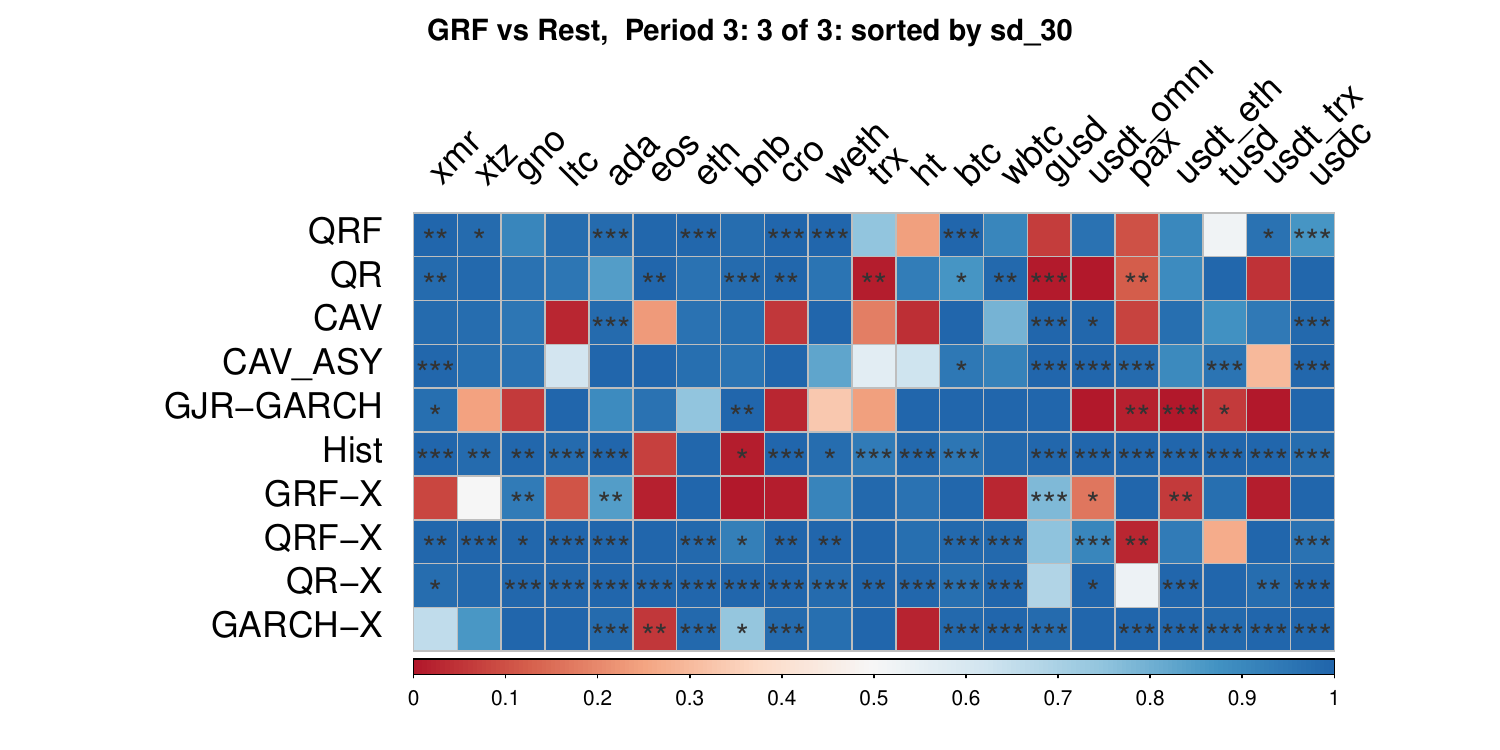}
    \end{subfigure}
    \caption*{\scriptsize The top part shows summary values that are shares over all cryptocurrencies in the respective time period. It describes the number of times that GRF had a better performance (i.e. more than 50\% of predicted losses by the CPA test were smaller for the GRF) relative to all crytpocurrencies (in that period), and the number of times that GRF was significantly better (at least at a 10\% level) as judged by the CPA test over all crytpocurrencies (in that period). The bottom part shows the detailed results of CPA-tests with the color of each box indicating the performance of GRF. Blue signifies a performance of 1, meaning that GRF has a smaller predicted loss in 100\% of cases.  *, **, *** shows significance on a level of 10\%, 5\%, and 1\%. The values are ordered by 30 day lagged standard deviation from highest to lowest (top/left to bottom/right).}
\end{table}

When considering the single time periods, it is notable how for Period 1 and 2 (bottom part of Table \ref{tab:CPA_p1_2_grf_base}), GRF(-X) is constantly outperforming the other methods for most cryptocurrencies, and only has somewhat worse performance for doge and the stablecoins, although these are insignificant (see below). For the first two periods, QR is the most competitive of the other methods, while there is a general tendency for the classic methods to perform worse with higher volatility of returns, which can be seen in Table \ref{tab:CPA_p1_2_grf_base} for Period 2 where the currencies are ordered from highest 30 day lagged SD on the left to the lowest on the right.
This becomes more apparent in Period 3, where we deal with much more cryptocurrencies (64). Here, methods such as GJR-GARCH and QR are on par with GRF or even better when looking at low-volatility (partly regulated) stablecoins such as pax, gusd, tusd, and usdt\_omni with its derivatives (usdt\_eth,usdt\_trx). CAV and CAV\_asy, on the other hand, are only rarely better here (as indicated by CPA tests), while being significantly outperformed for important and large assets such as btc, ada, xlm, and most stablecoins. The asymmetric version of CAV performs similar or worse in comparison to CAV most of the time, which is why we will only focus on the symmetric version in the following sub-analysis for ease of presentation.  

To highlight the specific properties of currencies where GRF outperforms the other methods, we split the assets into two groups. The first group (Group\_low) contains assets where GRF performance is low in comparison to the three other methods that were able to compete in some cases with GRF, namely QR, CAV, and GJR-GARCH. We add an asset into that group when at least two of the methods outperform GRF (in terms of loss difference) for that asset, separately for each time period. All other assets are sorted into the second group (Group\_high), indicating high performance of GRF. Note that 40\% of currencies belong to the Group\_high in Period 1. In Period 2 it is 36\%, in Period 3 it is 33\% of all currencies in Group\_high. The corresponding mean daily trading volumes in USD in Group\_high are  34 961 093\$ in Period 1,  419 414 899\$ in Period 2, and 1 629 301 034\$ in Period 3.

\begin{table}[!htbp]
\centering
\caption{Difference Between Covariates of Cryptos Where GRF is Better vs. Worse} 
\label{tab:crypto_dif_rel}
\resizebox{0.7\textwidth}{!}{ \def\arraystretch{0.5}
\begin{tabular}{rrrrr}
  \toprule 
 & Period 1 & Period 2 & Period 3 &  Full Data \\ 
  \midrule
  Ret & 0.83 & 0.96 & 4.24 & 1.83 \\ 
  Active\_Users & 0.30 & 0.23 & 3.00 & 2.40 \\ 
  Total\_Users & 0.48 & 0.18 & 0.32 & 0.27 \\ 
  Total\_Users\_USD100 & 0.12 & 0.13 & 0.50 & 0.50 \\ 
  Total\_Users\_USD\_10 & 0.18 & 0.14 & 0.79 & 0.71 \\ 
  SER & 0.74 & 0.49 & 0.47 & 0.48 \\ 
  Transactions & 0.38 & 0.04 & 5.78 & 3.79 \\ 
  Velocity & 4.11 & 3.46 & 0.93 & 0.76 \\ 
  sd\_3 & 0.91 & 0.64 & 0.88 & 0.97 \\ 
  sd\_7 & 0.92 & 0.64 & 0.88 & 0.95 \\ 
  sd\_30 & 0.93 & 0.66 & 0.87 & 0.95 \\ 
  sd\_60 & 0.92 & 0.63 & 0.88 & 0.96 \\ 
   \bottomrule
\end{tabular}
}
\caption*{\scriptsize The table shows shares of groups of cryptocurrencies where at least two of CAV, QR, and GJR-GARCH have better CPA-performance than GRF divided by the remaining rest. Raw values before division are mean values over all cryptocurrencies for the median of each covariate in the respective time period. A value smaller than 1 indicates that currencies where GRF performs better have a higher average median values of the respective covariate than the rest of the currencies.}
\end{table}
Table \ref{tab:crypto_dif_rel} summarizes the results over these groups for each time period and covariate. In each group, we take the mean over all cryptocurrencies of median values for each covariate. We then divide Group\_low by Group\_high. For example, an SER of 0.52 in Period 3 indicates that cryptocurrencies in Group\_low have, on average, a median SER that is 47\% to that of Group\_high, or in other words, the median SER for Group\_high is around $2.13=\dfrac{1}{0.47}$ times higher than that of Group\_low on average.
We see that covariates of cryptocurrencies for which GRF performs better have much higher volatility (especially for the second and third period), a much higher SER\footnote{Apart from the first period where the only currency belonging to Group\_low is doge.}, indicating a larger concentration of supply at a lot of small addresses, a higher market capitalization, a lower rate of turnover (Velocity), and more active and total users. To summarize, this confirms the observation that GRF performs better for assets with highly varying returns that are traded by a large amount of users, which could thus also be prone to speculation. On the other had, methods such as QR or GJR-GARCH are better with more stable currencies that are used more as a hedging device (e.g. stablecoins). This confirms our findings from the backtests in Section~\ref{sec:crypto_results_backetesting}, where GRF excels in particular for currencies with high SER values, high volatility, and a large number of active users.

\subsection{Extensions} \label{sec:crypto_results_case_study} 

\subsubsection{In-Depth Analysis of Specific Classes of Assets} 

To identify dynamics of the obtained results and the effect of specific events, we analyze the predicted loss series of CPA tests over the full horizon of availability for the three cryptocurrencies Bitcoin, Cardano, and Tether separately. We furthermore show how  the impact of  covariates on forecast performance changes dynamically over time using variable importance measures of GRF-X.

We choose Bitcoin since it is the largest currency by market cap, with the longest data availability, Tether as the largest stablecoin by daily volume and market cap, and Cardano as a fairly new (i.e. fewer observations), however large currency (again by market cap), which can be used for smart contracts, identity verification, or supply chain tracking\footnote{See e.g. \url{https://cardano.org/enterprise/}, accessed 19/05/2022.}. Since we deal with VaR-predictions, the initial loss function is the quantile loss with a quantile $\alpha=0.05$. 
\begin{figure}[htb]
    \centering
    \begin{subfigure}[b]{0.49\textwidth}
    \includegraphics[width=\textwidth]{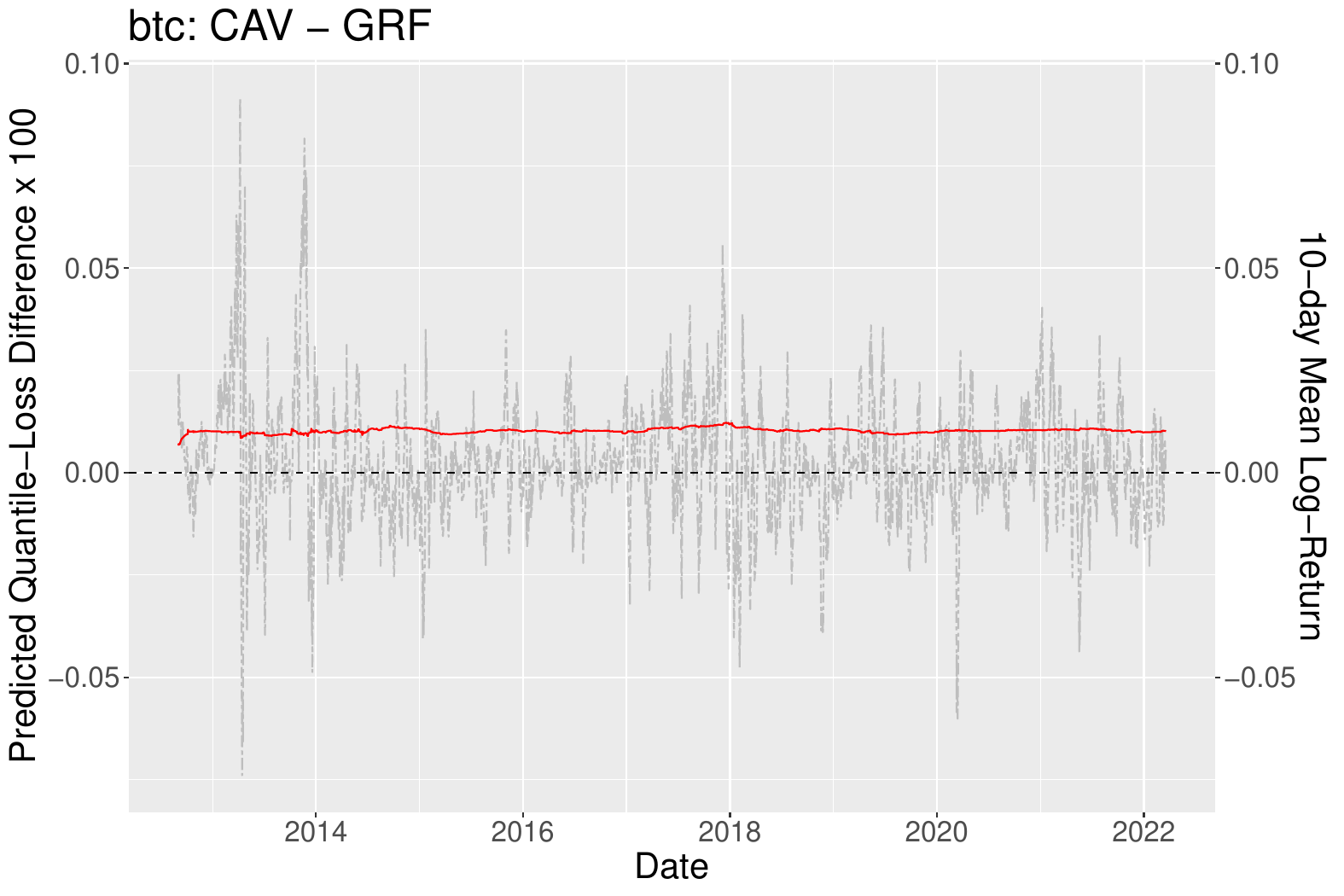}
    \end{subfigure}
    \hfill
    \begin{subfigure}[b]{0.49\textwidth}
    \includegraphics[width=\textwidth]{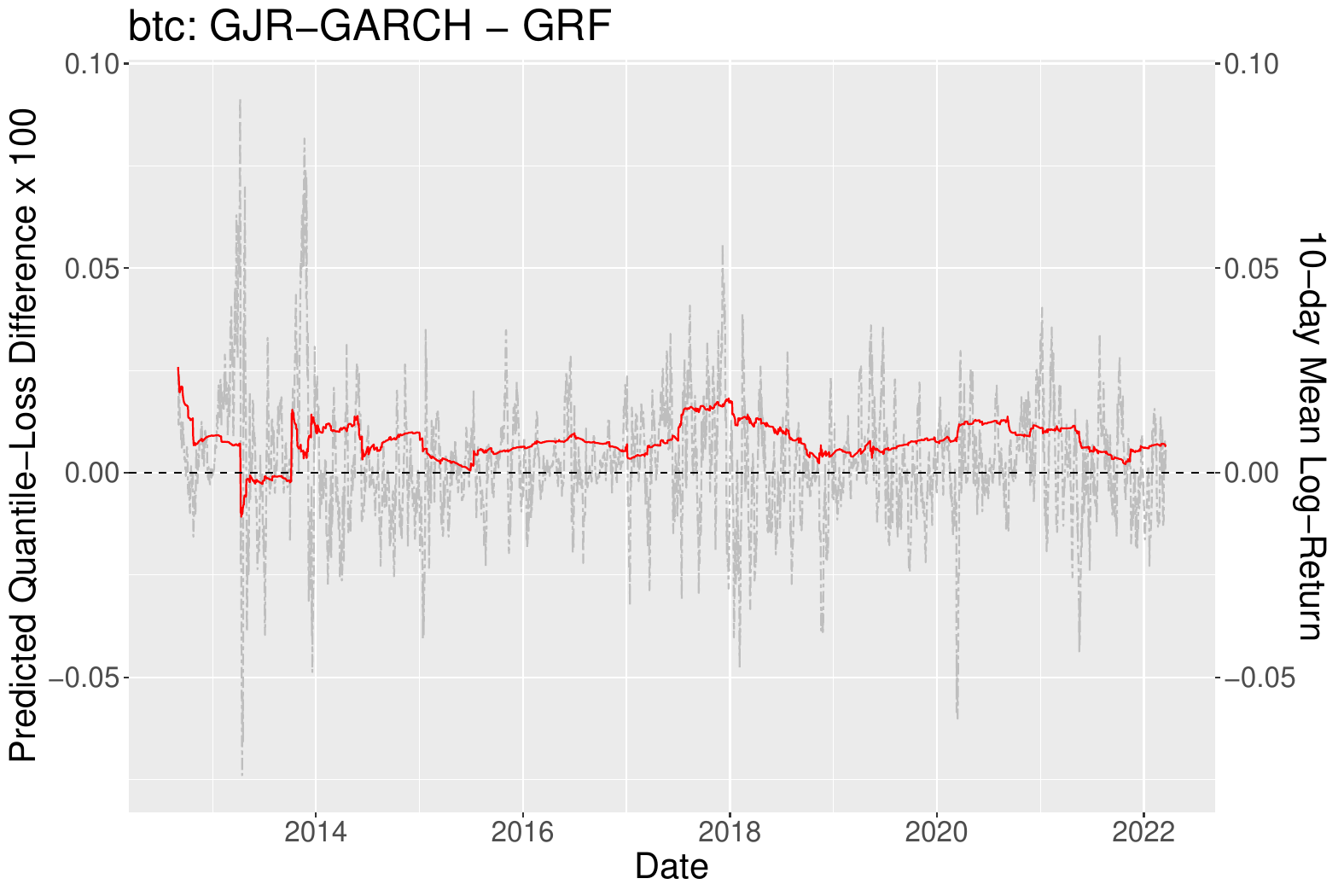}
    \end{subfigure}
    \begin{subfigure}[b]{0.49\textwidth}
    \includegraphics[width=\textwidth]{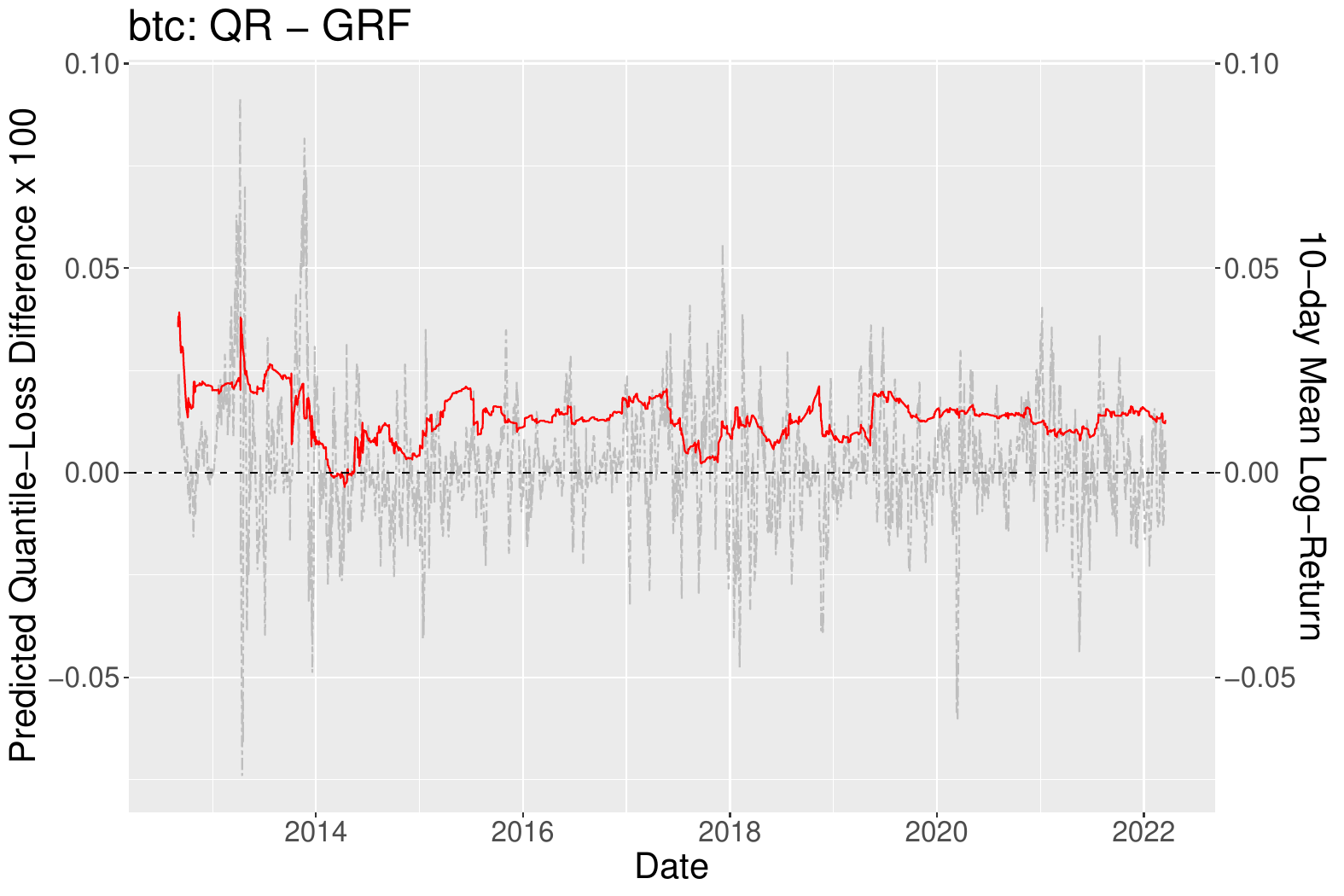}
    \end{subfigure}
    \hfill
    
    \caption{Rolling 180-day mean of predicted loss difference series ({\color{red}red}) of $ h_t \hat{\beta}_0$ of CPA tests on Bitcoin (btc) predicted 5\% VaR with $l=500$ for GRF vs. CAV (left), GJR-GARCH (right), and QR (bottom center) with rolling mean 10-day log-returns in grey. A positive predicted loss difference indicates that the prediction error of GRF is smaller than of the compared method.}
    \label{fig:Loss_difference_Bitcoin_3}
\end{figure}

\begin{figure}[htb]
    \centering
    \begin{subfigure}[b]{0.49\textwidth}
    \includegraphics[width=\textwidth]{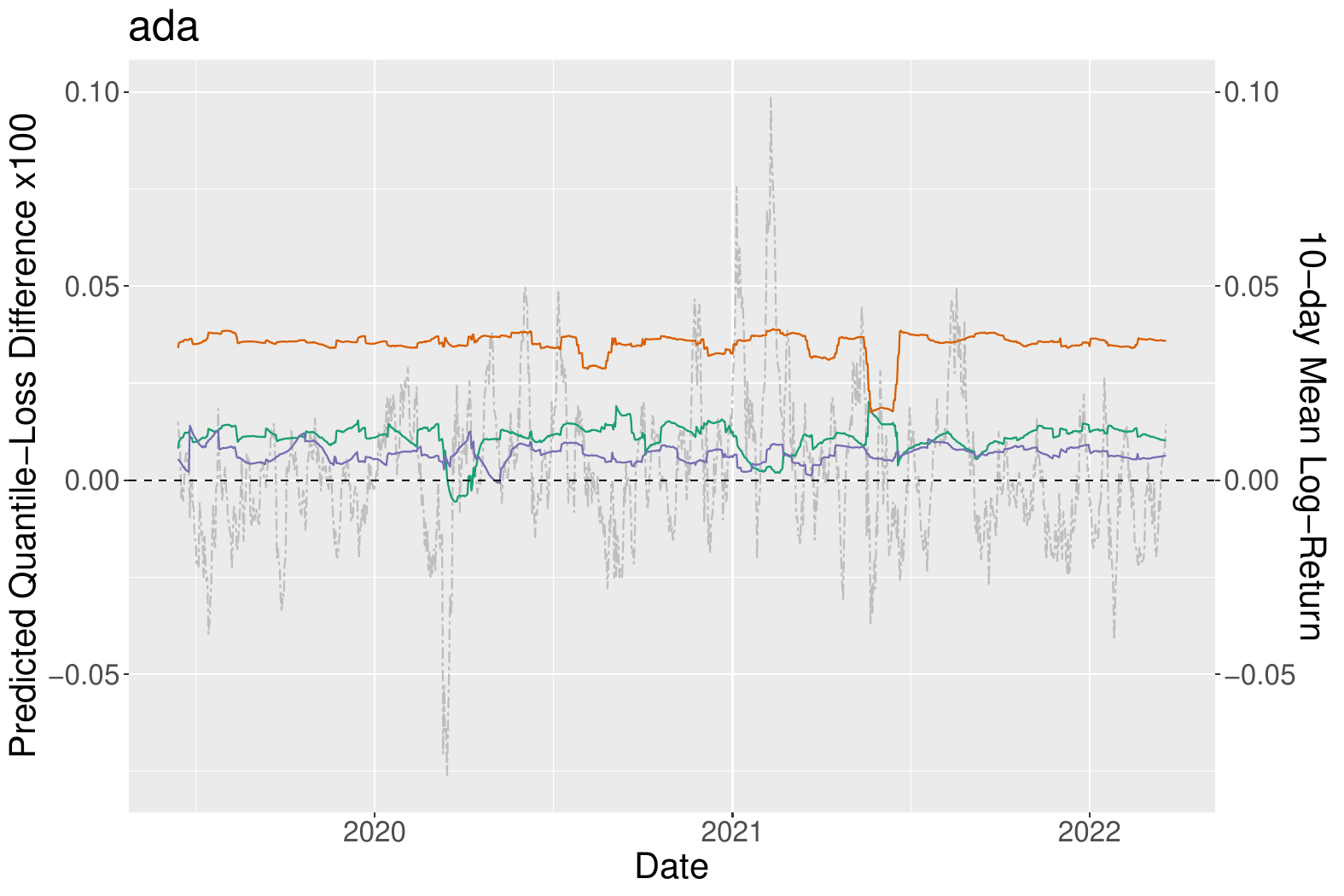}
    \end{subfigure}
    \hfill
    \begin{subfigure}[b]{0.49\textwidth}
    \includegraphics[width=\textwidth]{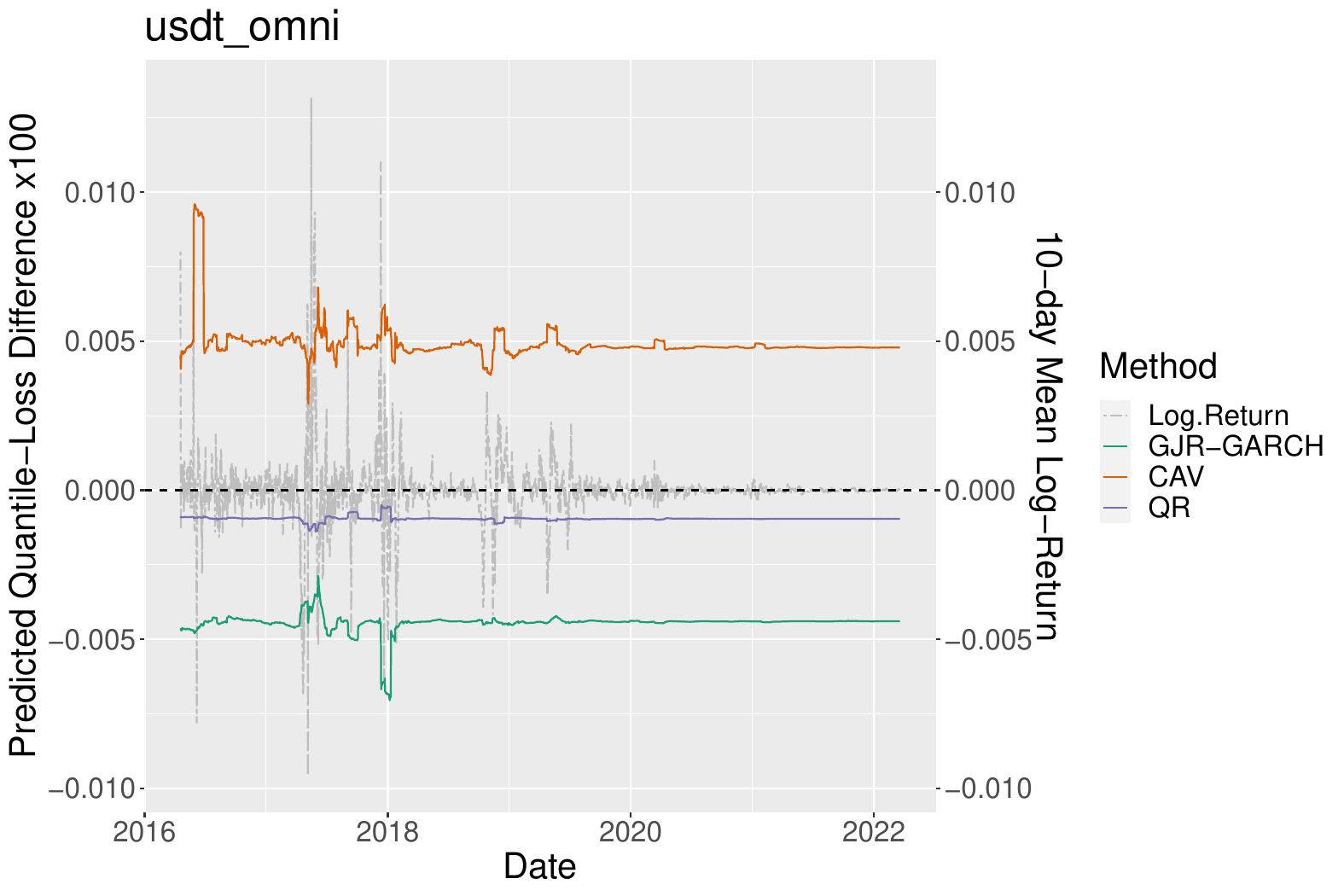}
    \end{subfigure}

    \caption{Rolling 30-day mean of predicted loss difference series of $ h_t \hat{\beta}_0$ of CPA tests on Cardano/Tether (ada/usdt\_omni) predicted 5\% VaR with $l=500$ for GRF vs. CAV (orange), GJR-GARCH (green), and QR (purple) with rolling mean 10-day log-returns in grey. A positive predicted loss difference indicates that the prediction error of GRF is smaller than of the compared method.}
    \label{fig:Loss_difference_ada_usdt}
\end{figure}

First, we look at Bitcoin (btc), the largest and most popular currency, where GRF largely outperforms QR, CAV, and GJR-GARCH in the CPA-tests. Figure \ref{fig:Loss_difference_Bitcoin_3} shows the predicted loss difference for each of the different methods. GRF outperforms the other methods consistently for most time frames. 
Secondly, as summarized on the left in Figure \ref{fig:Loss_difference_ada_usdt}, we look at Cardano (ada), a large and fairly new currency offering e.g. smart contracts or supply chain tracking. There, GRF is significantly outperforming GJR-GARCH and CAV, while being slightly better than QR, although not reaching a significant level. 
Finally, we also look more closely at Tether (usdt\_omni) as the largest stablecoin that is roughly bound to the USD\footnote{As it is backed by USD cash reserves, see \url{https://tether.to/en/}.}. The right part of Figure \ref{fig:Loss_difference_ada_usdt} shows the 30-day rolling means of the predicted loss differences. Notably, the rolling loss difference and rolling mean-return are already around 10 times smaller than those of btc or ada, indicating that usdt\_omni substantially differs from the other two currencies. While the losses of QR are very similar to those of GRF, CAV is significantly ouperformed. Only GJR-GARCH consistently has lower predicted losses than GRF, although they are deemed not significant by the CPA tests (see Figure \ref{fig:CPA_full_cpa_base} in the appendix for detailed results for the full time frame). 

\begin{figure}[htb]
    \centering
        \includegraphics[width=0.9\textwidth]{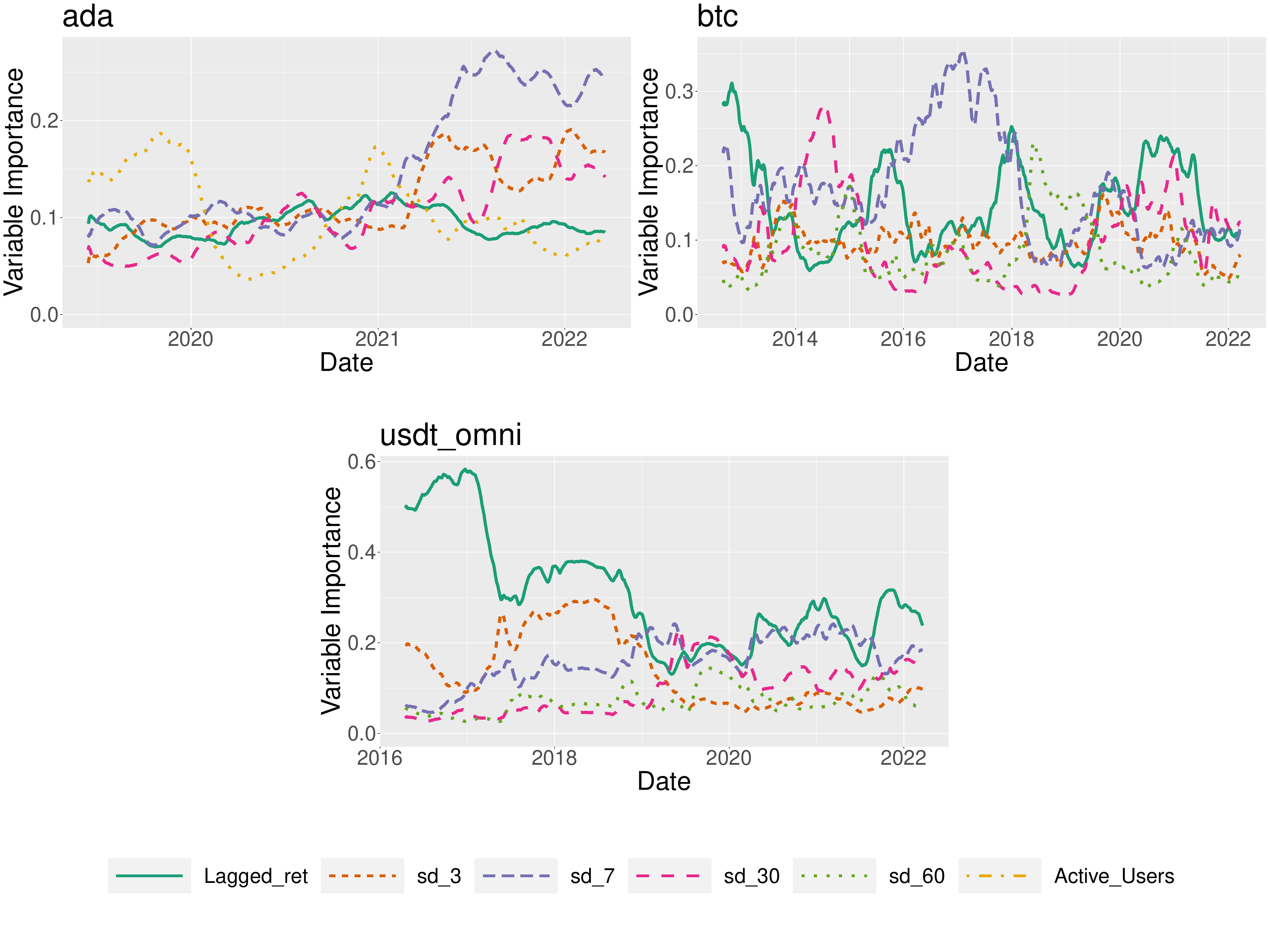}
       \caption{Rolling 30-day mean of the GRF variable importance in the out-of-sample period for the full data of ada, btc, and usdt\_omni, for predicted one-day ahead 5\% VaR with $l=500$. Plotted are the 5 most important variables for each cryptocurrency. Variable importance of covariate $x_p$ is measured as the proportion of splits on $x_p$ relative to all splits in a respective layer $l$ (over all trees in a trained forest), weighted by layer $l$. Variable description can be seen in Section \ref{Sec:Data}.}
    \label{fig:Var_imp_ada_btc_usdt}
   \end{figure}

To further understand the drivers of the GRF performance, we obtain variable importance measures that depict the frequency of inclusion in splits of the forest. The variable importance of a covariate $x_p$ is measured as the proportion of splits on $x_p$ relative to all splits in a respective layer $l$ (over all trees in a trained forest), weighted by layer $l$\footnote{We use a maximum depth of $d_{max}=5$ corresponding to the number of covariates and a weight decay of 2, meaning a split further down in each tree receives less weight $w_l$ in the final frequency as it is less important for the three specific currencies analyzed in the previous section. Specifically, for layer $l=1,\dots,5$, $w_l=\dfrac{l^{-2}}{\sum_{l=1}^5 l^{-2}}$.}. In Figure \ref{fig:Var_imp_ada_btc_usdt}, the importance difference of certain covariates over time for the three currencies is clearly visible.
Overall, the lagged return is very important for predicting VaR when returns are quite extreme relative to all returns in a specific asset, in the case of btc in times of hypes and crashes. Intuitively, this finding seems reasonable as in times of bubbles, when the volatility is driven by some short, bubble-like events and returns are highly variable, volatility lagged over a longer time horizon is less predictive for VaR and predictions are driven by events happening shortly before the prediction. In rather unstable times, but not in extreme cases, the lagged SD-measures gain importance, while the extra covariates only play a role for assets with relatively small volumes, e.g. when new currencies are created. This also explains why GRF-X performs much better for new, low-market-cap assets in Period 3 (see e.g. Figure \ref{fig:CPA_p3_cap} in the appendix).
   
Starting with ada, we see that it is the only asset of the three where the number of active addresses plays an important role, where for the other two assets, the 60-day lagged standard deviation is more important. The importance of variables can be split into two periods. The period until the beginning of 2021 is largely dominated by measures that somehow account for trading activity (Active\_Users,Total\_Users/\_USD100/\_USD10, Transactions). For reasons of clarity, we only plot the most important one of these measures, Active\_Users. The spike of the latter in importance at the beginning of 2021 is likely caused by the massive increase in price and market cap during that time, representing a period of hype with many actively trading users\footnote{See also Figure \ref{fig:log_returns_cryptos} in the appendix for an overview of the log-returns of the three currencies.}. For the rest of the time period, lagged SD, mostly 3-day lagged SD, followed by 30-day and 60-day SD, is dominating the predictions of GRF. This change of importance seems reasonable as the structure of the asset fundamentally changes with the price increasing tenfold and the volume increasing strongly at the same time.
 
 For btc, we have much more data covering 10 years, which is why the important variables change frequently in different periods. Lagged return is naturally important in phases of extreme hype and crashes that are characterized by large positive and negative returns, e.g. in the very beginning (where the price was still quite low), at the end of 2013 (the first time btc had a price of USD 1000), at the end of 2017 (with a price over USD 19,000), and from mid 2020 to the mid 2021, where there were multiple hypes and crashes during the COVID-19 pandemic. Between these hype periods, the lagged SDs are most important. In 2014-2015 and from the end of 2019 to mid 2021 , 30-day SD is contributing most to the GRF-predictions, followed by 7-day SD in 2016-2018 and 60-day SD 20 from 2018 to the end of 2019. This changing scheme is interesting, as 30-day SD seems to be a good predictor especially in very unstable times (return-wise), while 7-day and 60-day SD are more important in relatively stable times.
 
 Finally, the stablecoin usdt\_omni is an exception, being largely dominated by the lagged return. Given that by construction usdt\_omni is essentially bound to the US dollar, its dynamic properties also largely correspond to those of standard currencies. From mid 2019, the prices and returns are rather stable and the volume increases strongly, and the influence lagged 7-day SD increases slightly, while still being less important than lagged return. This is not surprising, as usdt\_omni is quite stable in comparison to btc and ada.

\subsubsection{The Post-Pandemic Period}\label{subsec:period4}
We additionally study the post-pandemic period 2023/01/01 - 2024/04/06 that is similar in terms of the number of traded currencies (83) to period 3 but has different exogenous constraints such as the war in Ukraine. Nevertheless, it is characterized by an overall lower level of volatility of all cryptocurrencies in the market. 

Here in the DQ-test, the GRF is  outperformed by the GJR-GARCH over all currencies, but for important subgroups the GRF-type still dominates. In particular, in Period 4, the impact of characteristics such as SER and number of traded assets is much more pronounced. 

\begin{table}[!htbp]
\centering
\caption{Period 4: Medians of P-Values for DQ-Tests} 
\label{tab:median_pvals_SER_p4}
\resizebox{\textwidth}{!}{
\begin{tabular}{lrrrrrrrrrrr}
  \toprule
Group & GRF & QRF & QR & CAV & CAV\_ASY & GJR-GARCH & Hist & GRF-X & QRF-X & QR-X & GARCH-X \\ 
  \midrule
  All  & 0.24 & 0.12 & 0.14 & 0.10 & 0.11 & 0.48 & 0.01 & 0.33 & 0.16 & 0.00 & 0.03 \\
  High & 0.22 & 0.07 & 0.23 & 0.12 & 0.15 & 0.33 & 0.00 & 0.32 & 0.09 & 0.00 & 0.00 \\ 
  Low & 0.25 & 0.14 & 0.11 & 0.06 & 0.10 & 0.49 & 0.01 & 0.34 & 0.25 & 0.00 & 0.09 \\   
   \bottomrule
\end{tabular}
}
\caption*{\scriptsize The table shows medians of p-values for DQ-tests. Higher median p-values indicate correct conditional coverage according to the DQ-test. The currencies are divided into two high and low groups based on the SER covariate as in Table \ref{tab:median_pvals_SER}. The remaining details correspond to Table \ref{tab:median_pvals}.}
\end{table}

This can be seen from the excellent DQ-performance results of GRF-X in Table \ref{tab:median_pvals_SER_p4} for currencies in the high SER-group that is supported by the impact results of Table \ref{tab:crypto_dif_rel_p4} where the two characteristics stand out in particular in relation to the periods before. Note that 24\% of currencies belong to the Group\_high in Period 4. The mean daily trading volume in USD in Group\_high is 334 296 265\$ in Period 4. In the pairwise CPA tests again GRF dominates overall but in terms of significant differences cannot be distinguished from GRF-X. For the pairwise CPA results for GRF-X  please see the Appendix.

\begin{table}[!htbp]
\centering
\caption{Difference Between Covariates of Cryptos Where GRF is Better vs. Worse} 
\label{tab:crypto_dif_rel_p4}
\def\arraystretch{0.5}
\resizebox{0.7\textwidth}{!}{
\begin{tabular}{rrrrrr}
  \toprule
 & Period 1 & Period 2 & Period 3 & Period 4 & Full Data \\ 
  \midrule
  Ret & 0.83 & 0.96 & 4.24 & 0.82 & 1.83 \\ 
  Active\_Users & 0.30 & 0.23 & 3.00 & 1.84 & 2.40 \\ 
  Total\_Users & 0.48 & 0.18 & 0.32 & 0.10 & 0.27 \\ 
  Total\_Users\_USD100 & 0.12 & 0.13 & 0.50 & 0.14 & 0.50 \\ 
  Total\_Users\_USD\_10 & 0.18 & 0.14 & 0.79 & 0.15 & 0.71 \\ 
  SER & 0.74 & 0.49 & 0.47 & 0.24 & 0.48 \\ 
  Transactions & 0.38 & 0.04 & 5.78 & 3.62 & 3.79 \\ 
  Velocity & 4.11 & 3.46 & 0.93 & 0.76 & 0.76 \\ 
  sd\_3 & 0.91 & 0.64 & 0.88 & 0.99 & 0.97 \\ 
  sd\_7 & 0.92 & 0.64 & 0.88 & 0.98 & 0.95 \\ 
  sd\_30 & 0.93 & 0.66 & 0.87 & 0.98 & 0.95 \\ 
  sd\_60 & 0.92 & 0.63 & 0.88 & 0.99 & 0.96 \\
   \bottomrule
\end{tabular}
}
\caption*{\scriptsize The table shows shares of groups of cryptocurrencies where at least two of CAV, QR, and GJR-GARCH have better CPA-performance than GRF divided by the remaining rest - calculated as in Table \ref{tab:crypto_dif_rel}. The numerators of the respective quotients are mean values over all cryptocurrencies for the median of each covariate in the respective time period. A value smaller than 1 indicates that currencies where GRF performs better have a higher average median value of the respective covariate than the rest of the currencies.}
\end{table}

\begin{table}[!htb]
\begin{minipage}{\linewidth}
\centering
\caption{Period 4: Performance and Significance of CPA-tests for GRF Without Additional Covariates} 
\label{tab:CPA_p1_2_grf_base_p4}
\resizebox{0.9\textwidth}{!}{
\begin{tabular}{rrrrrrrrrrr}
  \toprule
GRF vs.: & QRF & QR & CAV & CAV\_ASY & GJR-GARCH & Hist & GRF-X & QRF-X & QR-X & GARCH-X \\ 
  \midrule
  \multicolumn{10}{l}{\textit{Share of GRF With Better Performance}}  \\
  Period 4  & 0.90 & 0.70 & 0.70 & 0.75 & 0.65 & 0.94 & 0.33 & 0.94 & 0.98 & 0.73 \\
   \multicolumn{10}{l}{\textit{Share of GRF With Significantly Better Performance}}  \\
  Period 4 & 0.51 & 0.17 & 0.17 & 0.23 & 0.13 & 0.67 & 0.05 & 0.70 & 0.76 & 0.53 \\ 
     \bottomrule
\end{tabular}
}
\end{minipage}
    \caption*{\scriptsize The top part shows summary values that are shares over all cryptocurrencies in period 4. It describes the number of times that GRF had a better performance (i.e. more than 50\% of predicted losses by the CPA test were smaller for the GRF) relative to all crytpocurrencies (in that period), and the number of times that GRF was significantly better (at least at a 10\% level) as judged by the CPA test over all crytpocurrencies (period 4). }
\end{table}

\section{Conclusion}\label{Sec:Conclusion}
In this paper, we show that random forests can significantly improve the forecast performance for VaR-predictions when tailored to conditional quantiles. In both simulations and in analyzing return data of the 105 largest cryptocurrencies, the proposed GRF-type random forest shows superior prediction performance. In particular, the adaptive  non-linear form of GRF appears to capture time-variations of volatility and spike-behavior in cryptocurrency returns especially well in contrast to more conventional financial econometric methods. We further show that the GRF is superior in assessing the tail risk of cryptocurrencies in times where volatility in returns is high which often coincides with increased speculation in the market. In such periods, standard time-series based  procedures, even when augmented with external factors, are substantially inferior in VaR predictions suggesting that a comparison of predicted losses could serve as an easy, empirical pre-screening device to detect such speculative bubbles.

Our findings are highly relevant for the risk assessment of cryptocurrencies where we show that classic methods might lead to a miss-quantification of associated risks. Such results directly impact investment and hedging decisions for respective assets. Specifically, we detect superior performance of GRF especially for highly volatile cryptocurrencies that have a high number of active users and could thus be potentially prone to speculation and hypes. On the other hand, for the class of stablecoins that are usually tied to some large, classic currency such as the USD, and which are usually dominated by a smaller number of large accounts, other classic methods such as GJR-GARCH or quantile regression are more on par with GRF. This corresponds also to the simulation results mimicking the behavior of conventional assets such as as stock returns where the advantage of GRF decreases. Major gains of GRF, however, occur the more cryptocurrency-like the set-up becomes. The proposed random forest methodology allows identifying important external factors despite the non-linearity of the methodology. We show that such factors are time-varying and changing particularly in unstable times. For future research, this set of covariates could even be further augmented with other potentially driving real-time factors, such as for example properly extracted and filtered social media information. The relevance of such factors might also provide additional guidance for relevant exogenous information for portfolio formation and the design of e.g. trading strategies \citep[cp e.g.][]{platanakis2020should, huang2023diversification} .

\section*{Acknowlegements} 

We thank the audiences of the HKMetrics workshops and the COMPSTAT Conference 2024 for valuable comments and acknowledge excellent research assistance by Niklas Korn and Jonas Meirer.

\bibliography{References}

\begin{thebibliography}{40}
\expandafter\ifx\csname natexlab\endcsname\relax\def\natexlab#1{#1}\fi
\providecommand{\url}[1]{\texttt{#1}}
\providecommand{\href}[2]{#2}
\providecommand{\path}[1]{#1}
\providecommand{\DOIprefix}{doi:}
\providecommand{\ArXivprefix}{arXiv:}
\providecommand{\URLprefix}{URL: }
\providecommand{\Pubmedprefix}{pmid:}
\providecommand{\doi}[1]{\href{http://dx.doi.org/#1}{\path{#1}}}
\providecommand{\Pubmed}[1]{\href{pmid:#1}{\path{#1}}}
\providecommand{\bibinfo}[2]{#2}
\ifx\xfnm\relax \def\xfnm[#1]{\unskip,\space#1}\fi
\bibitem[{Ardia et~al.(2019)Ardia, Boudt \& Catania}]{Ardia2019}
\bibinfo{author}{Ardia, D.}, \bibinfo{author}{Boudt, K.}, \& \bibinfo{author}{Catania, L.} (\bibinfo{year}{2019}).
\newblock \bibinfo{title}{{Generalized autoregressive score models in R: The GAS package}}.
\newblock {\it \bibinfo{journal}{Journal of Statistical Software}\/},  {\it \bibinfo{volume}{88}\/}. \DOIprefix\doi{10.18637/jss.v088.i06}. \href{http://arxiv.org/abs/1609.02354}{\tt arXiv:1609.02354}.
\bibitem[{Athey et~al.(2019)Athey, Tibshirani \& Wager}]{Athey2019}
\bibinfo{author}{Athey, S.}, \bibinfo{author}{Tibshirani, J.}, \& \bibinfo{author}{Wager, S.} (\bibinfo{year}{2019}).
\newblock \bibinfo{title}{{Generalized random forests}}.
\newblock {\it \bibinfo{journal}{Annals of Statistics}\/},  {\it \bibinfo{volume}{47}\/}, \bibinfo{pages}{1179--1203}. \DOIprefix\doi{10.1214/18-AOS1709}. \href{http://arxiv.org/abs/1610.01271}{\tt arXiv:1610.01271}.
\bibitem[{Baur et~al.(2018)Baur, Hong \& Lee}]{Baur2018}
\bibinfo{author}{Baur, D.~G.}, \bibinfo{author}{Hong, K.~H.}, \& \bibinfo{author}{Lee, A.~D.} (\bibinfo{year}{2018}).
\newblock \bibinfo{title}{{Bitcoin: Medium of exchange or speculative assets?}}
\newblock {\it \bibinfo{journal}{Journal of International Financial Markets, Institutions and Money}\/},  {\it \bibinfo{volume}{54}\/}, \bibinfo{pages}{177--189}. \URLprefix \url{https://doi.org/10.1016/j.intfin.2017.12.004}. \DOIprefix\doi{10.1016/j.intfin.2017.12.004}.
\bibitem[{Bollerslev(1986)}]{Bollerslev1986}
\bibinfo{author}{Bollerslev, T.} (\bibinfo{year}{1986}).
\newblock \bibinfo{title}{{Generalized autoregressive conditional heteroskedasticity}}.
\newblock {\it \bibinfo{journal}{Journal of Econometrics}\/},  {\it \bibinfo{volume}{31}\/}, \bibinfo{pages}{307--327}. \DOIprefix\doi{10.1016/0304-4076(86)90063-1}.
\bibitem[{Breiman(1996)}]{breiman1996}
\bibinfo{author}{Breiman, L.} (\bibinfo{year}{1996}).
\newblock \bibinfo{title}{Some properties of splitting criteria}.
\newblock {\it \bibinfo{journal}{Machine Learning}\/},  {\it \bibinfo{volume}{24}\/}, \bibinfo{pages}{41--47}.
\bibitem[{Breiman(2001)}]{Breiman2001}
\bibinfo{author}{Breiman, L.} (\bibinfo{year}{2001}).
\newblock \bibinfo{title}{{Random Forests}}.
\newblock {\it \bibinfo{journal}{Machine Learning}\/},  {\it \bibinfo{volume}{45}\/}, \bibinfo{pages}{5--32}.
\bibitem[{Breiman et~al.(1984)Breiman, Friedman, Stone \& Olshen}]{breiman1984classification}
\bibinfo{author}{Breiman, L.}, \bibinfo{author}{Friedman, J.}, \bibinfo{author}{Stone, C.~J.}, \& \bibinfo{author}{Olshen, R.~A.} (\bibinfo{year}{1984}).
\newblock {\it \bibinfo{title}{{Classification and regression trees}}\/}.
\newblock \bibinfo{publisher}{CRC press}.
\bibitem[{Cheah \& Fry(2015)}]{CHEAH2015}
\bibinfo{author}{Cheah, E.~T.}, \& \bibinfo{author}{Fry, J.} (\bibinfo{year}{2015}).
\newblock \bibinfo{title}{{Speculative bubbles in Bitcoin markets? An empirical investigation into the fundamental value of Bitcoin}}.
\newblock {\it \bibinfo{journal}{Economics Letters}\/},  {\it \bibinfo{volume}{130}\/}, \bibinfo{pages}{32--36}. \URLprefix \url{https://www.sciencedirect.com/science/article/pii/S0165176515000890}. \DOIprefix\doi{10.1016/j.econlet.2015.02.029}.
\bibitem[{Christoffersen(1998)}]{Christoffersen1998}
\bibinfo{author}{Christoffersen, P.~F.} (\bibinfo{year}{1998}).
\newblock \bibinfo{title}{{Evaluating Interval Forecasts}}.
\newblock {\it \bibinfo{journal}{International Economic Review}\/},  {\it \bibinfo{volume}{39}\/}, \bibinfo{pages}{841}. \DOIprefix\doi{10.2307/2527341}.
\bibitem[{Chu et~al.(2017)Chu, Chan, Nadarajah \& Osterrieder}]{chu2017garch}
\bibinfo{author}{Chu, J.}, \bibinfo{author}{Chan, S.}, \bibinfo{author}{Nadarajah, S.}, \& \bibinfo{author}{Osterrieder, J.} (\bibinfo{year}{2017}).
\newblock \bibinfo{title}{{GARCH Modelling of Cryptocurrencies}}.
\newblock {\it \bibinfo{journal}{Journal of Risk and Financial Management}\/},  {\it \bibinfo{volume}{10}\/}, \bibinfo{pages}{17}. \DOIprefix\doi{10.3390/jrfm10040017}.
\bibitem[{Elendner et~al.(2017)Elendner, Trimborn, Ong \& Lee}]{elendner2018cross}
\bibinfo{author}{Elendner, H.}, \bibinfo{author}{Trimborn, S.}, \bibinfo{author}{Ong, B.}, \& \bibinfo{author}{Lee, T.~M.} (\bibinfo{year}{2017}).
\newblock \bibinfo{title}{{The Cross-Section of Crypto-Currencies as Financial Assets: Investing in Crypto-Currencies Beyond Bitcoin}}.
\newblock In {\it \bibinfo{booktitle}{Handbook of Blockchain, Digital Finance, and Inclusion, Volume 1: Cryptocurrency, FinTech, InsurTech, and Regulation}\/} (pp. \bibinfo{pages}{145--173}).
\newblock \bibinfo{publisher}{Elsevier}.
\newblock \DOIprefix\doi{10.1016/B978-0-12-810441-5.00007-5}.
\bibitem[{Engle \& Manganelli(2004)}]{Engle2004}
\bibinfo{author}{Engle, R.~F.}, \& \bibinfo{author}{Manganelli, S.} (\bibinfo{year}{2004}).
\newblock \bibinfo{title}{{CAViaR: Conditional autoregressive value at risk by regression quantiles}}.
\newblock {\it \bibinfo{journal}{Journal of Business and Economic Statistics}\/},  {\it \bibinfo{volume}{22}\/}, \bibinfo{pages}{367--381}. \URLprefix \url{https://www.tandfonline.com/doi/abs/10.1198/073500104000000370}. \DOIprefix\doi{10.1198/073500104000000370}.
\bibitem[{Ghysels \& Nguyen(2019)}]{Ghysels2019}
\bibinfo{author}{Ghysels, E.}, \& \bibinfo{author}{Nguyen, G.} (\bibinfo{year}{2019}).
\newblock \bibinfo{title}{{Price Discovery of a Speculative Asset: Evidence from a Bitcoin Exchange}}.
\newblock {\it \bibinfo{journal}{Journal of Risk and Financial Management}\/},  {\it \bibinfo{volume}{12}\/}, \bibinfo{pages}{164}. \DOIprefix\doi{10.3390/jrfm12040164}.
\bibitem[{Giacomini \& Komunjer(2005)}]{Giacomini2005}
\bibinfo{author}{Giacomini, R.}, \& \bibinfo{author}{Komunjer, I.} (\bibinfo{year}{2005}).
\newblock \bibinfo{title}{{Evaluation and combination of conditional quantile forecasts}}.
\newblock {\it \bibinfo{journal}{Journal of Business and Economic Statistics}\/},  {\it \bibinfo{volume}{23}\/}, \bibinfo{pages}{416--431}. \DOIprefix\doi{10.1198/073500105000000018}.
\bibitem[{Giacomini \& White(2006)}]{Giacomini2006}
\bibinfo{author}{Giacomini, R.}, \& \bibinfo{author}{White, H.} (\bibinfo{year}{2006}).
\newblock \bibinfo{title}{{Tests of conditional predictive ability}}.
\newblock {\it \bibinfo{journal}{Econometrica}\/},  {\it \bibinfo{volume}{74}\/}, \bibinfo{pages}{1545--1578}. \DOIprefix\doi{10.1111/j.1468-0262.2006.00718.x}.
\bibitem[{Gkillas \& Katsiampa(2018)}]{gkillas2018application}
\bibinfo{author}{Gkillas, K.}, \& \bibinfo{author}{Katsiampa, P.} (\bibinfo{year}{2018}).
\newblock \bibinfo{title}{{An application of extreme value theory to cryptocurrencies}}.
\newblock {\it \bibinfo{journal}{Economics Letters}\/},  {\it \bibinfo{volume}{164}\/}, \bibinfo{pages}{109--111}.
\bibitem[{Glaser et~al.(2014)Glaser, Zimmermann, Haferkorn, Weber \& Siering}]{Glaser2014}
\bibinfo{author}{Glaser, F.}, \bibinfo{author}{Zimmermann, K.}, \bibinfo{author}{Haferkorn, M.}, \bibinfo{author}{Weber, M.~C.}, \& \bibinfo{author}{Siering, M.} (\bibinfo{year}{2014}).
\newblock \bibinfo{title}{{Bitcoin - Asset or currency? Revealing users' hidden intentions}}.
\newblock {\it \bibinfo{journal}{ECIS 2014 Proceedings - 22nd European Conference on Information Systems}\/},  (pp. \bibinfo{pages}{1--14}).
\bibitem[{Glosten et~al.(1993)Glosten, Jagannathan \& Runkle}]{GLOSTEN_1993}
\bibinfo{author}{Glosten, L.~R.}, \bibinfo{author}{Jagannathan, R.}, \& \bibinfo{author}{Runkle, D.~E.} (\bibinfo{year}{1993}).
\newblock \bibinfo{title}{On the relation between the expected value and the volatility of the nominal excess return on stocks}.
\newblock {\it \bibinfo{journal}{The Journal of Finance}\/},  {\it \bibinfo{volume}{48}\/}, \bibinfo{pages}{1779--1801}. \DOIprefix\doi{10.1111/j.1540-6261.1993.tb05128.x}.
\bibitem[{Hafner(2020)}]{Hafner2020}
\bibinfo{author}{Hafner, C.~M.} (\bibinfo{year}{2020}).
\newblock \bibinfo{title}{{Testing for Bubbles in Cryptocurrencies with Time-Varying Volatility}}.
\newblock {\it \bibinfo{journal}{Journal of Financial Econometrics}\/},  {\it \bibinfo{volume}{18}\/}, \bibinfo{pages}{233--249}. \DOIprefix\doi{10.1093/jjfinec/nby023}.
\bibitem[{Hastie et~al.(2009)Hastie, Tibshirani \& Friedman}]{Hastie2009}
\bibinfo{author}{Hastie, T.}, \bibinfo{author}{Tibshirani, R.}, \& \bibinfo{author}{Friedman, J.} (\bibinfo{year}{2009}).
\newblock {\it \bibinfo{title}{{The Elements of Statistical Learning}}\/}.
\newblock (\bibinfo{edition}{2nd} ed.).
\newblock \bibinfo{publisher}{Springer}.
\bibitem[{Hencic \& Gouri{\'{e}}roux(2015)}]{hencic2015noncausal}
\bibinfo{author}{Hencic, A.}, \& \bibinfo{author}{Gouri{\'{e}}roux, C.} (\bibinfo{year}{2015}).
\newblock \bibinfo{title}{{Noncausal autoregressive model in application to bitcoin/USD exchange rates}}.
\newblock In {\it \bibinfo{booktitle}{Studies in Computational Intelligence}\/} (pp. \bibinfo{pages}{17--40}).
\newblock \bibinfo{publisher}{Springer} volume \bibinfo{volume}{583}.
\newblock \DOIprefix\doi{10.1007/978-3-319-13449-9_2}.
\bibitem[{Huang et~al.(2023)Huang, Han, Newton, Platanakis, Stafylas \& Sutcliffe}]{huang2023diversification}
\bibinfo{author}{Huang, X.}, \bibinfo{author}{Han, W.}, \bibinfo{author}{Newton, D.}, \bibinfo{author}{Platanakis, E.}, \bibinfo{author}{Stafylas, D.}, \& \bibinfo{author}{Sutcliffe, C.} (\bibinfo{year}{2023}).
\newblock \bibinfo{title}{The diversification benefits of cryptocurrency asset categories and estimation risk: pre and post covid-19}.
\newblock {\it \bibinfo{journal}{The European Journal of Finance}\/},  {\it \bibinfo{volume}{29}\/}, \bibinfo{pages}{800--825}.
\bibitem[{Koenker \& Bassett(1978)}]{Koenker1978}
\bibinfo{author}{Koenker, R.}, \& \bibinfo{author}{Bassett, G.} (\bibinfo{year}{1978}).
\newblock \bibinfo{title}{{Regression Quantiles}}.
\newblock {\it \bibinfo{journal}{Econometrica}\/},  {\it \bibinfo{volume}{46}\/}, \bibinfo{pages}{33}. \DOIprefix\doi{10.2307/1913643}.
\bibitem[{Koenker \& Hallock(2001)}]{koenker2001quantile}
\bibinfo{author}{Koenker, R.}, \& \bibinfo{author}{Hallock, K.~F.} (\bibinfo{year}{2001}).
\newblock \bibinfo{title}{Quantile regression}.
\newblock {\it \bibinfo{journal}{Journal of economic perspectives}\/},  {\it \bibinfo{volume}{15}\/}, \bibinfo{pages}{143--156}.
\bibitem[{Kupiec(1995)}]{Kupiec1995}
\bibinfo{author}{Kupiec, P.~H.} (\bibinfo{year}{1995}).
\newblock \bibinfo{title}{{Techniques for Verifying the Accuracy of Risk Measurement Models}}.
\newblock {\it \bibinfo{journal}{The Journal of Derivatives}\/},  {\it \bibinfo{volume}{3}\/}, \bibinfo{pages}{73--84}. \DOIprefix\doi{10.3905/jod.1995.407942}.
\bibitem[{Kwiatkowski et~al.(1992)Kwiatkowski, Phillips, Schmidt \& Shin}]{Kwiatkowski1992}
\bibinfo{author}{Kwiatkowski, D.}, \bibinfo{author}{Phillips, P.~C.}, \bibinfo{author}{Schmidt, P.}, \& \bibinfo{author}{Shin, Y.} (\bibinfo{year}{1992}).
\newblock \bibinfo{title}{{Testing the null hypothesis of stationarity against the alternative of a unit root. How sure are we that economic time series have a unit root?}}
\newblock {\it \bibinfo{journal}{Journal of Econometrics}\/},  {\it \bibinfo{volume}{54}\/}, \bibinfo{pages}{159--178}. \DOIprefix\doi{10.1016/0304-4076(92)90104-Y}.
\bibitem[{Liu et~al.(2020)Liu, Semeyutin, Lau \& Gozgor}]{liu2020forecasting}
\bibinfo{author}{Liu, W.}, \bibinfo{author}{Semeyutin, A.}, \bibinfo{author}{Lau, C. K.~M.}, \& \bibinfo{author}{Gozgor, G.} (\bibinfo{year}{2020}).
\newblock \bibinfo{title}{{Forecasting Value-at-Risk of Cryptocurrencies with RiskMetrics type models}}.
\newblock {\it \bibinfo{journal}{Research in International Business and Finance}\/},  {\it \bibinfo{volume}{54}\/}, \bibinfo{pages}{101259}. \DOIprefix\doi{10.1016/j.ribaf.2020.101259}.
\bibitem[{Liu \& Tsyvinski(2020)}]{Liu2020}
\bibinfo{author}{Liu, Y.}, \& \bibinfo{author}{Tsyvinski, A.} (\bibinfo{year}{2020}).
\newblock \bibinfo{title}{Risks and returns of cryptocurrency}.
\newblock {\it \bibinfo{journal}{The Review of Financial Studies}\/},  {\it \bibinfo{volume}{34}\/}, \bibinfo{pages}{2689--2727}. \DOIprefix\doi{10.1093/rfs/hhaa113}.
\bibitem[{Maciel(2020)}]{maciel2020cryptocurrencies}
\bibinfo{author}{Maciel, L.} (\bibinfo{year}{2020}).
\newblock \bibinfo{title}{{Cryptocurrencies value-at-risk and expected shortfall: Do regime-switching volatility models improve forecasting?}}
\newblock {\it \bibinfo{journal}{International Journal of Finance \& Economics}\/}, .
\bibitem[{Meinshausen(2006)}]{Meinshausen2006}
\bibinfo{author}{Meinshausen, N.} (\bibinfo{year}{2006}).
\newblock \bibinfo{title}{{Quantile Regression Forests}}.
\newblock {\it \bibinfo{journal}{Journal of Machine Learning Research}\/},  {\it \bibinfo{volume}{7}\/}, \bibinfo{pages}{983--999}.
\bibitem[{Pafka \& Kondor(2001)}]{Pafka2001}
\bibinfo{author}{Pafka, S.}, \& \bibinfo{author}{Kondor, I.} (\bibinfo{year}{2001}).
\newblock \bibinfo{title}{{Evaluating the RiskMetrics methodology in measuring volatility and Value-at-Risk in financial markets}}.
\newblock {\it \bibinfo{journal}{Physica A: Statistical Mechanics and its Applications}\/},  {\it \bibinfo{volume}{299}\/}, \bibinfo{pages}{305--310}. \DOIprefix\doi{10.1016/S0378-4371(01)00310-7}. \href{http://arxiv.org/abs/0103107}{\tt arXiv:0103107}.
\bibitem[{Petukhina et~al.(2021)Petukhina, Trimborn, H{\"{a}}rdle \& Elendner}]{Petukhina2021}
\bibinfo{author}{Petukhina, A.}, \bibinfo{author}{Trimborn, S.}, \bibinfo{author}{H{\"{a}}rdle, W.~K.}, \& \bibinfo{author}{Elendner, H.} (\bibinfo{year}{2021}).
\newblock \bibinfo{title}{{Investing with cryptocurrencies – evaluating their potential for portfolio allocation strategies}}.
\newblock {\it \bibinfo{journal}{Quantitative Finance}\/},  {\it \bibinfo{volume}{0}\/}, \bibinfo{pages}{1--29}. \URLprefix \url{https://doi.org/10.1080/14697688.2021.1880023}. \DOIprefix\doi{10.1080/14697688.2021.1880023}.
\bibitem[{Platanakis \& Urquhart(2019)}]{platanakis2019portfolio}
\bibinfo{author}{Platanakis, E.}, \& \bibinfo{author}{Urquhart, A.} (\bibinfo{year}{2019}).
\newblock \bibinfo{title}{{Portfolio management with cryptocurrencies: The role of estimation risk}}.
\newblock {\it \bibinfo{journal}{Economics Letters}\/},  {\it \bibinfo{volume}{177}\/}, \bibinfo{pages}{76--80}.
\bibitem[{Platanakis \& Urquhart(2020)}]{platanakis2020should}
\bibinfo{author}{Platanakis, E.}, \& \bibinfo{author}{Urquhart, A.} (\bibinfo{year}{2020}).
\newblock \bibinfo{title}{Should investors include bitcoin in their portfolios? a portfolio theory approach}.
\newblock {\it \bibinfo{journal}{The British accounting review}\/},  {\it \bibinfo{volume}{52}\/}, \bibinfo{pages}{100837}.
\bibitem[{Selmi et~al.(2018)Selmi, Tiwari \& Hammoudeh}]{Selmi2018}
\bibinfo{author}{Selmi, R.}, \bibinfo{author}{Tiwari, A.}, \& \bibinfo{author}{Hammoudeh, S.} (\bibinfo{year}{2018}).
\newblock \bibinfo{title}{{Efficiency or speculation? A dynamic analysis of the Bitcoin market}}.
\newblock {\it \bibinfo{journal}{Economics Bulletin}\/},  {\it \bibinfo{volume}{38}\/}, \bibinfo{pages}{2037--2046}.
\bibitem[{Silvennoinen \& Ter{\"a}svirta(2016)}]{silvennoinen_testing_2016}
\bibinfo{author}{Silvennoinen, A.}, \& \bibinfo{author}{Ter{\"a}svirta, T.} (\bibinfo{year}{2016}).
\newblock \bibinfo{title}{Testing constancy of unconditional variance in volatility models by misspecification and specification tests}.
\newblock {\it \bibinfo{journal}{Studies in Nonlinear Dynamics \& Econometrics}\/},  {\it \bibinfo{volume}{20}\/}. \URLprefix \url{https://www.degruyter.com/document/doi/10.1515/snde-2015-0033/html}. \DOIprefix\doi{10.1515/snde-2015-0033}.
\bibitem[{Takeda \& Sugiyama(2008)}]{Takeda2008}
\bibinfo{author}{Takeda, A.}, \& \bibinfo{author}{Sugiyama, M.} (\bibinfo{year}{2008}).
\newblock \bibinfo{title}{{N-Support Vector Machine As Conditional Value-At-Risk Minimization}}.
\newblock {\it \bibinfo{journal}{Proceedings of the 25th International Conference on Machine Learning}\/},  (pp. \bibinfo{pages}{1056--1063}). \DOIprefix\doi{10.1145/1390156.1390289}.
\bibitem[{Trimborn et~al.(2020)Trimborn, Li \& H{\"{a}}rdle}]{Trimborn2020}
\bibinfo{author}{Trimborn, S.}, \bibinfo{author}{Li, M.}, \& \bibinfo{author}{H{\"{a}}rdle, W.~K.} (\bibinfo{year}{2020}).
\newblock \bibinfo{title}{{Investing with Cryptocurrencies - A Liquidity Constrained Investment Approach}}.
\newblock {\it \bibinfo{journal}{Journal of Financial Econometrics}\/},  {\it \bibinfo{volume}{18}\/}, \bibinfo{pages}{280--306}. \DOIprefix\doi{10.1093/jjfinec/nbz016}.
\bibitem[{Trucios et~al.(2020)Trucios, Tiwari \& Alqahtani}]{trucios2020value}
\bibinfo{author}{Trucios, C.}, \bibinfo{author}{Tiwari, A.~K.}, \& \bibinfo{author}{Alqahtani, F.} (\bibinfo{year}{2020}).
\newblock \bibinfo{title}{{Value-at-risk and expected shortfall in cryptocurrencies' portfolio: A vine copula--based approach}}.
\newblock {\it \bibinfo{journal}{Applied Economics}\/},  {\it \bibinfo{volume}{52}\/}, \bibinfo{pages}{2580--2593}.
\bibitem[{Vigliotti \& Jones(2020)}]{vigliotti2020rise}
\bibinfo{author}{Vigliotti, M.~G.}, \& \bibinfo{author}{Jones, H.} (\bibinfo{year}{2020}).
\newblock \bibinfo{title}{{The Rise and Rise of Cryptocurrencies}}.
\newblock In {\it \bibinfo{booktitle}{The Executive Guide to Blockchain}\/} (pp. \bibinfo{pages}{71--91}).
\newblock \bibinfo{publisher}{Springer}.

\end{thebibliography}

\newpage


\appendix
\section{Additional Empirical Stylized Facts and the GRF Algorithm} \label{Sec:Appendix}
 \begin{algorithm}
 {\footnotesize
    \caption{Generalized Random Forest - Tree Building}
    \label{Algorithm_GRF}
  \begin{algorithmic}[1]
    \INPUT Set of ``honest'', subsampled observations $X_T$ and $R_T$; minimum node size $n_m$; quantile levels $\tau=(\tau_1,\dots,\tau_M)$
    \STATE \textbf{Growing the tree} Create root node $P_0$
    \STATE Initialize queue Q with $P_0$
    \WHILE{Queue is not empty}
      \STATE Take the oldest element from Q (Parent Node $P$) and remove it from Q
      \STATE Take a random subsample of $p$ variables index by set $P_{sub}=\{\Tilde{1},\dots,\Tilde{p}\}$ from $X_T$ on which to potentially split and take observations $x^{(P_{sub})}_i=(x^{(\Tilde{1})}_i,\dots,x^{(\Tilde{P})}_i)$ from $P$.
      \STATE Set $loss=\infty$
      \FOR{h in 1 to p}
      \STATE Compute quantiles $\theta_m$ of $r_t$ from parent node $P$ at $\tau_1,\dots,\tau_M$ and compute pseudo outcomes $\rho_t=\sum_{m=1}^M 1\{r_t>\theta_m \}$ for each $r_t \in P$.
      \STATE For each possible split point in $x^{(h)}$, compute the criterion from Equation \eqref{eq:gini}
      \STATE Save loss $s_h$ that minimizes this splitting criterion 
      \STATE Save optimal split point $split_h$
      \IF{$s_h<loss$}
      \STATE $loss \leftarrow s_h$
      \STATE $ind_h \leftarrow h$
      \ENDIF
      \ENDFOR
      \IF{Split on variable $h$ with $split_{ind_h}$ succeeded (based on hyperparameters)}
      \STATE Determine children $C_1,C_2$ according to optimal split
      \STATE Add both children $C_1,C_2$ to a new daughter node each with corresponding observations left and add these to Q
      \ENDIF
    \ENDWHILE
    \OUTPUT One tree of the forest
  \end{algorithmic}
}
\end{algorithm}

\begingroup
\renewcommand{\arraystretch}{0.5}

\begin{table}[ht]
\centering
\caption{Overview of All Employed Crypto Assets}
\label{tab: Appendix, overview assests}
\begin{tabular}{lllrrr}
  \hline
 id & Start.Date & End.Date & Obs. & X & Time.Periods \\ 
  \hline
  1inch & 2020-12-26 & 2024-04-07 &  1199 &  8 &            4 \\
     aave & 2020-10-10 & 2024-04-07 &  1276 &  8 &            4 \\
      ada & 2017-12-01 & 2024-04-07 &  2320 &  8 &         3, 4 \\
     algo & 2019-06-22 & 2024-04-07 &  1752 &  8 &         3, 4 \\
    alpha & 2020-10-11 & 2024-04-07 &  1275 &  8 &            4 \\
      ant & 2017-08-29 & 2024-04-07 &  2414 &  8 &         3, 4 \\
    avaxc & 2020-09-23 & 2024-04-06 &  1292 &  8 &            4 \\
    avaxp & 2020-09-23 & 2024-04-06 &  1292 &  8 &            4 \\
    avaxx & 2020-09-23 & 2024-04-06 &  1292 &  7 &            4 \\
      bal & 2020-06-25 & 2024-04-07 &  1383 &  8 &            4 \\
      bat & 2017-10-06 & 2024-04-07 &  2376 &  8 &         3, 4 \\
      bch & 2017-08-01 & 2024-04-07 &  2442 &  8 &         3, 4 \\
      bnb & 2017-07-15 & 2024-04-07 &  2459 &  8 &         3, 4 \\
  bnb\_eth & 2017-07-15 & 2019-04-21 &   646 &  8 &              \\
      bsv & 2018-11-15 & 2024-04-07 &  1971 &  8 &         3, 4 \\
      btc & 2010-07-18 & 2024-04-07 &  5013 &  8 &   1, 2, 3, 4 \\
      btg & 2017-10-25 & 2024-04-07 &  2357 &  8 &         3, 4 \\
     busd & 2019-09-20 & 2024-04-07 &  1662 &  8 &            4 \\
     comp & 2020-06-18 & 2024-04-07 &  1390 &  8 &            4 \\
      cro & 2019-03-20 & 2024-04-07 &  1846 &  8 &         3, 4 \\
      crv & 2020-08-15 & 2024-04-07 &  1332 &  8 &            4 \\
      cvc & 2017-09-11 & 2024-04-07 &  2401 &  8 &         3, 4 \\
      dai & 2019-11-20 & 2024-04-07 &  1601 &  8 &            4 \\
     dash & 2014-02-08 & 2024-04-07 &  3712 &  8 &   1, 2, 3, 4 \\
      dcr & 2016-05-17 & 2024-04-07 &  2883 &  8 &      2, 3, 4 \\
      dgb & 2015-02-10 & 2024-04-07 &  3345 &  8 &      2, 3, 4 \\
     doge & 2014-01-23 & 2024-04-07 &  3728 &  8 &   1, 2, 3, 4 \\
      dot & 2020-08-20 & 2024-04-07 &  1327 &  8 &            4 \\
     drgn & 2018-01-03 & 2024-04-07 &  2287 &  8 &         3, 4 \\
      elf & 2017-12-22 & 2024-04-07 &  2299 &  8 &         3, 4 \\
      eos & 2018-06-09 & 2024-04-07 &  2130 &  2 &         3, 4 \\
  eos\_eth & 2017-06-29 & 2018-06-01 &   338 &  8 &              \\
      etc & 2016-07-25 & 2024-04-07 &  2814 &  8 &      2, 3, 4 \\
      eth & 2015-08-08 & 2024-04-07 &  3166 &  8 &      2, 3, 4 \\
      ftt & 2019-08-20 & 2024-04-07 &  1693 &  8 &            4 \\
      fun & 2017-09-02 & 2024-04-07 &  2410 &  8 &         3, 4 \\
      gas & 2017-08-08 & 2024-04-07 &  2435 &  8 &         3, 4 \\
      gno & 2017-05-02 & 2024-04-07 &  2533 &  8 &         3, 4 \\
      gnt & 2017-02-19 & 2024-04-06 &  2604 &  8 &         3, 4 \\
     grin & 2019-01-29 & 2024-04-07 &  1896 &  2 &         3, 4 \\
     gusd & 2018-09-16 & 2024-04-07 &  2031 &  8 &         3, 4 \\
     hbtc & 2019-12-09 & 2024-04-07 &  1582 &  8 &            4 \\
     hedg & 2019-11-02 & 2022-01-27 &   818 &  8 &              \\
       ht & 2019-03-06 & 2024-04-07 &  1860 &  8 &         3, 4 \\
     husd & 2019-07-18 & 2022-11-17 &  1219 &  8 &              \\
      icp & 2021-05-11 & 2024-04-07 &  1063 &  8 &            4 \\
      kcs & 2020-04-04 & 2024-04-07 &  1465 &  5 &            4 \\
      knc & 2017-09-27 & 2024-04-07 &  2385 &  8 &         3, 4 \\
     lend & 2017-12-09 & 2024-04-06 &  2311 &  8 &         3, 4 \\
  leo\_eos & 2019-05-21 & 2024-04-06 &  1783 &  4 &         3, 4 \\
  leo\_eth & 2019-05-21 & 2024-04-06 &  1783 &  8 &         3, 4 \\
     link & 2017-09-29 & 2024-04-07 &  2383 &  8 &         3, 4 \\
     loom & 2018-05-03 & 2024-04-07 &  2167 &  8 &         3, 4 \\
  \hline
\end{tabular}
\end{table}

\begin{table}[ht]
\centering
\caption*{\textbf{Table \ref{tab: Appendix, overview assests}:} Continued}
\begin{tabular}{lllrrr}

  \hline
  id & Start.Date & End.Date & Obs. & X & Time.Periods \\ 
  \hline
  lpt & 2018-12-20 & 2024-04-07 &  1936 &  8 &         3, 4 \\
      ltc & 2013-04-01 & 2024-04-07 &  4025 &  8 &   1, 2, 3, 4 \\
     maid & 2014-07-10 & 2024-04-07 &  3560 &  8 &      2, 3, 4 \\
     mana & 2017-08-25 & 2024-04-07 &  2418 &  8 &         3, 4 \\
matic\_eth & 2019-04-27 & 2024-04-06 &  1807 &  8 &         3, 4 \\
      mkr & 2017-12-26 & 2024-04-07 &  2295 &  8 &         3, 4 \\
      neo & 2017-07-15 & 2024-04-07 &  2459 &  8 &         3, 4 \\
      nxm & 2020-08-26 & 2024-04-06 &  1320 &  8 &            4 \\
      omg & 2017-07-15 & 2024-04-07 &  2459 &  8 &         3, 4 \\
      pax & 2018-11-30 & 2024-04-07 &  1956 &  8 &         3, 4 \\
     paxg & 2020-02-15 & 2024-04-07 &  1514 &  8 &            4 \\
      pay & 2017-10-03 & 2024-04-07 &  2379 &  8 &         3, 4 \\
     perp & 2021-02-04 & 2024-04-07 &  1159 &  8 &            4 \\
     poly & 2018-06-15 & 2024-04-07 &  2124 &  8 &         3, 4 \\
     powr & 2017-11-02 & 2024-04-07 &  2349 &  8 &         3, 4 \\
      ppt & 2017-09-20 & 2022-09-12 &  1819 &  8 &              \\
     qash & 2017-11-06 & 2024-04-07 &  2345 &  8 &         3, 4 \\
      qnt & 2019-03-16 & 2024-04-07 &  1850 &  8 &         3, 4 \\
      ren & 2018-12-07 & 2024-04-07 &  1949 &  8 &         3, 4 \\
   renbtc & 2020-05-13 & 2024-04-06 &  1425 &  8 &            4 \\
      rep & 2016-10-04 & 2024-04-07 &  2743 &  8 &         3, 4 \\
  rev\_eth & 2020-03-26 & 2024-03-27 &  1463 &  8 &              \\
      sai & 2017-12-23 & 2019-11-30 &   708 &  8 &              \\
      snt & 2017-06-19 & 2024-04-07 &  2485 &  8 &         3, 4 \\
      snx & 2020-04-09 & 2024-04-07 &  1460 &  8 &            4 \\
      srm & 2020-08-11 & 2024-04-07 &  1336 &  8 &            4 \\
    sushi & 2020-09-01 & 2024-04-07 &  1315 &  8 &            4 \\
     swrv & 2020-09-22 & 2024-04-07 &  1294 &  8 &            4 \\
      trx & 2018-06-25 & 2024-04-07 &  2114 &  3 &         3, 4 \\
  trx\_eth & 2017-10-07 & 2018-06-24 &   261 &  8 &              \\
     tusd & 2018-07-06 & 2024-04-07 &  2103 &  7 &         3, 4 \\
      uma & 2020-09-08 & 2024-04-07 &  1308 &  8 &            4 \\
      uni & 2020-09-18 & 2024-04-07 &  1298 &  8 &            4 \\
     usdc & 2018-09-28 & 2024-04-07 &  2019 &  8 &         3, 4 \\
     usdk & 2020-06-13 & 2023-04-17 &  1039 &  8 &              \\
 usdt\_eth & 2017-11-28 & 2024-04-06 &  2322 &  8 &         3, 4 \\
usdt\_omni & 2014-10-06 & 2024-04-06 &  3471 &  8 &      2, 3, 4 \\
 usdt\_trx & 2019-04-16 & 2024-04-06 &  1818 &  8 &         3, 4 \\
      vtc & 2014-01-29 & 2023-12-10 &  3603 &  8 &      1, 2, 3 \\
     wbtc & 2018-11-24 & 2024-04-07 &  1962 &  8 &         3, 4 \\
     weth & 2017-12-12 & 2024-04-07 &  2309 &  8 &         3, 4 \\
     wnxm & 2020-08-26 & 2024-04-07 &  1321 &  8 &            4 \\
      wtc & 2017-08-28 & 2024-04-07 &  2415 &  8 &         3, 4 \\
     xaut & 2020-01-25 & 2024-04-07 &  1535 &  8 &            4 \\
      xem & 2015-04-01 & 2024-04-07 &  3295 &  3 &      2, 3, 4 \\
      xlm & 2015-09-30 & 2024-04-07 &  3113 &  8 &      2, 3, 4 \\
      xmr & 2014-05-20 & 2024-04-07 &  3611 &  2 &      2, 3, 4 \\
      xrp & 2014-08-15 & 2024-04-07 &  3524 &  8 &      2, 3, 4 \\
      xtz & 2018-06-30 & 2024-04-07 &  2109 &  8 &         3, 4 \\
      xvg & 2017-09-30 & 2024-04-07 &  2382 &  8 &         3, 4 \\
      yfi & 2020-07-25 & 2024-04-07 &  1353 &  8 &            4 \\
      zec & 2016-10-29 & 2024-04-07 &  2718 &  8 &         3, 4 \\
      zrx & 2017-08-11 & 2024-04-07 &  2432 &  8 &         3, 4 \\ 
   \hline
\end{tabular}
\caption{All employed cryptocurrencies where $X$ denotes the number of available covariates out of 8. Numbers in ``time periods'' mark periods where the respective asset is fully observed. All cryptocurrencies in this table, with the exception of vtc, exist until the end of period 4.}
\end{table}

\endgroup

\begin{table}[ht]
\centering
\caption{Overview of Non-Stationarity Tests} 
\label{tab:crypto_tests}
\def\arraystretch{0.5}
\begin{tabular}{rrrr}
  \toprule
 & KPSS\_level & KPSS\_trend & ADF \\ 
  \midrule
1inch & 0.10 & 0.10 & 0.01 \\ 
  ada & 0.10 & 0.10 & 0.01 \\ 
  algo & 0.10 & 0.03 & 0.01 \\ 
  alpha & 0.10 & 0.01 & 0.01 \\ 
  avaxc & 0.10 & 0.04 & 0.01 \\ 
  avaxp & 0.10 & 0.04 & 0.01 \\ 
  avaxx & 0.10 & 0.04 & 0.01 \\ 
  bnb & 0.06 & 0.07 & 0.01 \\ 
  bnb\_eth & 0.06 & 0.05 & 0.01 \\ 
  btc & 0.02 & 0.10 & 0.01 \\ 
  dash & 0.02 & 0.10 & 0.01 \\ 
  dot & 0.10 & 0.01 & 0.01 \\ 
  drgn & 0.08 & 0.10 & 0.01 \\ 
  hbtc & 0.10 & 0.02 & 0.01 \\ 
  ht & 0.05 & 0.10 & 0.01 \\ 
  icp & 0.01 & 0.10 & 0.01 \\ 
  lend & 0.10 & 0.04 & 0.01 \\ 
  omg & 0.10 & 0.10 & 0.01 \\ 
  perp & 0.10 & 0.08 & 0.01 \\ 
  poly & 0.10 & 0.03 & 0.01 \\ 
  ppt & 0.10 & 0.06 & 0.01 \\ 
  renbtc & 0.10 & 0.01 & 0.01 \\ 
  snx & 0.10 & 0.02 & 0.01 \\ 
  uni & 0.10 & 0.07 & 0.01 \\ 
  wbtc & 0.10 & 0.10 & 0.01 \\ 
  weth & 0.10 & 0.05 & 0.01 \\ 
  xem & 0.03 & 0.10 & 0.01 \\ 
  xmr & 0.10 & 0.09 & 0.01 \\ 
  yfi & 0.10 & 0.04 & 0.01 \\ 
   \bottomrule
\end{tabular} 
\caption*{\footnotesize Values in columns show p-values for KPSS-test and ADF-tests as described in Section \ref{Sec:Data}. We only show values for assets that have values smaller than $0.1$ for the KPSS-tests. No asset had p-values larger than $0.01$ for the ADF-tests.}
\end{table}

\begin{table}[ht]
\centering
\caption{External Covariates and Descriptions} 
\label{tab:crypto_cov_descr}
\begin{tabular}{llp{0.4\textwidth}}
  \toprule
Variable Name & Coding Coinmetrics & Description \\ 
  \midrule
Active\_Users & AdrActCnt & The number of unique active daily addresses \\ 
  Total\_Users & AdrBalCnt & The number of unique addresses that hold any amount of native units of that currency \\ 
  Total\_Users\_USD100 & AdrBalUSD100Cnt & The number of unique addresses that hold at least 100 USD of native units of that currency \\ 
  Total\_Users\_USD10 & AdrBalUSD10Cnt & The number of unique addresses that hold at least 10 USD of native units of that currency \\ 
  SER & SER & The supply equality ratio, i.e. the ratio of supply held by addresses with less than 1 over 10 millionth of the current supply to the top one percent of addresses with the highest current supply \\ 
  Transactions & TxCnt & The number of daily initiated transactions \\ 
  Velocity & VelCur1yr & The velocity of supply in the current year, which describes the the ratio of current supply to the sum of the value transferred in the last year \\ 
   \bottomrule
\end{tabular}
\caption*{\scriptsize The table contains all used external covariates. The variable coding corresponds to \url{https://docs.coinmetrics.io/}. Detailed variable descriptions are available on \url{https://docs.coinmetrics.io/info/metrics}.}
\end{table}

\section{Tables and Figures with Additional Results} \label{Sec:Appendixb}
\begin{table}[htbp]
\centering
\caption{Medians of P-Values for DQ-Tests Sorted by Covariate Values in Different Time Periods} 
\label{tab:median_pvals_sorted}
\resizebox{\textwidth}{!}{
\begin{tabular}{rlrrrrrrrrrrr}
  \toprule
Period & Sorting & GRF & QRF & QR & CAV & CAV\_ASY & GJR-GARCH & Hist & GRF-X & QRF-X & QR-X & GARCH-X \\ 
  \midrule
  \multicolumn{12}{l}{\textbf{High Group}} \\
  \multicolumn{12}{l}{\textit{SER}} \\
 1 & 1 Highest & 0.17 & 0.81 & 0.00 & 0.00 & 0.00 & 0.12 & 0.00 & 0.01 & 0.00 & 0.00 & 0.00 \\  
  2 & 5 Highest & 0.15 & 0.03 & 0.20 & 0.05 & 0.41 & 0.10 & 0.09 & 0.13 & 0.08 & 0.00 & 0.00 \\ 
  3 & 15 Highest & 0.41 & 0.04 & 0.08 & 0.06 & 0.08 & 0.16 & 0.00 & 0.22 & 0.22 & 0.00 & 0.00 \\
  \multicolumn{12}{l}{\textit{Number of Active Users}} \\
   1 & 1 Highest & 0.17 & 0.81 & 0.00 & 0.00 & 0.00 & 0.12 & 0.00 & 0.01 & 0.00 & 0.00 & 0.00 \\ 
  2 & 5 Highest & 0.15 & 0.06 & 0.40 & 0.10 & 0.41 & 0.11 & 0.09 & 0.14 & 0.06 & 0.00 & 0.00 \\ 
  3 & 15 Highest & 0.37 & 0.05 & 0.06 & 0.06 & 0.08 & 0.16 & 0.00 & 0.21 & 0.22 & 0.00 & 0.00 \\
   \multicolumn{12}{l}{\textit{7-Day Lagged Standard Deviation}} \\
    1 & 1 Highest & 0.14 & 0.02 & 0.03 & 0.00 & 0.00 & 0.01 & 0.00 & 0.00 & 0.00 & 0.00 & 0.01 \\ 
  2 & 5 Highest & 0.53 & 0.01 & 0.27 & 0.22 & 0.10 & 0.23 & 0.20 & 0.45 & 0.17 & 0.00 & 0.33 \\
  3 & 15 Highest & 0.04 & 0.00 & 0.01 & 0.00 & 0.00 & 0.05 & 0.00 & 0.01 & 0.00 & 0.00 & 0.02 \\ 
   \multicolumn{12}{l}{\textbf{Low Group}} \\
    \multicolumn{12}{l}{\textit{SER}} \\
    1 & 4 Lowest & 0.11 & 0.20 & 0.08 & 0.09 & 0.10 & 0.02 & 0.00 & 0.02 & 0.01 & 0.00 & 0.00 \\ 
  2 & 9 Lowest & 0.31 & 0.04 & 0.08 & 0.17 & 0.29 & 0.27 & 0.17 & 0.45 & 0.07 & 0.00 & 0.11 \\
  3 & 49 Lowest & 0.04 & 0.02 & 0.03 & 0.01 & 0.00 & 0.15 & 0.00 & 0.10 & 0.03 & 0.00 & 0.02 \\
   \multicolumn{12}{l}{\textit{Number of Active Users}} \\
    1 & 4 Lowest & 0.11 & 0.20 & 0.08 & 0.09 & 0.10 & 0.02 & 0.00 & 0.02 & 0.01 & 0.00 & 0.00 \\ 
  2 & 9 Lowest & 0.53 & 0.01 & 0.08 & 0.15 & 0.10 & 0.23 & 0.17 & 0.45 & 0.08 & 0.00 & 0.11 \\
  3 & 49 Lowest & 0.06 & 0.02 & 0.04 & 0.01 & 0.00 & 0.15 & 0.00 & 0.10 & 0.03 & 0.00 & 0.02 \\
  \multicolumn{12}{l}{\textit{7-Day Lagged Standard Deviation}} \\
    1 & 4 Lowest & 0.12 & 0.43 & 0.08 & 0.09 & 0.10 & 0.07 & 0.00 & 0.02 & 0.01 & 0.00 & 0.00 \\
  2 & 9 Lowest & 0.28 & 0.04 & 0.06 & 0.10 & 0.33 & 0.11 & 0.09 & 0.14 & 0.06 & 0.00 & 0.00 \\
  3 & 49 Lowest & 0.21 & 0.02 & 0.06 & 0.02 & 0.01 & 0.26 & 0.00 & 0.21 & 0.11 & 0.00 & 0.00 \\
  \bottomrule
\end{tabular}
}
\caption*{\scriptsize The table shows the medians p-values of DQ-Tests sorted by covariate values in different time periods. The currencies are divided into two groups based on specific covariate values for each time period. The groups are constructed to best separate the sorted covariate values of the currencies. They are therefore separated at their steepest decay to obtain two groups with homogeneous covariate values where the top group in each period contains at least the top quintile but not more than the top sixtile of the data.The remaining details correspond to Table \ref{tab:median_pvals}.}
\end{table}

\begin{table}[htb]
\begin{minipage}{\linewidth}
\centering
\caption{Performance and Significance of CPA-tests with GRF-X Over Different Time Periods} 
\label{tab:CPA_p1_2_grf-x}
\resizebox{0.7\textwidth}{!}{
\begin{tabular}{lrrrrrrrrrr}
  \toprule
 GRF-X vs.: & GRF & QRF & QR & CAV & CAV\_ASY & GJR-GARCH & Hist & QRF-X & QR-X & GARCH-X \\ 
  \midrule
  \multicolumn{10}{l}{\textit{Share of GRF With Better Performance}}  \\
  Period 1 & 0.40 & 1.00 & 0.80 & 0.60 & 0.80 & 1.00 & 1.00 & 1.00 & 1.00 & 1.00 \\ 
  Period 2 & 0.64 & 1.00 & 0.86 & 0.86 & 0.93 & 0.93 & 1.00 & 1.00 & 1.00 & 1.00 \\ 
  Period 3 & 0.56 & 0.92 & 0.73 & 0.81 & 0.80 & 0.77 & 0.98 & 0.97 & 0.95 & 0.81 \\ 
  Period 4 & 0.61 & 0.93 & 0.71 & 0.80 & 0.80 & 0.70 & 0.95 & 0.94 & 0.95 & 0.75 \\ 
  Full Data & 0.61 & 0.92 & 0.70 & 0.80 & 0.78 & 0.63 & 0.96 & 0.92 & 0.96 & 0.75 \\ 
 \multicolumn{10}{l}{\textit{Share of GRF With Significantly Better Performance}}  \\
  Period 1 & 0.00 & 0.40 & 0.20 & 0.00 & 0.20 & 0.00 & 1.00 & 1.00 & 1.00 & 0.60 \\ 
  Period 2 & 0.07 & 0.71 & 0.21 & 0.29 & 0.50 & 0.29 & 0.79 & 0.93 & 0.93 & 0.71 \\ 
  Period 3 & 0.08 & 0.61 & 0.17 & 0.23 & 0.34 & 0.20 & 0.83 & 0.80 & 0.88 & 0.55 \\ 
  Period 4 & 0.06 & 0.54 & 0.16 & 0.23 & 0.29 & 0.19 & 0.75 & 0.76 & 0.80 & 0.53 \\ 
  Full Data & 0.08 & 0.53 & 0.13 & 0.24 & 0.28 & 0.17 & 0.76 & 0.72 & 0.76 & 0.53 \\  
   \bottomrule
\end{tabular}
}
\vspace{1cm}
\end{minipage}
\begin{subfigure}{0.48\textwidth}
    \centering
    \includegraphics[width=\textwidth]{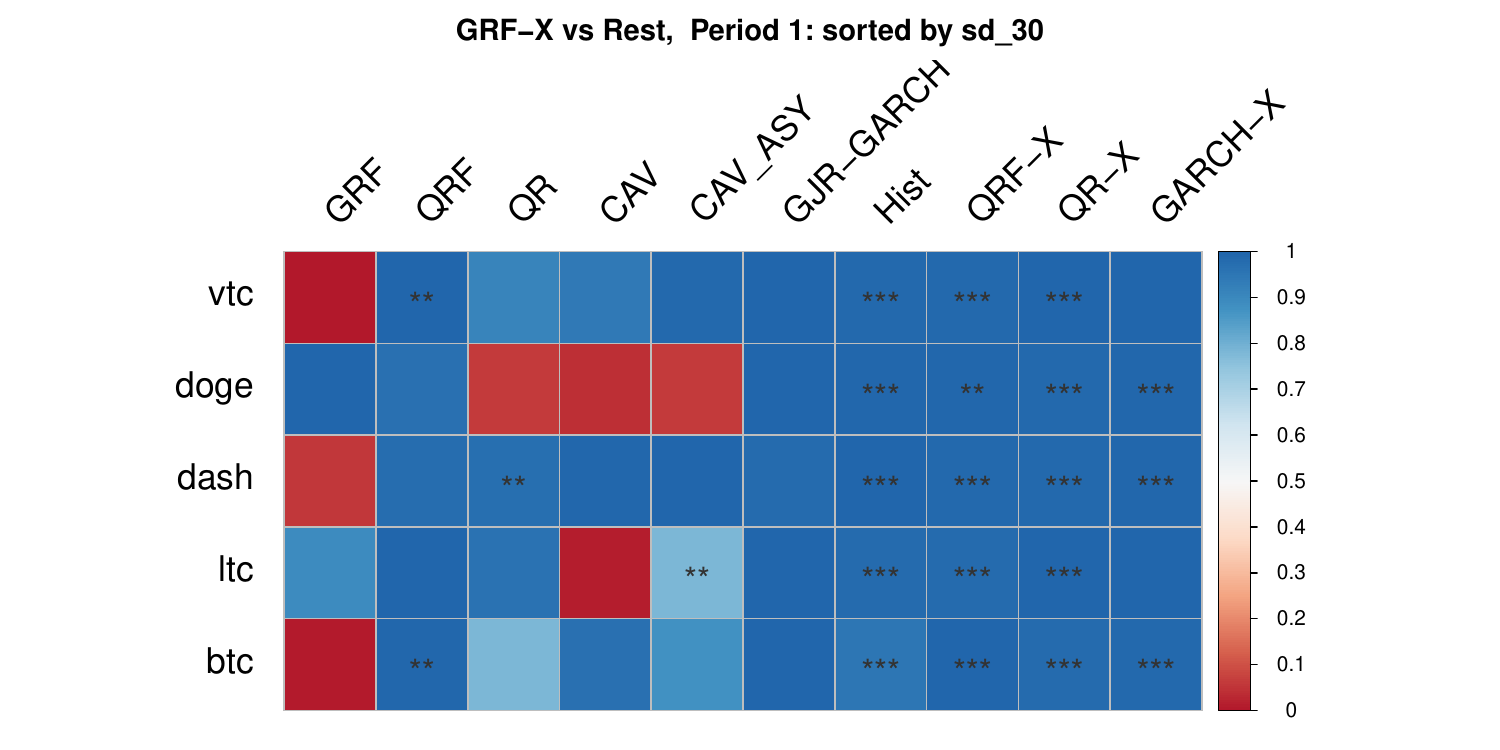}
\end{subfigure}
\begin{subfigure}{0.48\textwidth}
    \centering
    \includegraphics[width=\textwidth]{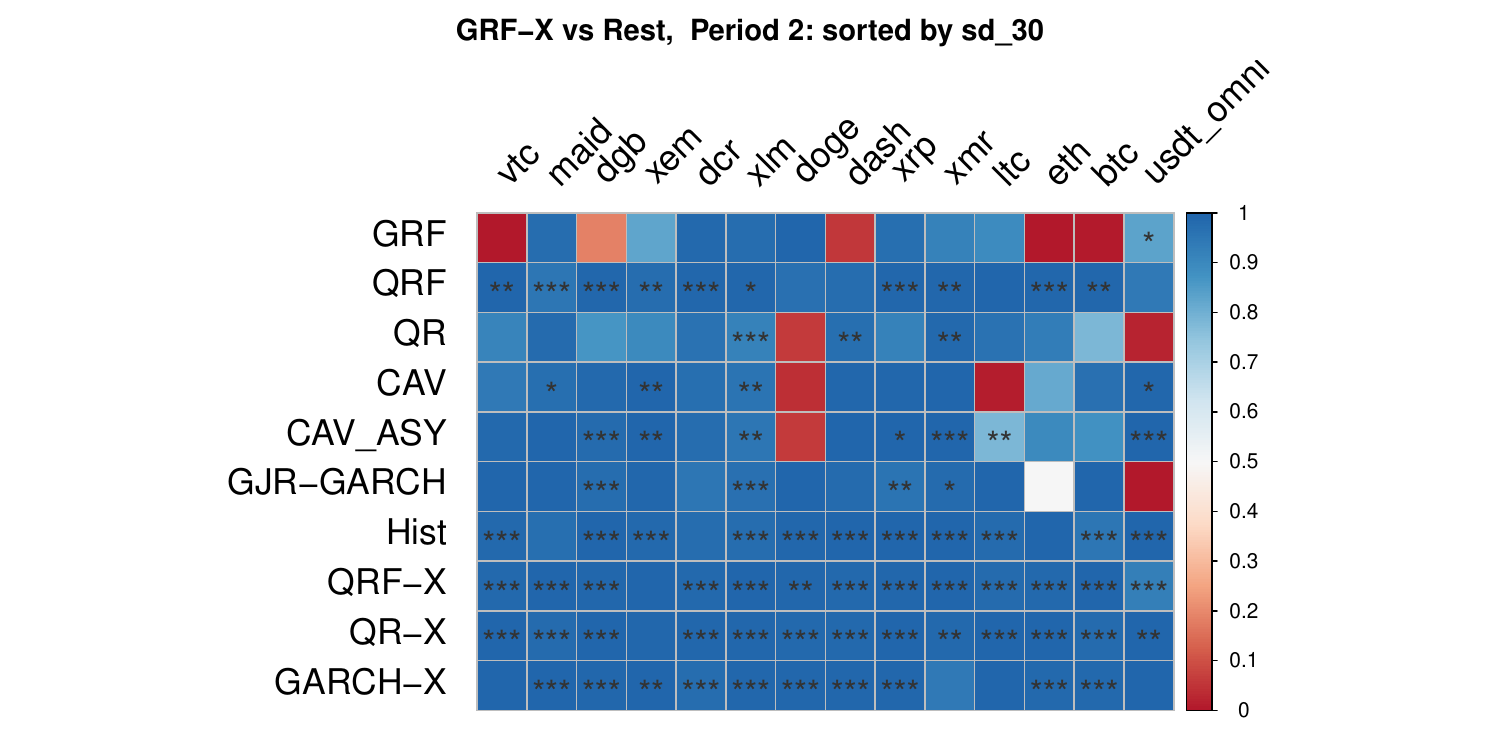}
\end{subfigure}
    \caption*{\scriptsize The top part shows summary values that are shares over all cryptocurrencies in the respective time period. It describes the number of times that GRF-X had a better performance (i.e. more than 50\% of predicted losses by the CPA test were smaller for the GRF-X) relative to all crytpocurrencies (in that period), and the number of times that GRF-X was significantly better (at least at a 10\% level) as judged by the CPA test over all crytpocurrencies (in that period). The bottom part shows the detailed results of CPA-tests with the color of each box indicating the performance of GRF-X. Blue signifies a performance of 1, meaning that GRF-X has a smaller predicted loss in 100\% of cases.  *, **, *** shows significance on a level of 10\%, 5\%, and 1\%. The values are ordered by 30 day lagged standard deviation from highest to lowest.}
\end{table}

\begin{figure}[htb]
    \centering
    \begin{subfigure}[b]{0.49\textwidth}
    \includegraphics[width=\textwidth]{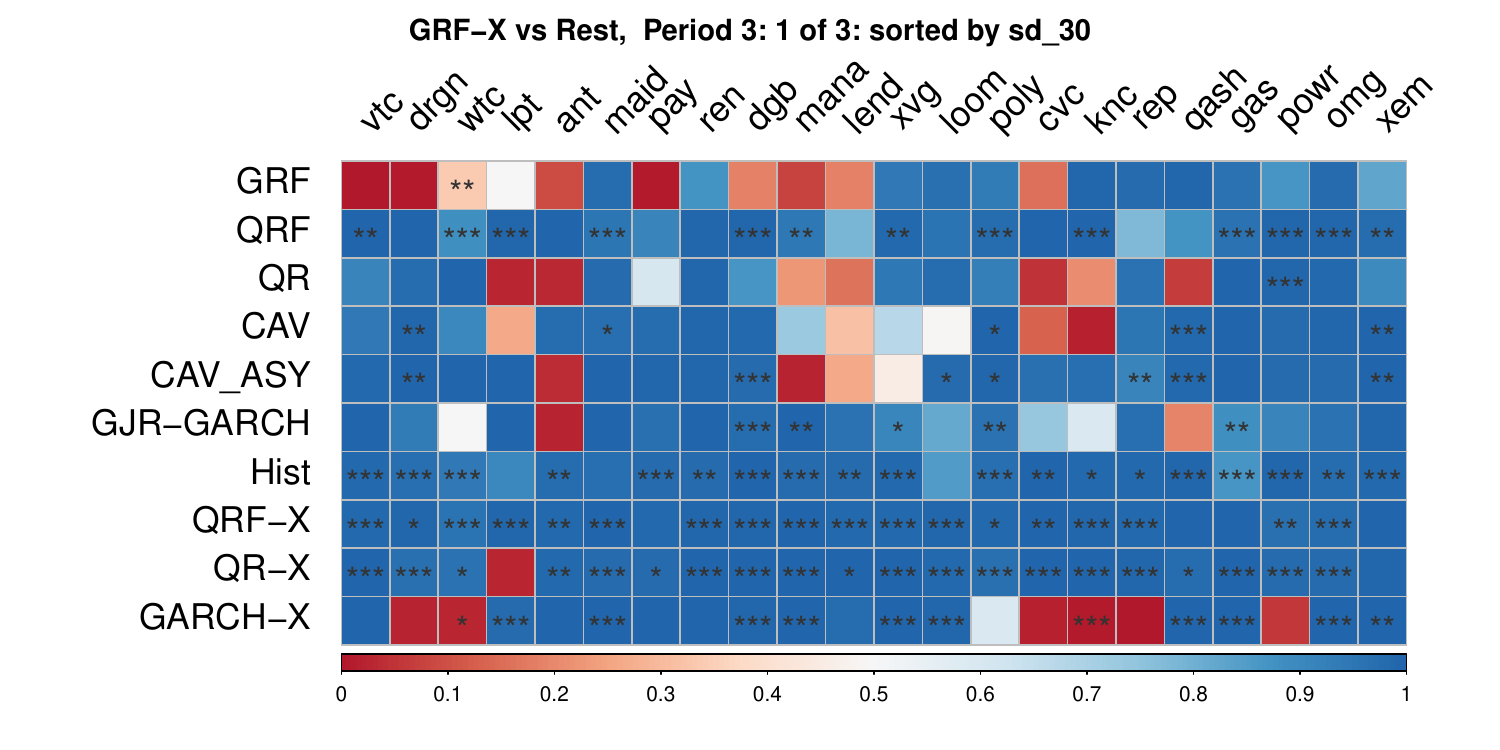}
    \end{subfigure}
    \hfill
    \begin{subfigure}[b]{0.49\textwidth}
    \includegraphics[width=\textwidth]{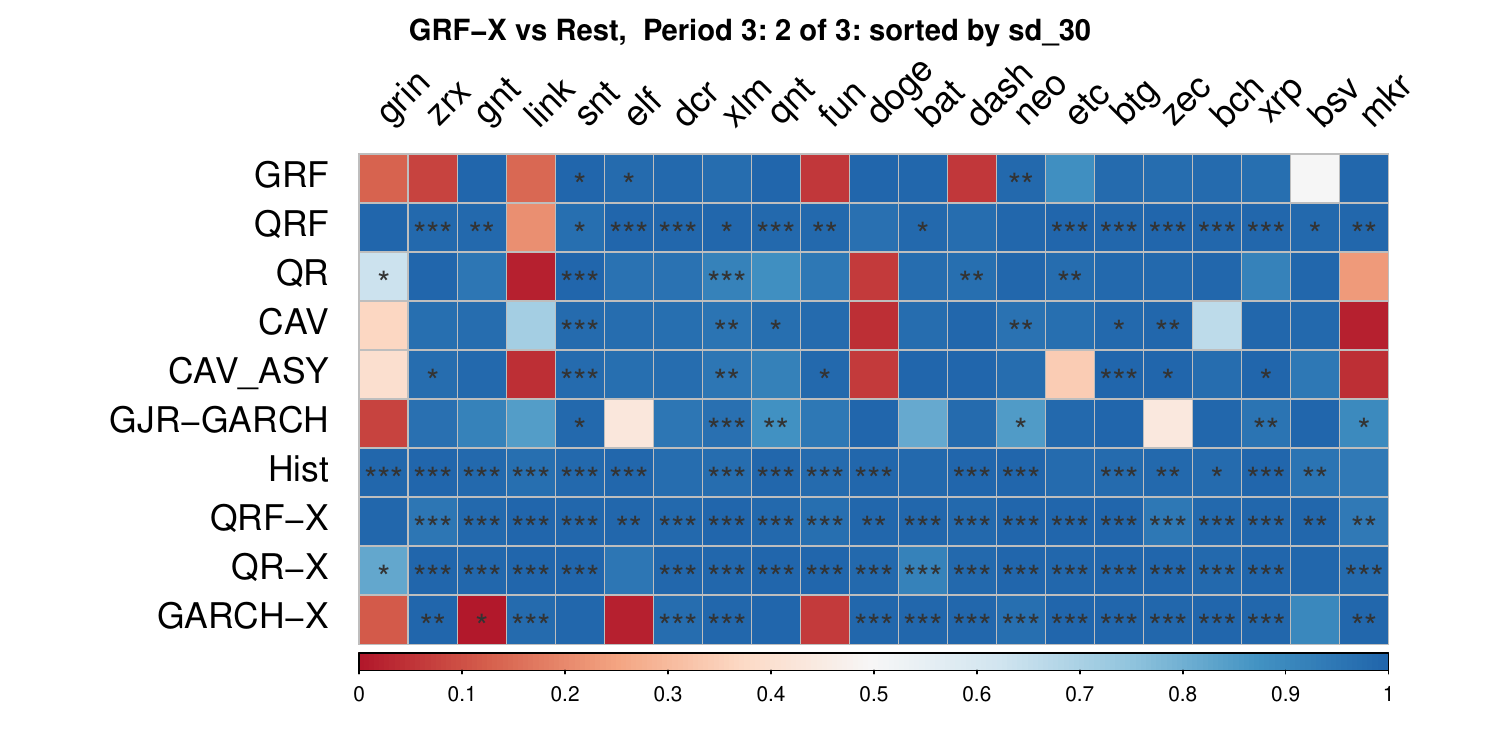}
    \end{subfigure}
        \begin{subfigure}[b]{0.49\textwidth}
    \includegraphics[width=\textwidth]{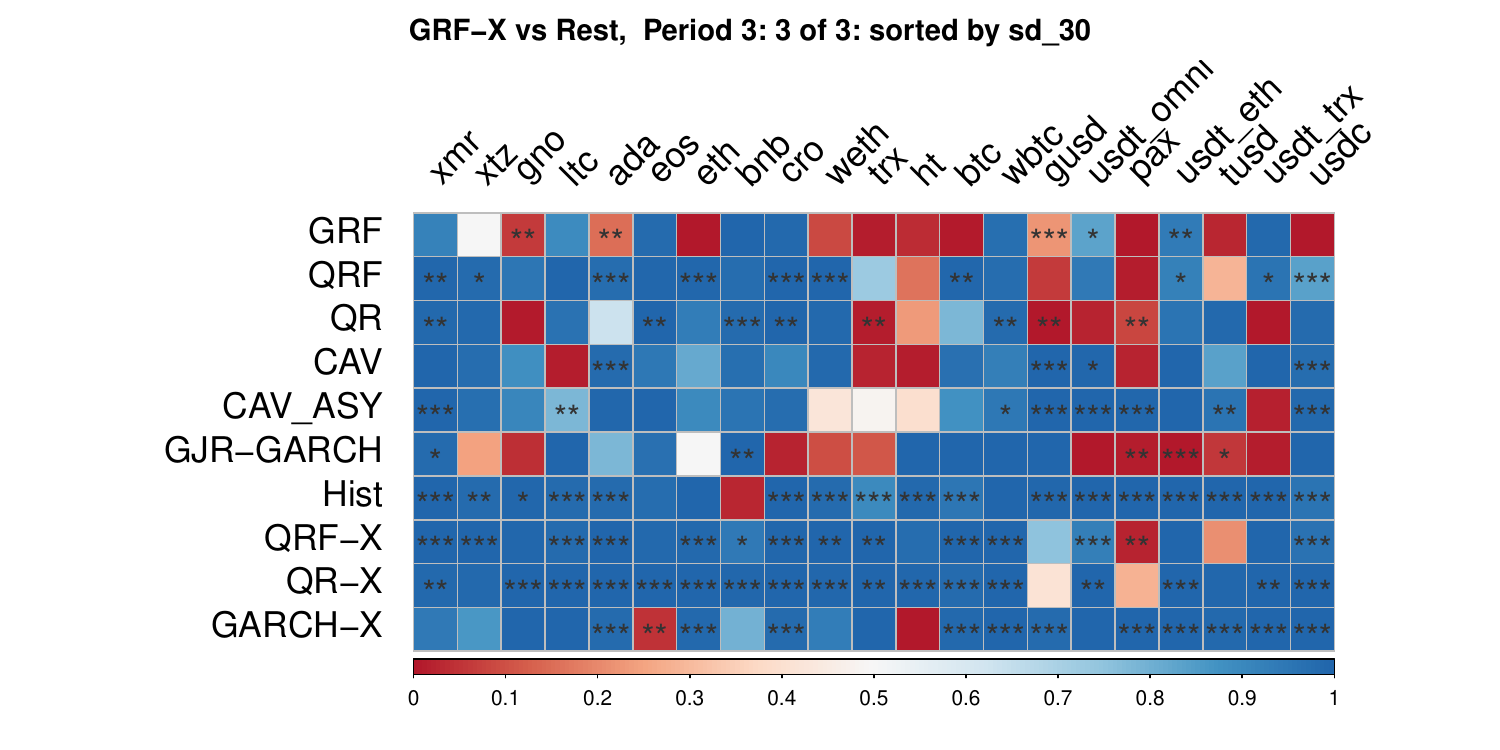}
    \end{subfigure}
    \caption[CPA tests for Period 3 for GRF-X]{Overview of results for CPA-tests of GRF-X vs. all other methods for cryptocurrencies in the third period ordered by 30 day lagged standard deviation from highest to lowest, left to right. The color of each box indicates the performance of GRF-X, with 1 indicating that GRF-X has a smaller predicted loss in 100\% of cases. *, **, *** indicate significance on a level of 10\%, 5\%, and 1\%.}
    \label{fig:CPA_p3_grf-x}
\end{figure}


\begin{table}[ht]
\centering
\caption{Difference Between Covariates of Cryptos Where GRF-X is Better vs. Worse} 
\label{tab:crypto_dif_rel_grf-x}
\resizebox{0.5\textwidth}{!}{
\begin{tabular}{rrrrrr}
  \toprule
 & Period 1 & Period 2 & Period 3 & Period 4 & Full Data \\ 
  \midrule
Ret & 0.83 & 1.64 & -1.16 & -0.34 & 0.16 \\ 
  Active\_Users & 0.30 & 0.58 & 2.68 & 2.99 & 2.18 \\ 
  Total\_Users & 0.48 & 0.54 & 0.14 & 0.12 & 0.14 \\ 
  Total\_Users\_USD100 & 0.12 & 0.21 & 0.08 & 0.11 & 0.18 \\ 
  Total\_Users\_USD\_10 & 0.18 & 0.32 & 0.12 & 0.15 & 0.20 \\ 
  SER & 0.74 & 1.45 & 0.38 & 0.29 & 0.42 \\ 
  Transactions & 0.38 & 0.12 & 2.67 & 2.33 & 1.97 \\ 
  Velocity & 4.11 & 9.26 & 1.40 & 1.32 & 1.02 \\ 
  sd\_3 & 0.91 & 0.94 & 0.86 & 0.81 & 0.99 \\ 
  sd\_7 & 0.92 & 0.96 & 0.87 & 0.82 & 0.98 \\ 
  sd\_30 & 0.93 & 0.98 & 0.87 & 0.82 & 0.99 \\ 
  sd\_60 & 0.92 & 0.98 & 0.87 & 0.84 & 1.00 \\ 
   \bottomrule
\end{tabular}
}
\caption*{\scriptsize The table shows shares of groups of cryptocurrencies where at least two of CAV, QR, and GJR-GARCH have better CPA-performance than GRF-X divided by the remaining rest. Raw values before division are mean values over all cryptocurrencies for the median of each covariate in the respective time period.}
\end{table}


\begin{figure}[htb]
    \centering
    \begin{minipage}{0.5\textwidth}
        \centering
        \includegraphics[width=\linewidth]{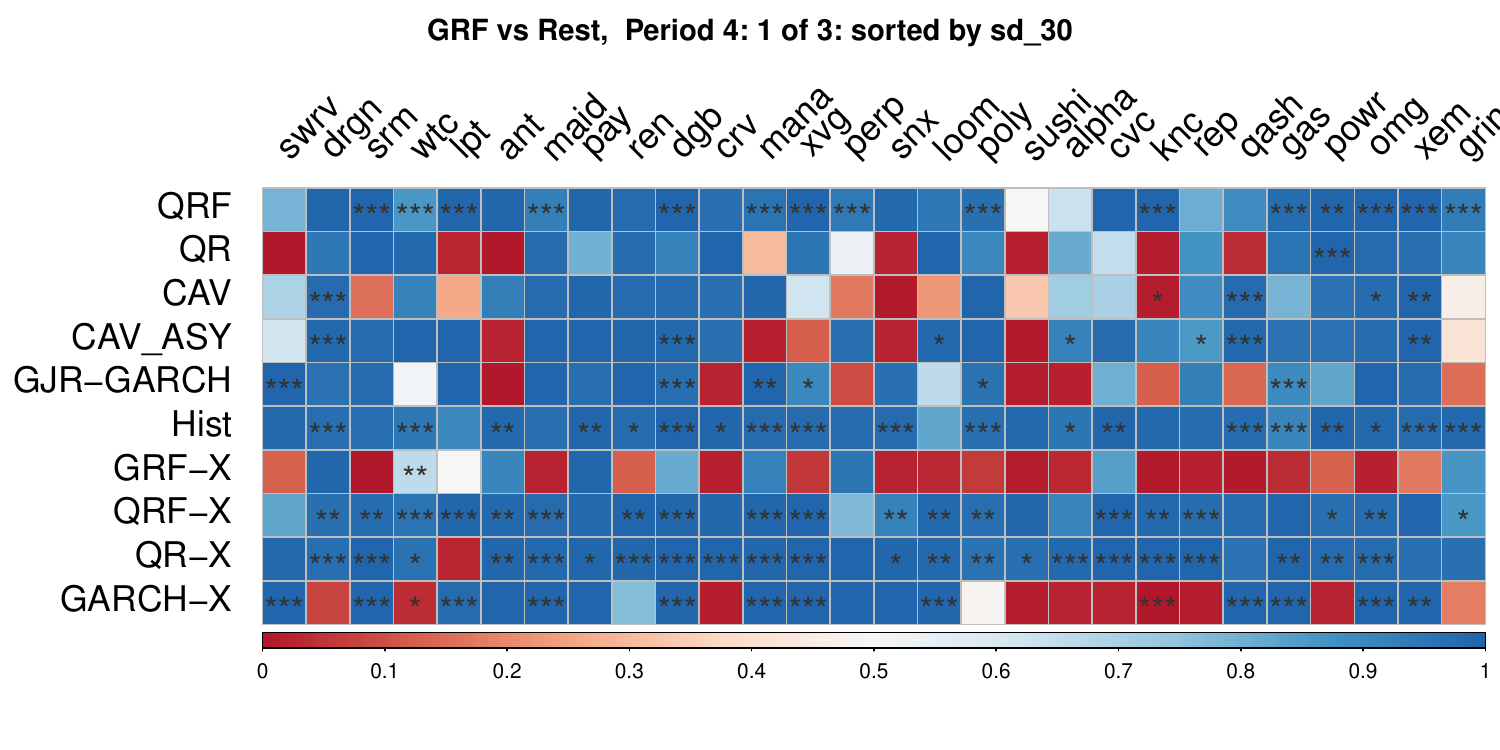}
        \label{fig:mein_bild1}
    \end{minipage}\hfill
    \begin{minipage}{0.5\textwidth}
        \centering
        \includegraphics[width=\linewidth]{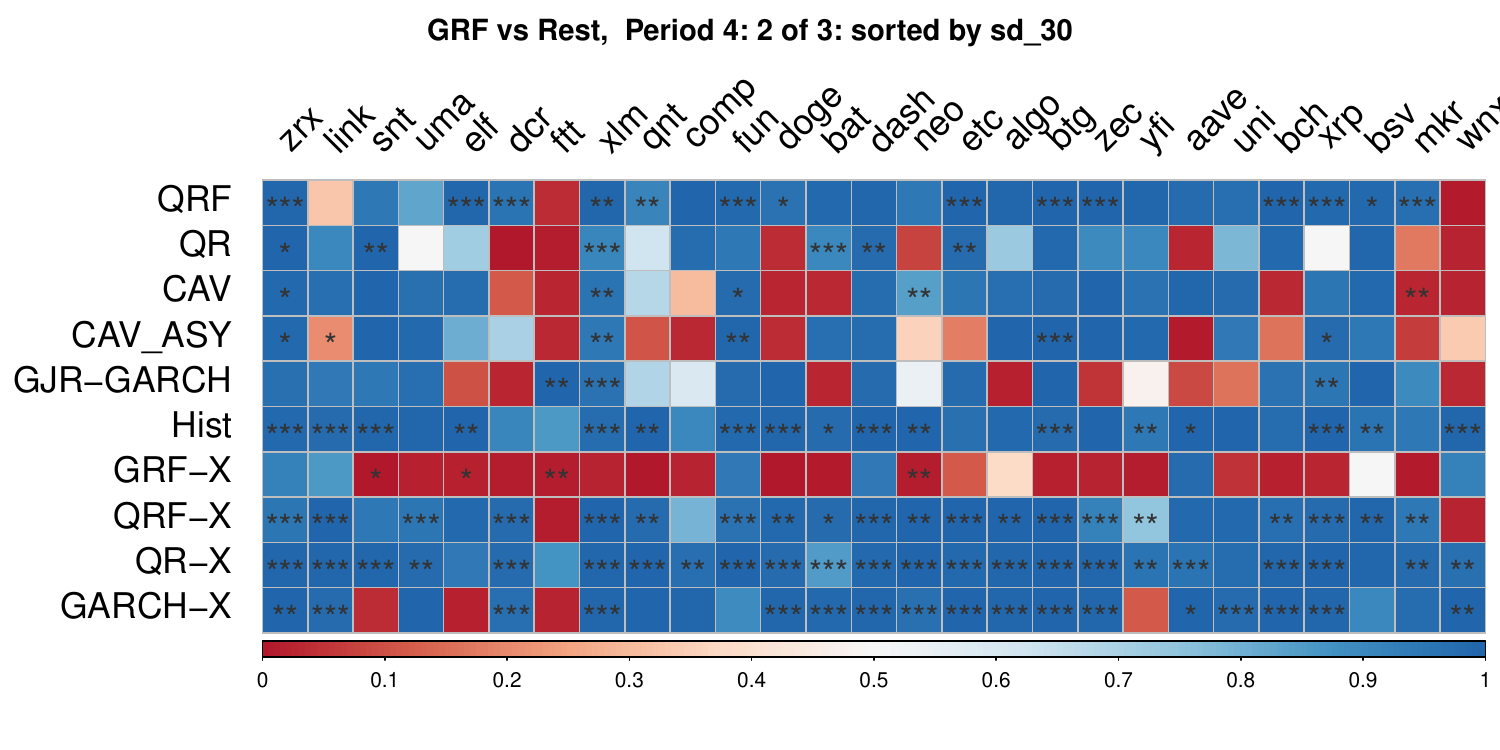}
        \label{fig:mein_bild2}
    \end{minipage}
    \label{fig: Results, CPA Period 4}
    \begin{minipage}{0.5\textwidth}
        \centering
        \includegraphics[width=\linewidth]{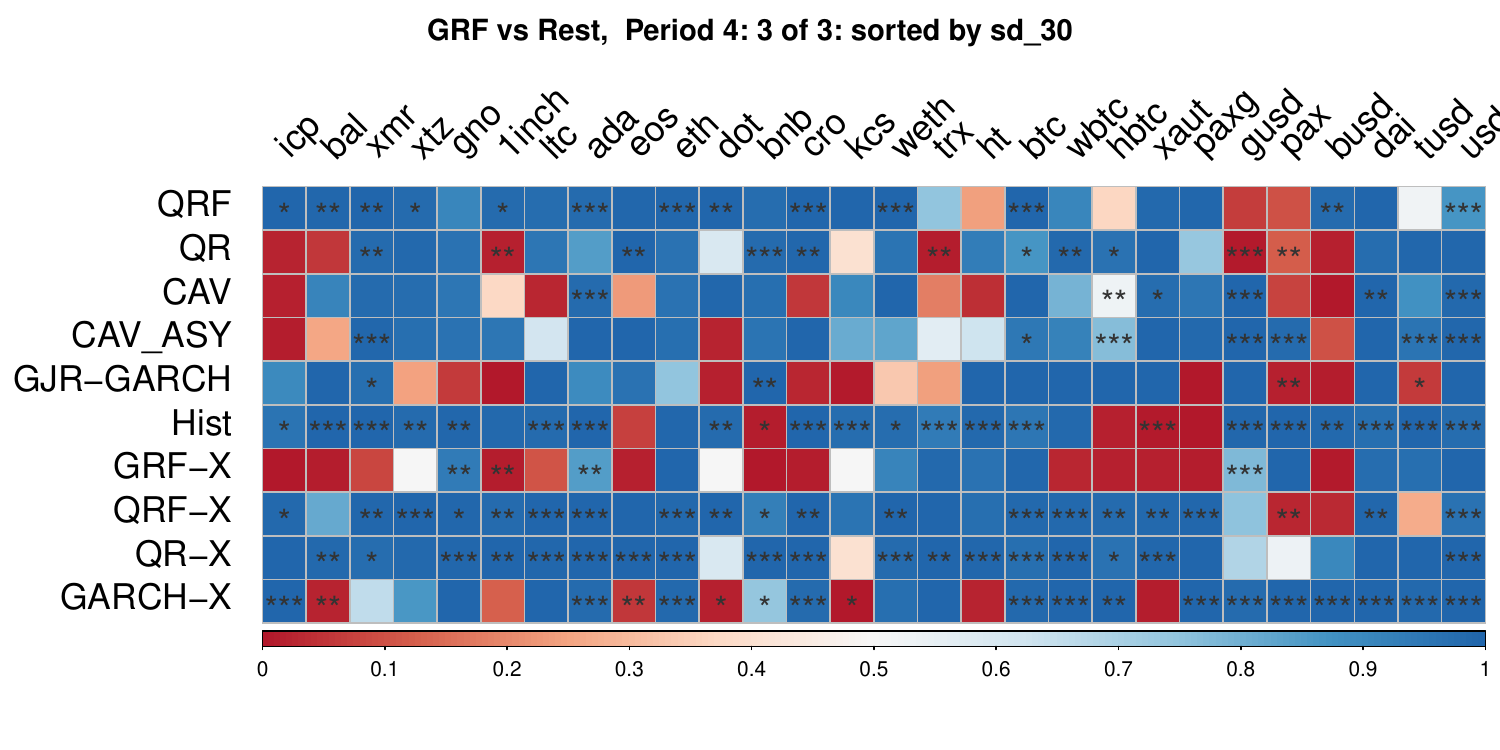}
        \label{fig:mein_bild1}
    \end{minipage}\hfill
\caption{Overview of results for CPA-tests of GRF vs. all other methods for cryptocurrencies
in the fourth period ordered by 30 day lagged standard deviation from highest to lowest, left to
right. The color of each box indicates the performance of GRF, with 1 (blue) indicating that
GRF has a smaller predicted loss in 100\% of cases. *, **, *** indicate signifcance on a level of
10\%, 5\%, and 1\%.}
\end{figure}

\begin{figure}[htb]
    \centering
    \begin{subfigure}[b]{0.49\textwidth}
     \includegraphics[width=0.8\textwidth]{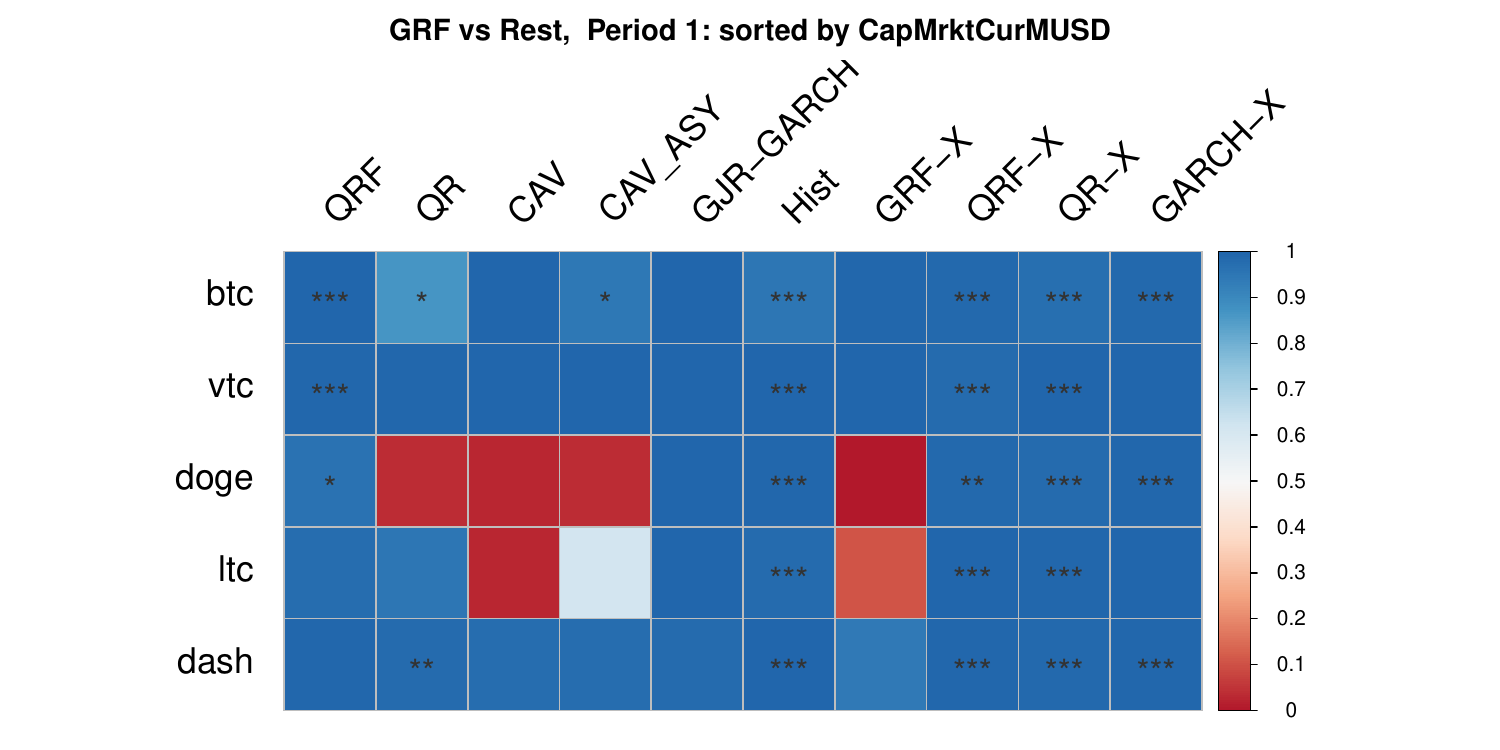}
    \end{subfigure}
    \hfill
    \centering
    \begin{subfigure}[b]{0.49\textwidth}
    \includegraphics[width=0.8\textwidth]{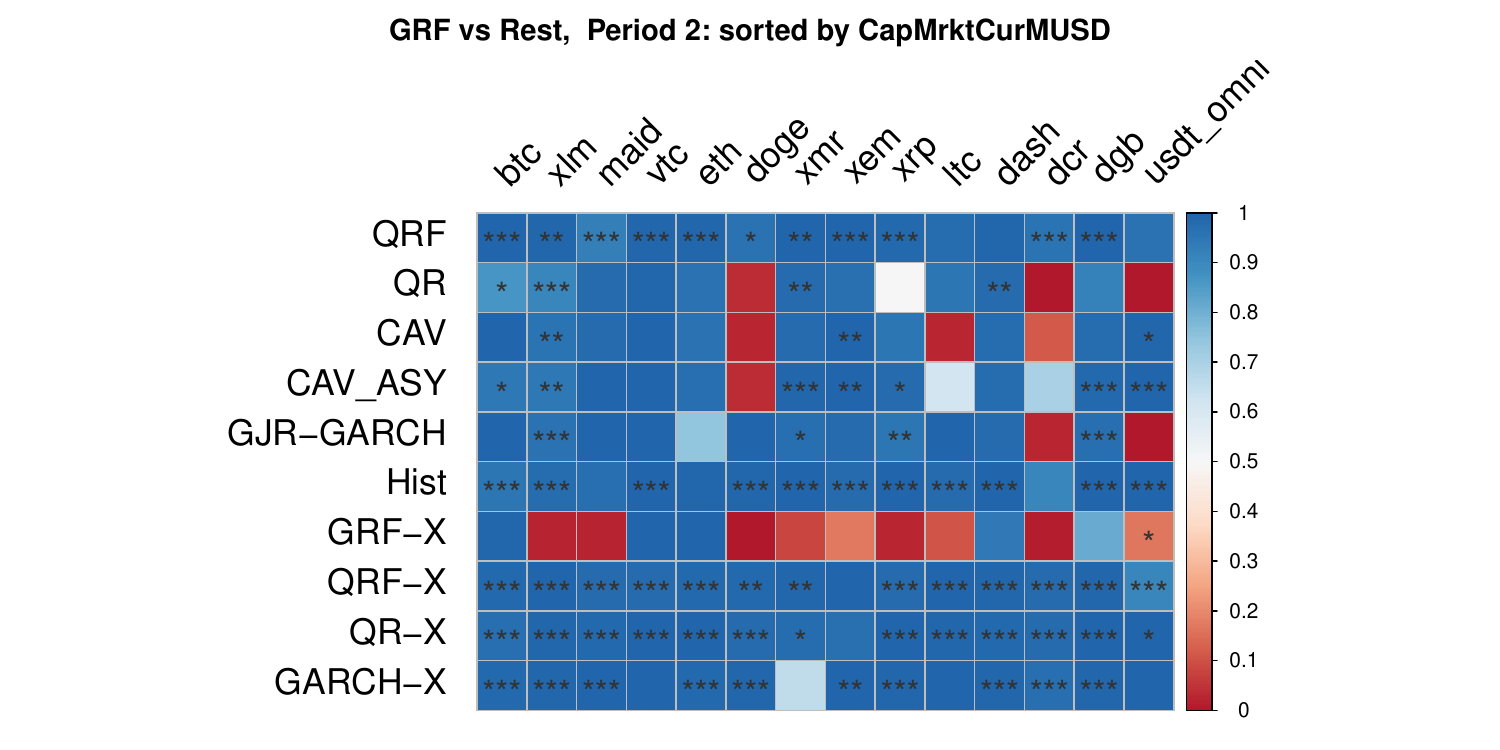}
    \end{subfigure}
    \caption{Overview of results for CPA-tests of GRF vs. all other methods for cryptocurrencies in the first and second period ordered by market cap from highest to lowest. The color of each box indicates the performance of GRF, with 1 indicating that GRF has a smaller predicted loss in 100\% of cases.  *, **, *** indicate significance on a level of 10\%, 5\%, and 1\%.}
    \label{fig:CPA_p1_cap}
\end{figure}

\begin{figure}[htb]
    \centering
    \begin{subfigure}[b]{0.49\textwidth}
    \includegraphics[width=\textwidth]{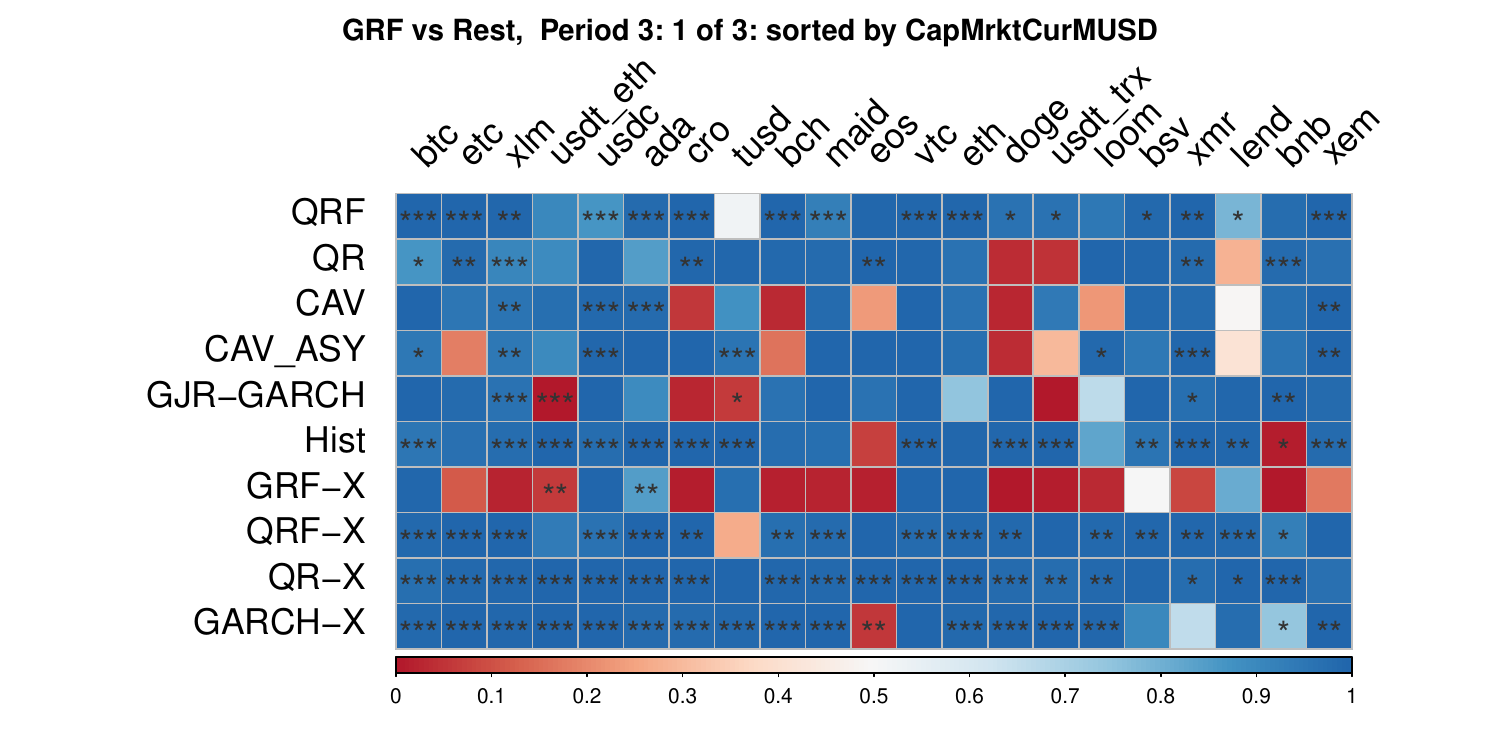}
    \end{subfigure}
    \hfill
    \begin{subfigure}[b]{0.49\textwidth}
    \includegraphics[width=\textwidth]{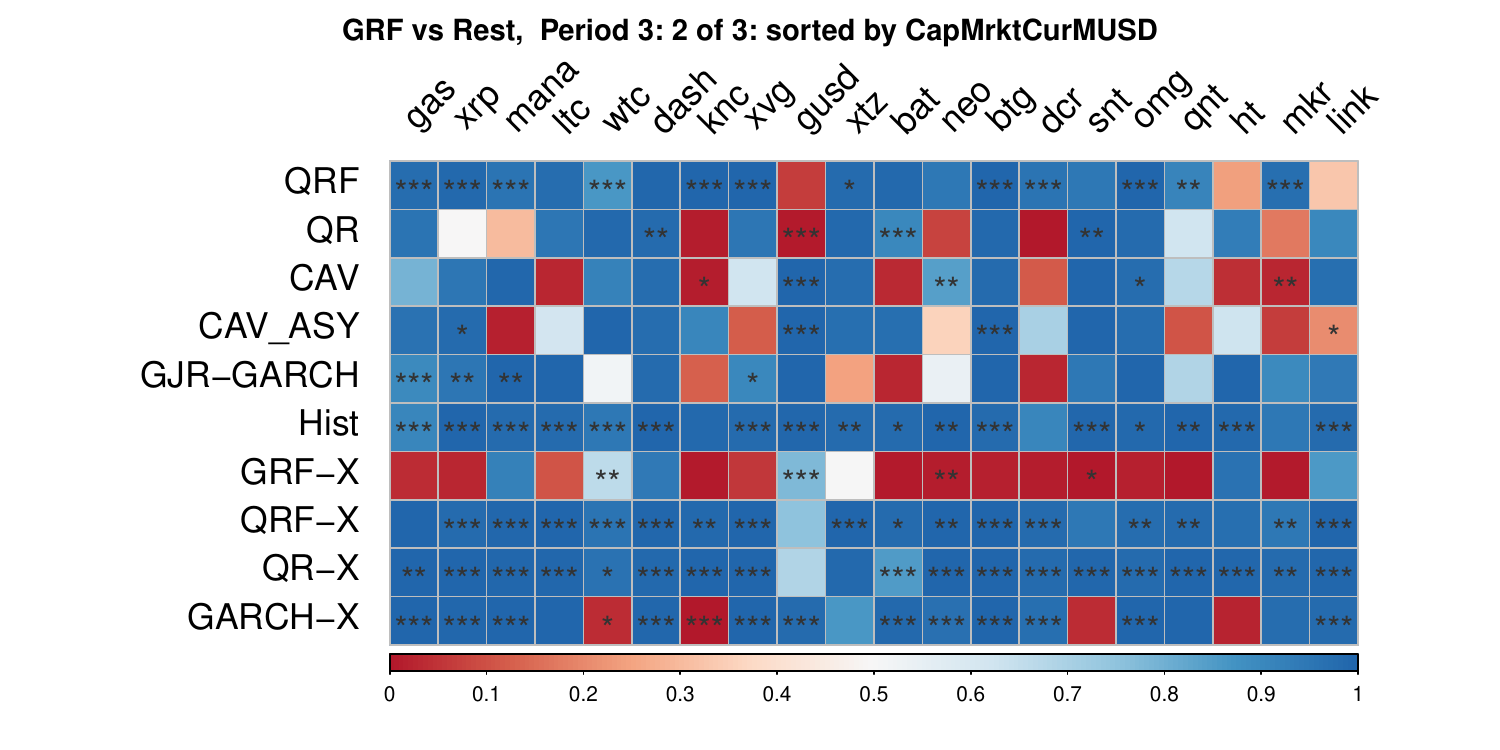}
    \end{subfigure}
        \begin{subfigure}[b]{0.49\textwidth}
    \includegraphics[width=\textwidth]{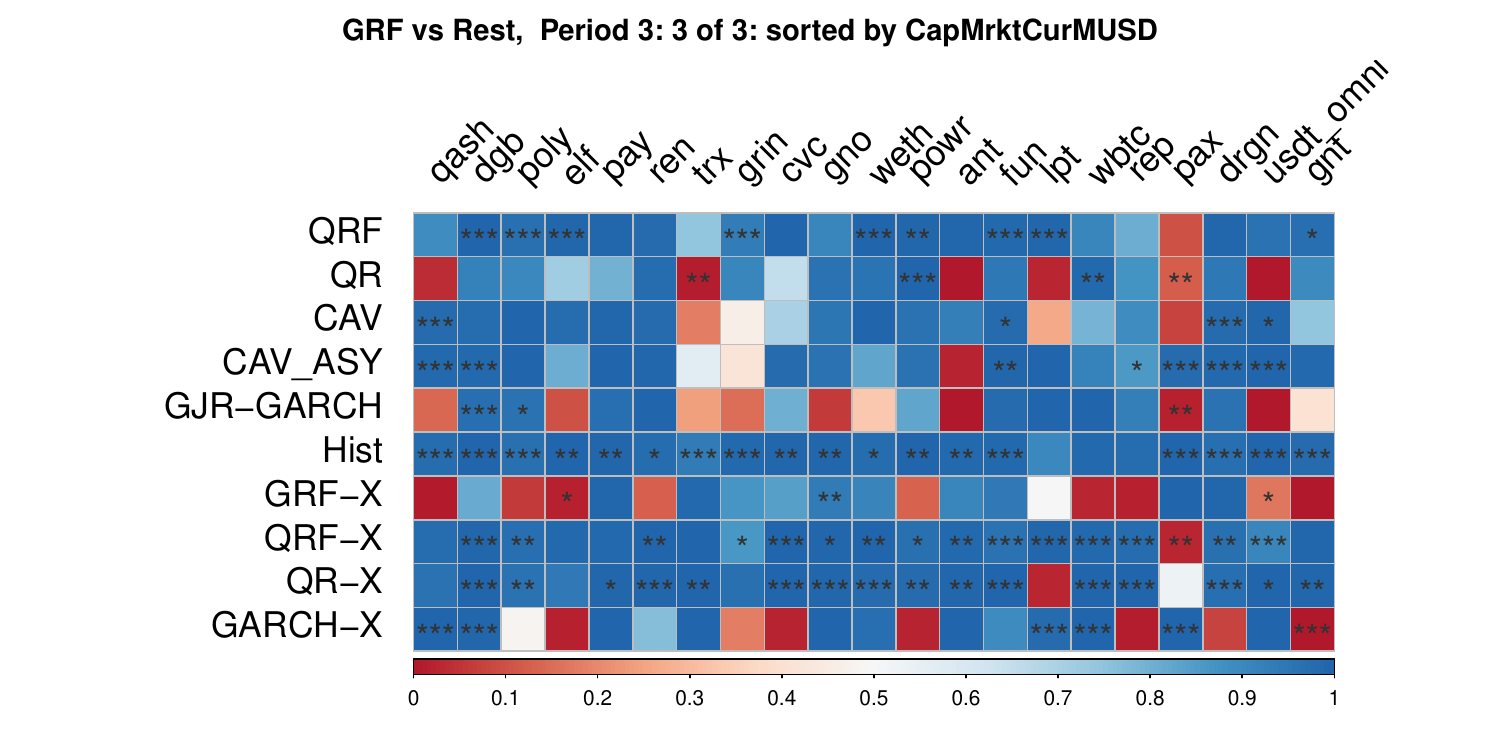}
    \end{subfigure}
    \caption{Overview of results for CPA-tests of GRF vs. all other methods for cryptocurrencies in the third period ordered by market cap from highest to lowest, left to right. The color of each box indicates the performance of GRF, with 1 indicating that GRF has a smaller predicted loss in 100\% of cases. *, **, *** indicate significance on a level of 10\%, 5\%, and 1\%.}
    \label{fig:CPA_p3_cap}
\end{figure}

\begin{figure}[htb]
    \centering
    \begin{subfigure}[b]{0.49\textwidth}
    \includegraphics[width=\textwidth]{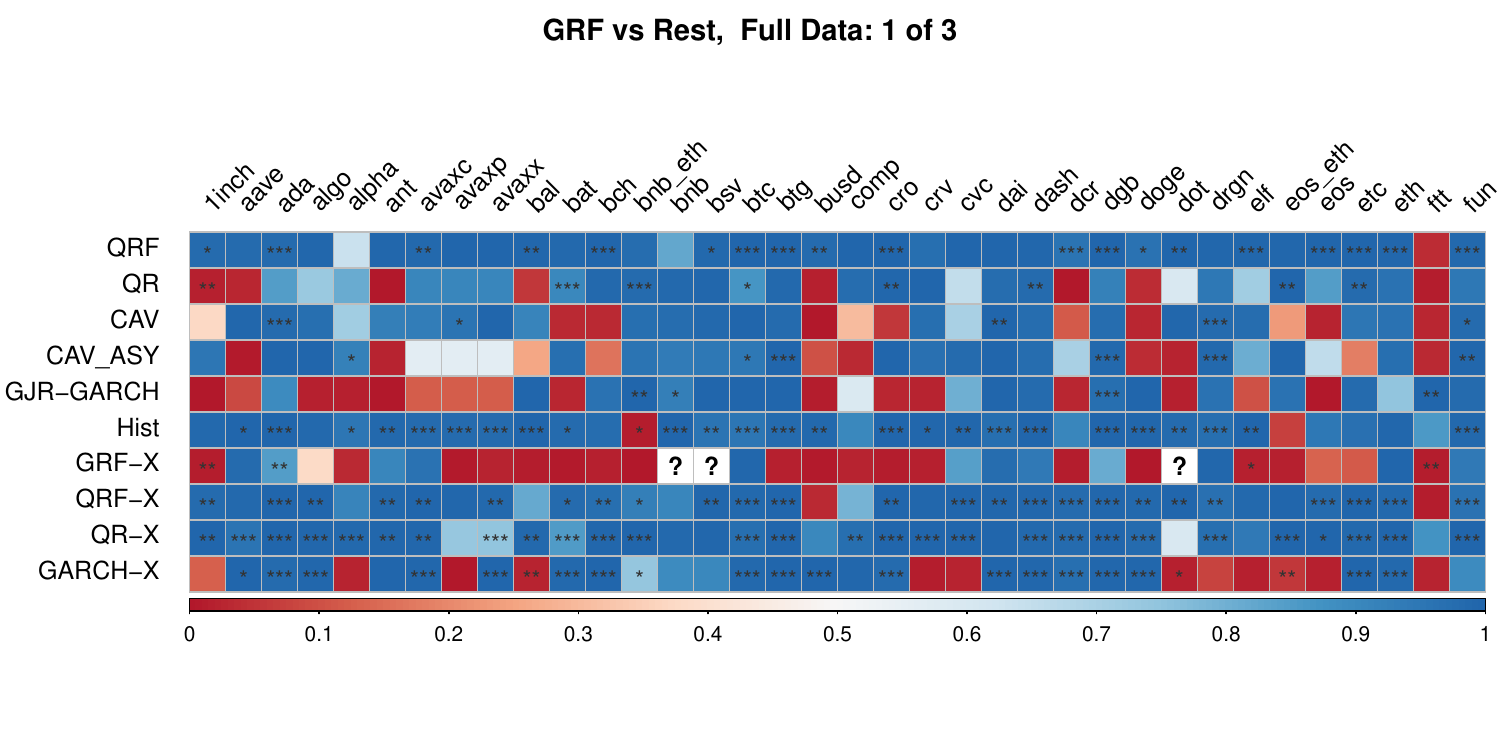}
    \end{subfigure}
    \hfill
    \begin{subfigure}[b]{0.49\textwidth}
    \includegraphics[width=\textwidth]{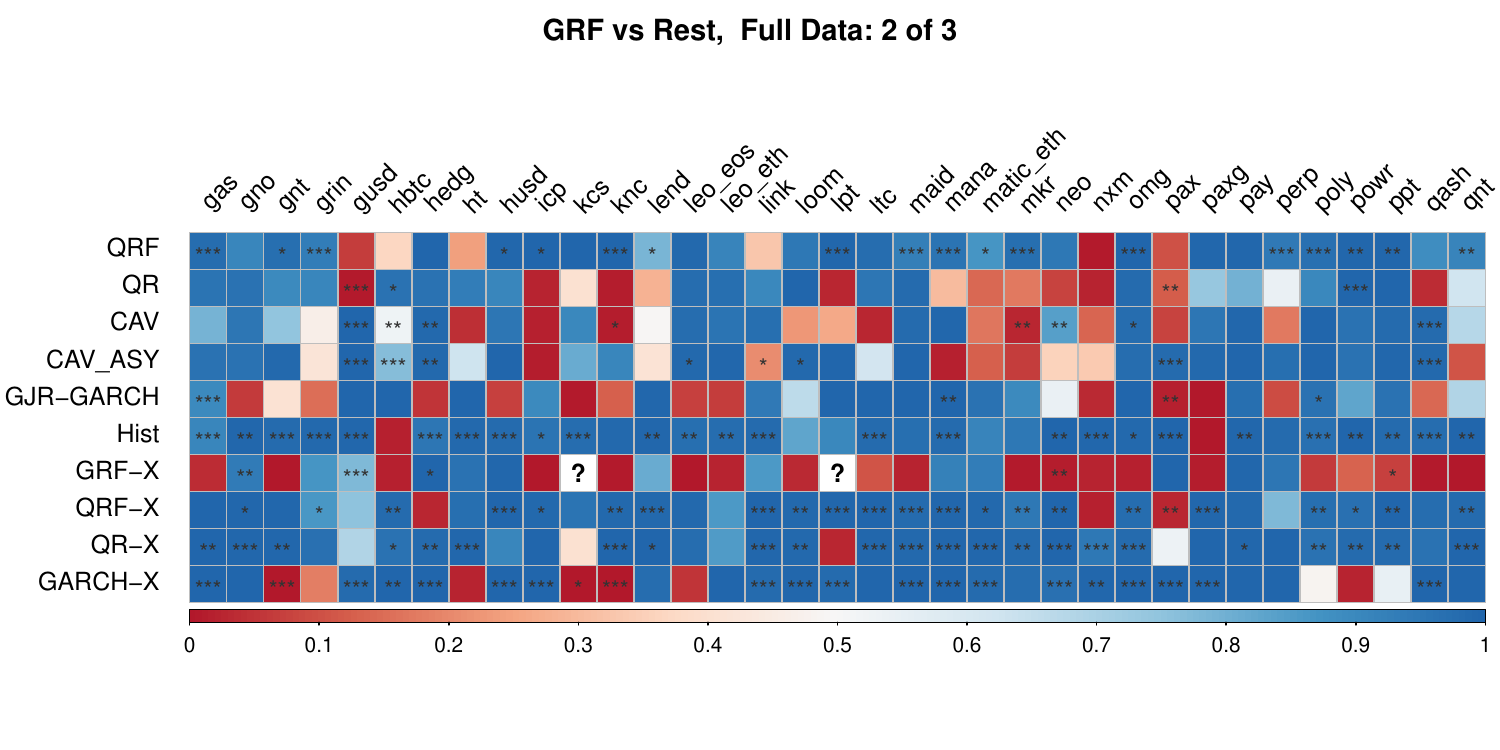}
    \end{subfigure}
        \begin{subfigure}[b]{0.49\textwidth}
    \includegraphics[width=\textwidth]{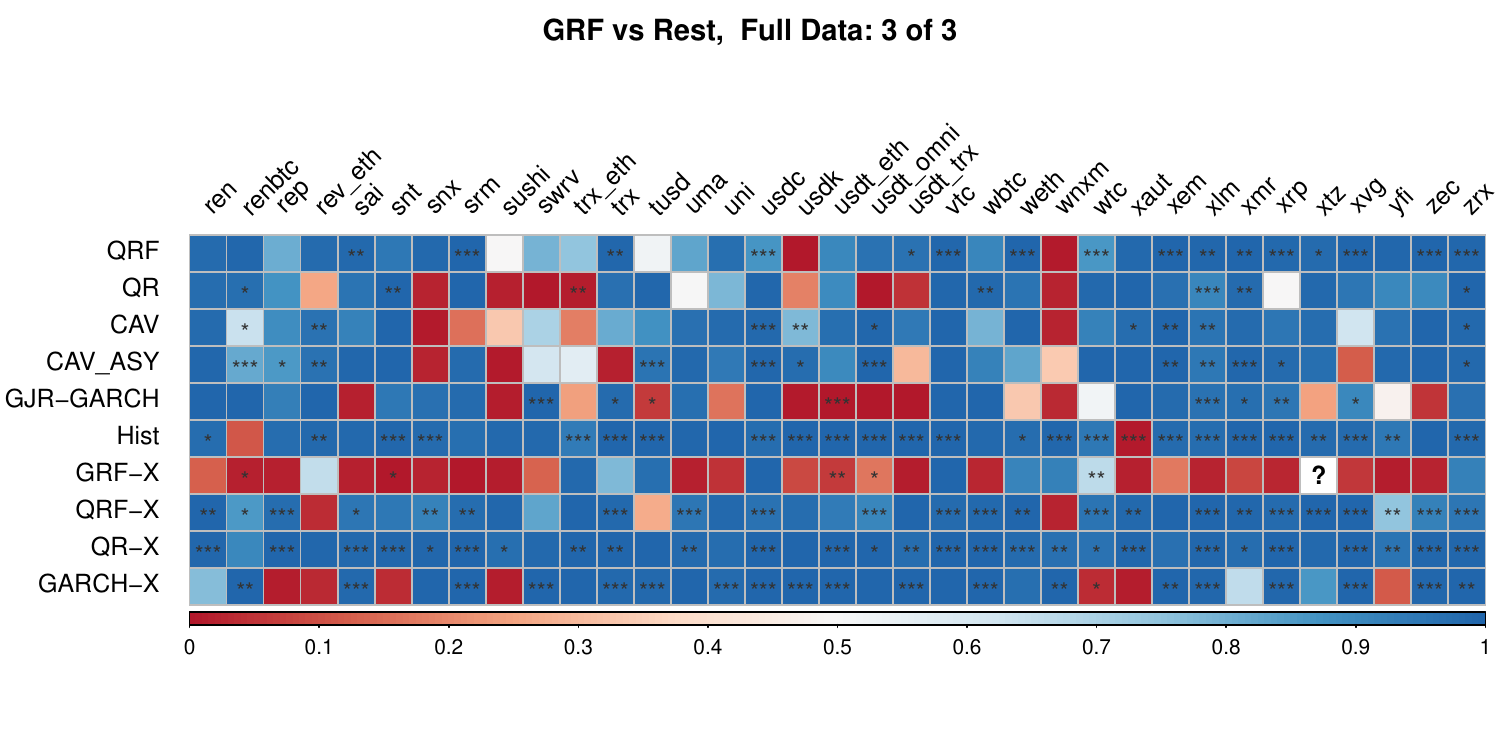}
    \end{subfigure}
    \caption{Overview of results for CPA-tests of GRF vs. all other methods for cryptocurrencies for the full data set ordered alphabetically. The color of each box indicates the performance of GRF, with 1 indicating that GRF has a smaller predicted loss in 100\% of cases. *, **, *** indicate significance on a level of 10\%, 5\%, and 1\%.}
    \label{fig:CPA_full_cpa_base}
\end{figure}

\begin{figure}[htb]
    \centering
    \includegraphics[width=\textwidth]{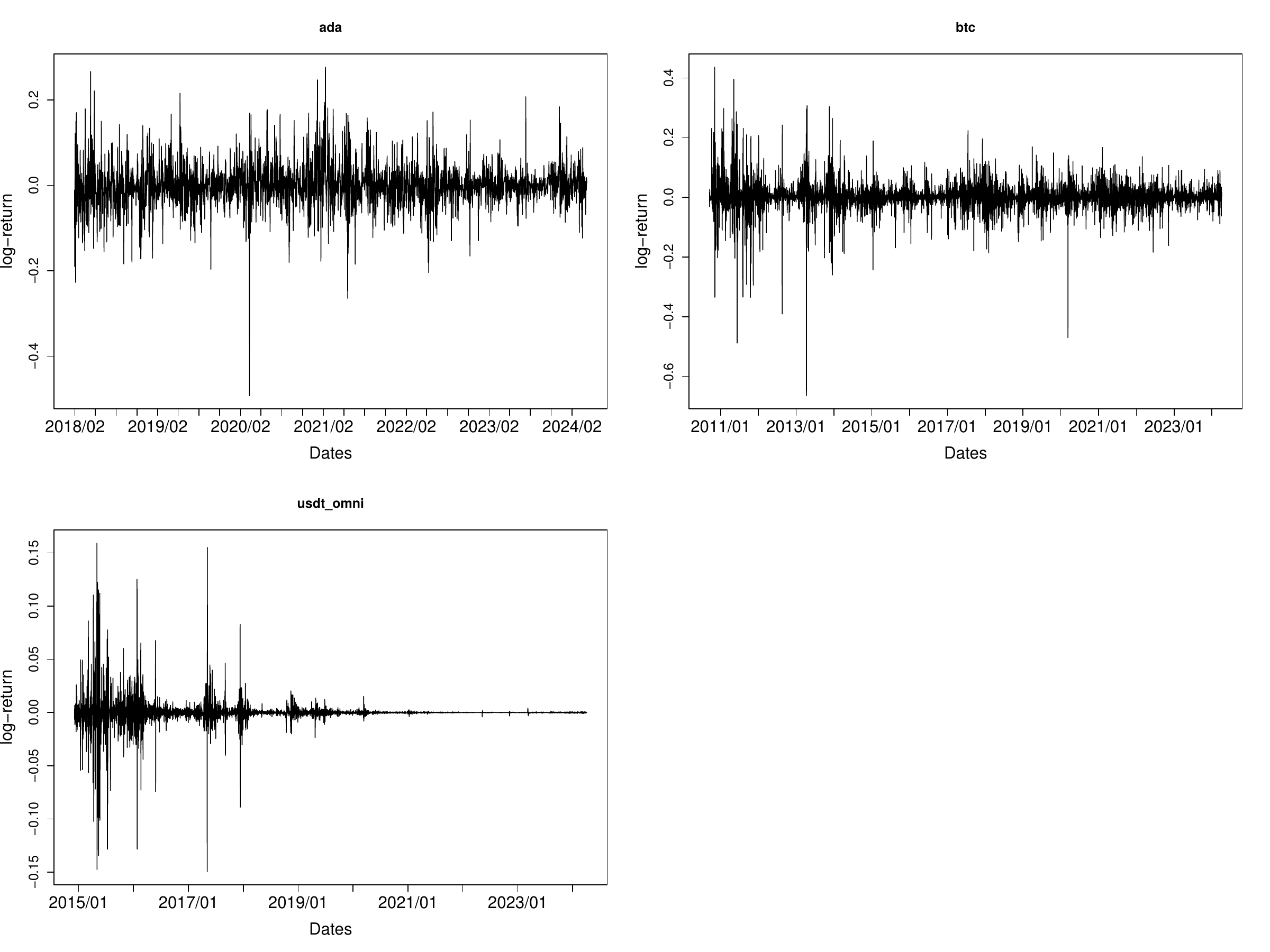}
    \caption{Log-returns of the specific cryptocurrencies analyzed in Subsection \ref{sec:crypto_results_case_study}.}
    \label{fig:log_returns_cryptos}
\end{figure}

\section{Data and Code Availability}
The code to replicate the results of this paper is available at https://github.com/KITmetricslab/crypto-VaR-predictions. The cryptocurrency data can be downloaded from https://coinmetrics.io.
  
\end{document}